\documentclass[aps,prb,twocolumn,amsmath,amssymb,superscriptaddress,floatfix,eqsecnum,showpacs,nofootinbib]{revtex4}
\usepackage{amsmath}
\usepackage{graphicx}
\usepackage{latexsym}
\usepackage{amsmath,amssymb}
\usepackage{bm} 
\usepackage{color}
\usepackage{stackrel}
\usepackage{accents}

\DeclareMathOperator{\sgn}{sgn}

\newcommand{\ord}[1]{\bm{\mathit{O}}\left(#1\right)}
\newcommand{\ordb}[1]{\bm{\mathit{O}}(#1)}

\newcommand{\vex}[1]{\bm{\mathrm{#1}}}

\newcommand{\sss}[1]{\scriptscriptstyle{#1}}
\newcommand{\msf}[1]{\mathsf{#1}}

\newcommand{\ek}{\tilde{\varepsilon}}

\newcommand{\tr}{\mathsf{Tr}}

\newcommand{\ket}[1]{\left| {#1} \right\rangle}
\newcommand{\bra}[1]{\left\langle {#1} \right|}

\newcommand{\pup}[1]{\scriptscriptstyle{({#1})}}

\newcommand{\ls}{\mathfrak{L}}

\newcommand{\bsub}{\begin{subequations}}
\newcommand{\esub}{\end{subequations}}
\newcommand{\ts}[1]{{\textstyle{{#1}}}}

\newcommand{\e}{\varepsilon}
\newcommand{\uoq}{u_0}
\newcommand{\uoqpm}{u_0^{\scriptscriptstyle{(\pm)}}}
\newcommand{\uoqa}[1]{u_0^{\scriptscriptstyle{({#1})}}}

\newcommand{\uqa}[1]{u^{\scriptscriptstyle{({#1})}}}
\newcommand{\uqpm}{u^{\scriptscriptstyle{(\pm)}}}

\newcommand{\emax}{\varepsilon_\Lambda}
\newcommand{\Dasy}{\Delta_{\sss{\infty}}}
\newcommand{\masy}{\mu_{\sss{\infty}}}
\newcommand{\Easy}{E_{\sss{\infty}}}
\newcommand{\Do}{\Delta_{\sss{0}}}

\newcommand{\Dcoh}{\Delta_{\msf{Coh}}}
\newcommand{\Dmr}{\Delta_{\msf{MR}}}
\newcommand{\Dqcp}{\Delta_{\msf{QCP}}}
\newcommand{\ef}{\mathcal{E}_F}
\newcommand{\inte}{\int_0^{2(\Lambda + \mu)}\!\!\!\!\!\!\!\!\!\!\!\! d \e \,\,\,\,}
\newcommand{\intei}{\int_0^{2[\Lambda + \mui]}\!\!\!\!\!\!\!\!\!\!\!\!\! d \e \,\,\,\,\,\,}
\newcommand{\intem}{\int_0^{\emax} d \e\,}
\newcommand{\Vmin}{V_{\msf{min}}}
\newcommand{\qp}{\mathcal{Q}_{2 N}}

\newcommand{\q}{\mathcal{Q}}

\newcommand{\W}{\mathcal{W}}
\newcommand{\G}{\mathcal{G}}
\newcommand{\kc}{\msf{k}}

\newcommand{\Di}{\Delta_{\sss{0}}^{{\sss{(}}i{\sss{)}}}}
\newcommand{\mui}{\mu_{\sss{0}}^{{\sss{(}}i{\sss{)}}}}
\newcommand{\ffi}{f^{{\sss{(}}i{\sss{)}}}}
\newcommand{\Ei}{E^{{\sss{(}}i{\sss{)}}}}
\newcommand{\Df}{\Delta_{\sss{0}}^{{\sss{(}}f{\sss{)}}}}
\newcommand{\muf}{\mu_{\sss{0}}^{{\sss{(}}f{\sss{)}}}}
\newcommand{\Gi}{G_{{i}}}
\newcommand{\Gf}{G_{{f}}}
\newcommand{\gi}{g_{{i}}}
\newcommand{\gf}{g_{{f}}}
\newcommand{\Eone}{\mathcal{E}_1^{{\sss{(}}i{\sss{)}}}}
\newcommand{\Etwo}{\mathcal{E}_2^{{\sss{(}}i{\sss{)}}}}
\newcommand{\Eot}{\mathcal{E}_{1,2}^{{\sss{(}}i{\sss{)}}}}

\newcommand{\dos}{\nu_0}

\newcommand{\bcpm}{\beta_{\msf{c}}^{\pup{\pm}}}
\newcommand{\bcp}{\beta_{\msf{c}}^{\pup{+}}}
\newcommand{\bcm}{\beta_{\msf{c}}^{\pup{-}}}
\newcommand{\bcoh}{\beta_{\msf{Coh}}}
\newcommand{\bmr}{\beta_{\msf{MR}}}
\newcommand{\bqcp}{\beta_{\msf{QCP}}}
\newcommand{\uc}[1]{u_{\msf{c}}^{\pup{#1}}}

\newcommand{\Rhot}{\msf{R}}
\newcommand{\et}{\tilde{\e}}
\newcommand{\muco}{\msf{m}}
\newcommand{\loa}{\msf{R}_d}
\DeclareMathOperator{\cn}{cn}

\DeclareMathOperator{\dn}{dn}
\newcommand{\rr}{\mathfrak{r}}
\newcommand{\ii}{\mathfrak{i}}
\newcommand{\pp}{\mathcal{J}}
\newcommand{\qq}{\mathcal{K}}
\newcommand{\nn}{\mathcal{N}}
\newcommand{\Aa}{\mathcal{A}}

\newcommand{\wRhot}{\widetilde{\mathsf{R}}}

\newcommand{\tpb}{t_{\msf{pb}}}
\newcommand{\tq}{t_{\msf{quench}}}
\newcommand{\Emin}{E_{\msf{min}}}
\newcommand{\esp}{\varepsilon_0}
\newcommand{\tsfb}[1]{{{\textsf{#1}}}}

\newcommand{\eRF}{\mathcal{E}_{3,2}}

\begin{document}

\title{
	Quantum quench in a $p+i p$ superfluid:  Winding numbers and topological states far from equilibrium
}
\author{Matthew S. Foster}
\email{matthew.foster@rice.edu} 
\affiliation{Department of Physics and Astronomy, 
	     Rice University, 
             Houston, 
             Texas 77005,
	     USA}
\author{Maxim Dzero}
\affiliation{Department of Physics,
	     Kent State University, 
	     Kent, Ohio 44242,
	     USA}
\author{Victor Gurarie}
\affiliation{
	     Department of Physics, 
	     University of Colorado, 
	     Boulder, CO 80309,
	     USA}
\author{Emil A.\ Yuzbashyan}
\affiliation{Center for Materials Theory, Department of Physics and Astronomy, 
	     Rutgers University, 
	     Piscataway, 
	     New Jersey 08854, 
	     USA}

\date{\today}

\begin{abstract}
We study the non-adiabatic dynamics of a 2D $p+ip$ superfluid following an instantaneous quantum quench
of the BCS coupling constant. The model describes a topological superconductor with a non-trivial BCS (trivial BEC) 
phase appearing at weak (strong) coupling strengths.
We extract the exact long-time asymptotics of the order parameter $\Delta(t)$ by exploiting the integrability of the classical
p-wave Hamiltonian, which we establish via a Lax construction.  
Three different types of asymptotic behavior can occur depending upon the strength and 
direction of the interaction quench.
We refer to these as the non-equilibrium phases \{{\bf I}, {\bf II}, {\bf III}\}, characterized as follows.
In phase {\bf I}, the order parameter asymptotes to zero due to dephasing.
In phase {\bf II}, $\Delta \rightarrow \Dasy$, a non-zero constant.
Phase {\bf III} is characterized by persistent oscillations of $\Delta(t)$. 
For quenches within phases {\bf I} and {\bf II}, we determine the topological character of the asymptotic states. 
We show that two different formulations of the bulk topological winding number, although equivalent in the
BCS or BEC ground states, must be regarded as independent out of equilibrium.
The first winding number $Q$ characterizes the Anderson pseudospin texture of the initial state;
we show that $Q$ is generically conserved. 
For $Q \neq 0$, this leads to the prediction of a ``gapless topological'' state when $\Delta$ 
asymptotes to zero. 
The presence or absence of Majorana edge modes in a sample with a boundary
is encoded in the second winding number $W$, which is formulated in terms of the retarded Green's function.
We establish that $W$ can change following a quench across the quantum critical point.	
When the order parameter asymptotes to a non-zero constant, the final value of $W$ is well-defined and quantized. 
We discuss the implications for the (dis)appearance of Majorana edge modes.
Finally, we show that the parity of zeros in the 
\emph{bulk}
out-of-equilibrium Cooper pair distribution function 
constitutes a $\mathbb{Z}_2$-valued quantum number, which is non-zero whenever 
$W \neq Q$.
The pair distribution can in principle be measured using RF spectroscopy in an ultracold 
atom realization, allowing direct experimental detection of the $\mathbb{Z}_2$ number. 
This has the following interesting implication: topological information that is experimentally inaccessible 
in the bulk ground state can be transferred to an observable distribution function when the system
is driven far from equilibrium.
\end{abstract}


\maketitle
\tableofcontents

\section{Introduction \label{Sec: Intro}}

Topology has emerged as a key tool to characterize phases of quantum many-body particle systems. A recent application is
the classification of topological insulators and superconductors.\cite{TISC,TopClassesDirty} 
These are distinguished by a topological winding number in the bulk; when this number is quantized
to a non-zero integer value, it implies the presence of gapless, delocalized states at the sample surface. Both the bulk topological
invariant and the gapless surface states are argued to be protected against generic local perturbations.

A natural means to generate non-trivial dynamics in a topological system is via a \emph{global} deformation of the system Hamiltonian, 
otherwise known as a quantum 
quench.\cite{Bloch02,Weiss06,Stamper-Kurn06,
Barankov04,amin,simons,WarnerLeggett05,YuzbashyanAltshuler05,YKA05,YuzbashyanAltshuler06,BarankovLevitov06,DzeroYuzbashyan06,Chien2010,
Rigol,CardyCalabrese06,Kollath07,Cincio07,Polk11} 
Through the evolution induced by a quench, one can probe the stability of the system 
topology--when and how can it change? Under what circumstances does it remain well-defined when the system is coherently driven 
far away from its ground state? Finally, can the quench be employed as an experimental tool to reveal the bulk topology?

In the setup for a 
quench, a many-particle system is initially prepared in a particular pure state;
this can be taken as the ground state of some initial Hamiltonian. In addition, we assume that there is a gap to
excitations. Performing the quench, a parameter of the Hamiltonian (such as the interparticle interaction strength) 
is changed over a time interval much shorter than the inverse excitation gap. The system subsequently evolves as a highly excited, coherent 
admixture of many-body eigenstates of the final, post-quench Hamiltonian. Quantum quenches have become a standard 
protocol to investigate ultracold atomic systems.\cite{BlochRMP08,Bloch02,Weiss06,Stamper-Kurn06} Ultracold gases are engineered to
be well-isolated from any outside environment or heat bath, and typically exhibit a high degree of external tunability. 
The long-time out-of-equilibrium dynamics induced by a quench in an isolated many particle system can show 
different \emph{dynamical phases} as function of the quench 
parameters.\cite{Barankov04,amin,simons,WarnerLeggett05,YuzbashyanAltshuler05,YKA05,YuzbashyanAltshuler06,BarankovLevitov06,DzeroYuzbashyan06,galperin,shumeiko} 

In this work, we probe the response of a 2D topological $p+i p$ superfluid\cite{Volovik,ReadGreen2000,TISC} to an instantaneous 
quantum quench. We envisage an ultracold fermionic atom\cite{GurarieFB-1,GurarieFB-2,Gaebler07,Fuchs08,Inada08,Jone08,LCG08,ZhangTewari08,Sato09,SauSensarma11,Zhu11,LiuJiang12}
or molecule\cite{CooperShlyapnikov11} realization of the system, such that the effective pairing interaction strength
can be tuned externally, e.g.\ by manipulating a Feshbach resonance.\cite{GurarieFB-1,GurarieFB-2,Gaebler07,Fuchs08,Inada08,Jone08,LCG08} Initially, the system occupies the ground 
state of the pre-quench Hamiltonian, residing within either the topologically non-trivial BCS or trivial BEC phase (see below). 
Subsequently, the BCS interaction coupling is deformed instantaneously to stronger or weaker pairing.
We consider quenches both within and between the BCS and BEC phases. We calculate the asymptotic time 
evolution\cite{Barankov04,amin,simons,WarnerLeggett05,YuzbashyanAltshuler05,YKA05,YuzbashyanAltshuler06,BarankovLevitov06,DzeroYuzbashyan06,Chien2010} using an integrable 
version of the p-wave BCS Hamiltonian.\cite{Richardson02,Skrypnyk09,Sierra09,Sierra10,Ortiz10} Our treatment is exact in the thermodynamic 
limit when pair-breaking can be neglected.\cite{Barankov04,YKA05,YuzbashyanAltshuler05} 

An overview of our main results is provided in Sec.~\ref{Sec: Results}; these include the following:
First, we compute the out-of-equilibrium phase diagram shown in Fig.~\ref{Fig--PhaseDiagBasic}, 
as determined by the exact long-time dynamics of the 
order parameter
$\Delta$
(which is \emph{not} the same as the quasiparticle gap for p-wave---see 
Secs.~\ref{Sec: PwaveReview}, \ref{Sec: GNDRoots},  
and Appendix \ref{Sec: APP--GND} for a brief review).
Similar to the s-wave case,\cite{Barankov04,amin,simons,WarnerLeggett05,YuzbashyanAltshuler05,YKA05,YuzbashyanAltshuler06,BarankovLevitov06,DzeroYuzbashyan06}
we find that $\Delta$ exhibits one of three behaviors in the long time limit $t \rightarrow \infty$:
for strong-to-weak pairing quenches within phase {\bf I} of Fig.~\ref{Fig--PhaseDiagBasic}, $\Delta(t) \rightarrow 0$ 
due to 
dephasing.\cite{BarankovLevitov06,DzeroYuzbashyan06} 
In phase {\bf II}, which includes the case of zero quench, $\Delta(t) \rightarrow \Dasy$, a non-zero constant.\cite{WarnerLeggett05,YuzbashyanAltshuler05,YuzbashyanAltshuler06}
Finally, 
for weak-to-strong pairing quenches within phase {\bf III}, $\Delta(t)$ exhibits coherent, undamped oscillations\cite{Barankov04,YuzbashyanAltshuler06,BarankovLevitov06} 
as $t \rightarrow \infty$, see Fig.~\ref{Fig--RegionIIIOsc1}.

In the ground state, the bulk topology and edge states are encoded in a $\mathbb{Z}$-valued winding number. 
We consider two formulations: the winding of the Anderson pseudospin texture $Q$,\cite{Volovik,ReadGreen2000}
and of the retarded single-particle Green's function $W$.\cite{Volovik,GurarieRGF,EssinGurarie11} 
In the ground state $W = Q$, and $Q = 1$ ($Q = 0$) in the weak pairing BCS (strong pairing BEC) phase.
We show that $Q$ does not evolve from its initial value following a quench, Fig.~\ref{Fig--PhaseDiagQ}.
We identify a ``gapless topological'' phase, characterized by $Q = 1$ and 
${\displaystyle{\lim_{t \rightarrow \infty}}} \Delta(t) = 0$. 
Although equivalent in equilibrium, we find that $Q$ and $W$ must be regarded as independent
following a quench. In particular, the presence or absence of Majorana edge modes in a sample
with a boundary is encoded in $W$, not $Q$. Moreover, a quench across the topological quantum
phase transition (e.g., from BCS to BEC) induces a change in $W$.	
Whenever ${\displaystyle{\lim_{t \rightarrow \infty}}} \Delta(t) = \Dasy \neq 0$, 
$W$ nevertheless assumes a quantized value in the asymptotic steady-state, Fig.~\ref{Fig--PhaseDiagW}. 
We discuss implications for the (dis)appearance of Majorana edge modes.

\begin{figure}[b]
   \includegraphics[width=0.35\textwidth]{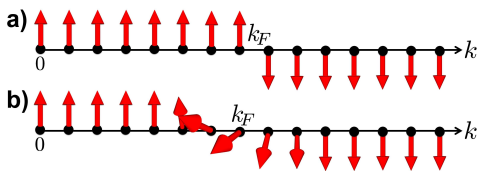}
   \caption{Anderson pseudospin description of
	(a) a normal Fermi liquid 
	(b) an s-wave superconductor,
	in any number of spatial dimensions at zero temperature.
	In this figure, $k$ measures the radial coordinate along any direction in momentum space.
	}
   \label{Fig--DWalls}
\end{figure}

Some of the results discussed in this paper also appear in an abbreviated form 
in Ref.~\onlinecite{PwaveLett}. The quench phase diagram and
the asymptotic values of the winding number $W$ for quenches in phase {\bf II} are 
presented in that work, as well as the link between $W$ and Majorana edge modes.
In Ref.~\onlinecite{PwaveLett}, we show that a topologically non-trivial 
cold atomic p-wave superfluid could be induced by quenching from very weak 
initial coupling to strong pairing, using a Feshbach resonance. In the non-trivial
case, the order parameter oscillates periodically in time (phase {\bf III}), and 
the presence of edge modes is established using a Floquet analysis.\cite{Floquet-1,Floquet-2,Floquet-3,Floquet-4}
We do not discuss Floquet in this paper. 
Instead, we provide the detailed derivation
of the phase diagram and the topological characterization of phases {\bf I} and {\bf II}.
We consider both formulations $Q$ and $W$ of the bulk winding number. 
We compute the long-time dynamics of the order parameter exactly, using a variant
of the 
Lax construction
employed in the s-wave case.\cite{YKA05,YuzbashyanAltshuler05,YuzbashyanAltshuler06,BarankovLevitov06,DzeroYuzbashyan06}
Finally, in this paper we search for \emph{bulk} signatures of the system topology.	

Because the topology resides in a quantum mechanical Berry phase, it is typically difficult to measure a 
bulk invariant directly. We show that when $\Delta$ asymptotes to a non-zero constant and $Q \neq W$, as 
occurs for a quench across the quantum critical point, the number of zeroes in the out-of-equilibrium 
\emph{Cooper pair distribution function} is odd, as demonstrated in Figs.~\ref{Fig--PhaseDiagSec} and  
\ref{Fig--ZeroParity1}. The parity of these zeroes constitutes a non-equilibrium $\mathbb{Z}_2$-valued 
bulk winding number. We show that this number can in principle be detected through the modulation of the 
absorption amplitude in RF spectroscopy. This is unique to the non-equilibrium preparation; the winding 
number distinguishing the BCS and BEC ground states cannot be ascertained via a bulk RF measurement.

In the remainder of this Introduction, we briefly review the topological character of 2D $p + ip$ superfluids. 
We close with an outline for the organization of this paper.


\subsection{Topological superfluidity in 2D \label{Sec: PwaveReview}}

The topological properties of 2D $p+i p$ superconductors were originally obtained by Volovik\cite{Volovik}
in the context of ${}^3$He-A, and subsequently discussed in the context of the fractional quantum Hall effect
by Read and Green.\cite{ReadGreen2000} The simplest p-wave channel BCS Hamiltonian for spinless (or spin-polarized)
fermions is\cite{Volovik} 
\begin{align}\label{Hactual}
	H
	=
	\sum'_{\vex{k}}
	\frac{k^2}{m}
	s^z_{\vex{k}}
	-
	\frac{2 G}{m} 	
	\sum'_{\vex{k},\vex{q}}
	\vex{k}\cdot\vex{q} \,
	s_{\vex{k}}^{+} s_{\vex{q}}^{-},
\end{align}	
where $G > 0$ is a dimensionless, attractive BCS interaction strength.
Eq.~(\ref{Hactual}) is expressed in terms of $SU(2)$ Anderson pseudospin\cite{AndBCS58} operators,
defined as follows:
\begin{align}\label{Pseudospins}
\begin{aligned}
	s_{\vex{k}}^z 
	\equiv&\,
	{\textstyle{\frac{1}{2}}}
	(
	c^\dagger_{\vex{k}}
	c_{\vex{k}}
	+
	c^\dagger_{-\vex{k}}
	c_{-\vex{k}}
	-
	1
	),
	\\
	s_{\vex{k}}^+
	\equiv&\,
	c^\dagger_{\vex{k}} c^\dagger_{-\vex{k}},
	\\
	s_{\vex{k}}^-
	\equiv&\,
	c_{-\vex{k}} c_{\vex{k}},
\end{aligned}
\end{align}
where $c_{\vex{k}} c^\dagger_{\vex{q}} +  c^\dagger_{\vex{q}} c_{\vex{k}} = \delta_{\vex{k},\vex{q}}$.
The primed sums in Eq.~(\ref{Hactual}) run over 2D momenta in the half plane 
$\vex{k} = \{k^x \in \mathbb{R}, k^y \geq 0 \}$;
with $\{\vex{k},\vex{q}\}$ restricted to this range,
the pseudospins satisfy 
$[s^a_{\vex{k}},s^b_{\vex{q}}] = i \epsilon^{a b c} \delta_{\vex{k},\vex{q}} \, s^c_{\vex{k}}$.
In Eq.~(\ref{Hactual}), we have assumed that Cooper pairs are created and destroyed
with zero center-of-mass momentum only (``reduced BCS'' theory).\cite{Schrieffer} 
This neglects pair-breaking processes; 
we address the limitations of this approximation 
in the conclusion Sec.~\ref{Sec: End}.

\begin{figure}
   \includegraphics[width=0.45\textwidth]{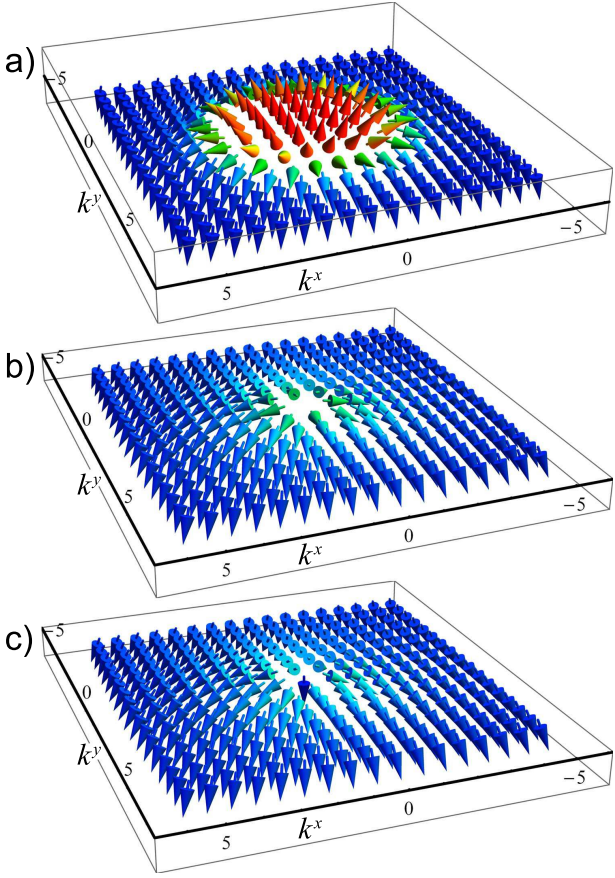}
   \caption{Momentum space pseudospin texture of the $p+ip$ ground state
	(a) in the topological BCS phase, $Q = 1$
	(b) at the BCS-BEC quantum phase transition, and 
	(c) in the topologically-trivial BEC phase, $Q = 0$.
	Here $Q$ denotes the pseudospin winding number (skyrmion charge),
	defined via Eq.~(\ref{Q}). For a $p+ip$ state,
	$Q = 1$ ($Q = 0$) when the $\vex{k} = 0$ spin $\vec{s}_{0}$ is up (down).
	By contrast, $Q$ is ill-defined at the quantum phase transition.
	}
   \label{Fig--Skyrm}
\end{figure}

The Anderson pseudospins provide a simple way to visualize the ground state of a
BCS superconductor. 
The expectation $\langle s^z_{\vex{k}} \rangle$ measures the double-occupancy
of a pair of states related by time-reversal symmetry; in Eq.~(\ref{Pseudospins}), $\langle s^z_{\vex{k}} \rangle= 1/2$ ($-1/2$)
implies that the states $\{\vex{k},-\vex{k}\}$ are occupied (vacant). 
The Fermi liquid ground state is a discontinuous domain wall, 
as depicted in Fig.~\ref{Fig--DWalls}(a). By contrast, an s-wave paired state exhibits 
smooth pseudospin canting near $k = k_F$, Fig.~\ref{Fig--DWalls}(b).
In the thermodynamic limit, $H$ in Eq.~(\ref{Hactual}) has a $p+ip$ ground state, defined in terms
of the 
order parameter
\begin{align}\label{GapGND}
	\Delta(\vex{k}) 
	\equiv 
	- 
	2 G	
	\sum'_{\vex{q}} \vex{k}\cdot\vex{q} \langle s_{\vex{q}}^- \rangle 
	=
	\Do \,
	(k^x - i k^y).
\end{align}
The 
amplitude $\Do$ is non-zero for any $G > 0$.
The pseudospin texture for the weak-pairing BCS ground state is shown in 
Fig.~\ref{Fig--Skyrm}(a). The $p+ip$ texture differs from the s-wave one
in that the canting angle in spin space is correlated to the polar angle 
$\phi_k$ in momentum space; the result is a topologically non-trivial (skyrmion) 
configuration,\cite{Volovik,ReadGreen2000} discussed in more detail below.

\begin{figure}[b]
   \includegraphics[width=0.3\textwidth]{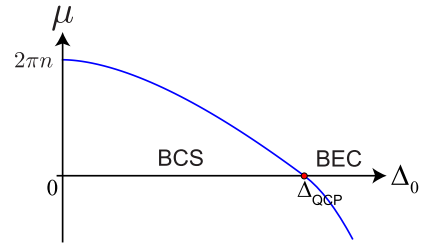}
   \caption{Zero temperature chemical potential $\mu$ as a function of the ground state 
	order parameter amplitude
	$\Do$ in the $p+i p$ ground state, for fixed particle density $n$. 
	The point $\{\Do,\mu\} = \{\Dqcp,0\}$ is a quantum phase transition
	between the topologically non-trivial BCS and trivial BEC phases. The quasiparticle spectrum 
	has a gapless Dirac node in the bulk at $\vex{k} = 0$ when $\Do = \Dqcp$. 
	}
   \label{Fig--MuPlot}
\end{figure}

The quasiparticle energy in the $p+ip$ paired state is given by
\begin{align}\label{Ek}
	E_{k} = \sqrt{\left(\frac{k^2}{2} - \mu\right)^2 + k^2 \Do^2},
\end{align}
where $\mu$ is the chemical potential. 
Here we have set $m = 1$, since the mass can be factored from $H$ in Eq.~(\ref{Hactual});
$\mu$ and $(\Do)^2$ both carry units of density.
The spectrum is fully-gapped for any $\mu \neq 0$.
In a system with fixed density $n$, the chemical potential is a monotonically 
decreasing function of the 
pairing
amplitude $\Do$:
\begin{align}\label{muGND}
	\mu 
	=&\, 
	\left[2 \pi n - \frac{\Do^2}{2} \log\left( \frac{2 \Lambda}{e \Do^2} \right) \right]\theta(\Dqcp - \Delta)
	\nonumber\\&\,
	+
	\left[\frac{\Do^2}{2} - \Lambda \exp\left(-\frac{4 \pi n}{\Do^2} - 1\right) \right]\theta(\Delta - \Dqcp),
\end{align}
where $\Lambda$ is a high-energy cutoff (see Appendix~\ref{Sec: APP--GND} for details),
and $\theta(\Delta)$ denotes the unit step function.
A plot of $\mu$ versus $\Do$ is shown in Fig.~\ref{Fig--MuPlot}.
At the special value $\Do = \Dqcp$ [defined via Eq.~(\ref{DDefs})], $\mu = 0$.
Here, the bulk quasiparticle spectrum develops a massless Dirac node at $\vex{k} = 0$. 

$\Dqcp$ marks a \emph{topological} quantum phase transition between
the topologically non-trivial, weak-pairing BCS phase ($0 < \Do < \Dqcp$) and the
trivial, strong-pairing BEC phase ($\Do > \Dqcp$).\cite{ReadGreen2000}
These can be distinguished by a bulk topological invariant.
There are several equivalent formulations of the invariant in equilibrium. We will employ
two different definitions. The first measures the winding of the pseudospin texture,\cite{Volovik,ReadGreen2000}
\begin{align}\label{Q}
	Q 
	\equiv 
	8 \pi \epsilon_{a b c}
	\int \frac{d^2 \vex{k}}{(2 \pi)^2} 
	\frac{1}{k} 
	\langle s_{\vex{k}}^a \rangle
	\partial_k
	\langle s_{\vex{k}}^b \rangle
	\partial_{\phi_k}
	\langle s_{\vex{k}}^c \rangle,
\end{align}
where $\phi_k$ denotes the polar angle in momentum space.
A generalized $p+ip$ state can be defined by the pseudospin configuration
\begin{align}\label{PseudoTex}
\begin{aligned}
	\langle s_{\vex{k}}^- \rangle \equiv&\, \ts{\frac{1}{2}}\sqrt{1 - \varrho^2(k)} \exp\left[-i \phi_k + i \Theta(k)\right],\\
	\langle s_{\vex{k}}^z \rangle \equiv&\, \ts{\frac{1}{2}}\varrho(k), 
\end{aligned}
\end{align}
where $\varrho(k)$ and $\Theta(k)$ are real, continuous functions of $k$, independent of $\phi_k$, 
$|\varrho(k)| \leq 1$, 
and $|\varrho(0)| = - \varrho(k\rightarrow\infty) = 1$.
Then the integrand reduces to a total derivative, leading to 
$Q = \ts{\frac{1}{2}}\left\{\sgn[\varrho(0)] + 1 \right\}$.
In the $p+ip$ ground state,
\[
	\varrho(k) = \frac{2\mu - k^2}{2 E_k}, \;\; \Theta(k) = 0,
\]
so that
\begin{align}
	Q 
	=
	\left\{
	\begin{array}{ll}
	1, & \mu > 0 \;\; (\text{BCS}),\\
	0, & \mu < 0 \;\; (\text{BEC}).
	\end{array}
	\right.
\end{align}	
At $\Do = \Dqcp$ ($\mu = 0$), both $\langle \vec{s}_{\vex{k} = 0}\rangle$ and $Q$ are undefined. 
Pseudospin textures at the critical point and in the BEC phase are depicted in Figs.~\ref{Fig--Skyrm}(b,c).

A different formulation of the invariant based upon the TKNN formula\cite{TKNN}
was derived by Volovik,\cite{Volovik} and utilizes the retarded single particle 
Green's function 
\begin{align}\label{GDef}
	\G_{\vex{k}}(t,t')
	\equiv&\,	
	-i
	\begin{bmatrix}
	\langle \{c^\dagger_{-\vex{k}}(t),c_{-\vex{k}}(t')\} \rangle 
	&
	\langle \{c^\dagger_{-\vex{k}}(t), c_{\vex{k}}^\dagger(t')\} \rangle 
 	\\
	\langle \{c_{\vex{k}}(t),c_{-\vex{k}}(t')\} \rangle 
	& 
	\langle \{c_{\vex{k}}(t),c_{\vex{k}}^\dagger(t')\} \rangle 
	\end{bmatrix}
	\nonumber\\
	&\,
	\phantom{-}
	\times
	\theta(t - t').
\end{align}
The winding number $W$ is\cite{Volovik,GurarieRGF,EssinGurarie11} 
\begin{align}\label{W}
	W
	\equiv&\,
	\frac{\epsilon_{\alpha \beta \gamma}}{3!}
	\int_{-\infty}^{\infty}
	d \omega
	\int 
	\frac{d^2 \vex{k}}{(2 \pi)^2}
	\tr\left[
	\begin{aligned}
	&\G^{-1} \left(\partial_\alpha \G\right)\\
	&\!\times\! \G^{-1} \left(\partial_\beta \G\right)\\
	&\!\times\! \G^{-1} \left(\partial_\gamma \G\right)
	\end{aligned}
	\right],
\end{align}
where $\tr$ denotes the trace in Nambu (particle-hole) space,
and $\G \equiv \G_{\vex{k}}(i \omega)$ is the Fourier
transform of $\G_{\vex{k}}(t,0)$, analytically-continued to imaginary frequency.
In both the BCS and BEC ground states, $W = Q$.

In the BCS phase, the advent of a non-zero bulk winding number implies the presence
of 1D chiral Majorana edge states at the boundary of a superfluid droplet.\cite{Volovik,ReadGreen2000}
Gapless Majorana edge channels are the hallmark of a topological superconductor.\cite{TISC}
When a temperature gradient is applied across the droplet, these states carry a perpendicular, 
dissipationless energy current, with a quantized thermal Hall conductance\cite{KaneFisher,ReadGreen2000,Capelli02}
\[
	\kappa_{x y} = \frac{\pi^2 k_B T}{12 \pi \hbar}.
\]
Here $T$ is the average temperature of the bulk. Additional ``Majorana'' signatures include
isolated zero modes in type II vortices.\cite{ReadGreen2000}


\subsection{Outline}

This paper is organized as follows.
In Sec.~\ref{Sec: Results}, we provide an overview of our main results concerning the
order parameter
dynamics and asymptotic winding numbers following a quench. 
We discuss implications for Majorana edge states and RF spectroscopy.
These results are derived in the remaining sections. 
In Sec.~\ref{Sec: QPD}, we derive the quench phase diagram from the exact solution
to the long-time 
dynamics. We exploit a new Lax vector
construction for the integrable p-wave problem. The solution obtains
by classifying the isolated roots of the system's spectral polynomial. 
In Sec.~\ref{Sec: GapDyn}, we establish the precise relation between the roots
of the spectral polynomial and the steady-state 
order parameter
dynamics. 
We compute the exact form of the persistent oscillations in $\Delta(t)$ 
for weak-to-strong quenches, and we present formulae relating the period and amplitude of
these to the isolated roots. 
In Sec.~\ref{Sec: WindingObs}, we derive the out-of-equilibrium pseudospin and Cooper pair distribution functions.
Using these results, for quenches wherein 
$\Delta(t)$ 
asymptotes to a finite constant
(which may be zero), we derive the power-law approach 
to this value.
We also compute the asymptotic values of the winding numbers $W$ and $Q$, defined above. 
We conclude with open questions in Sec.~\ref{Sec: End}.

Various technical details are relegated to the appendices.
Ground state properties, including the tunneling density of states, are
reviewed in Appendix \ref{Sec: APP--GND}.
Appendix \ref{Sec: APP--IIIDyn} supplies additional results on persistent 
order parameter
oscillations
in a narrow sliver of the quench phase diagram. 
Finally, in Appendix \ref{Sec: APP--GFs} we compute the coherence factors and single particle 
Green's functions following the quench.

\section{P+ip superfluid quench: Key results \label{Sec: Results}}


\subsection{Chiral p-wave BCS model}

To study quench dynamics in a 2D $p+ip$ superfluid, we consider a ``chiral'' variant\cite{Sierra09} 
of the model in Eq.~(\ref{Hactual}), defined via
\begin{align}\label{Hchiral}
	H
	=
	\sum'_{\vex{k}}
	k^2 \,
	s^z_{\vex{k}}
	-
	G
	\sum'_{\vex{k},\vex{q}}
	(k^x - i k^y)(q^x + i q^y) \,
	s_{\vex{k}}^{+} s_{\vex{q}}^{-},
\end{align}	
where the mass $m = 1$.
The relation between Eqs.~(\ref{Hactual}) and (\ref{Hchiral}) follows from
\[
	\vex{k}\cdot{\vex{q}} = \frac{1}{2}\left[(k^x - i k^y)(q^x + i q^y) + (k^x + i k^y)(q^x - i q^y)\right],	
\]
discarding the second term.	
In the thermodynamic limit, $H$ in Eq.~(\ref{Hchiral}) possesses the same $p+ip$ ground state 
(BCS product wavefunction)\cite{Schrieffer} as 
Eq.~(\ref{Hactual}), in both the topologically non-trivial BCS and trivial BEC phases.
However, the model in Eq.~(\ref{Hchiral}) breaks time-reversal symmetry explicitly, and preferentially selects $k^x - i k^y$ over 
$k^x + i k^y$ pairing. These are degenerate in the time-reversal invariant Hamiltonian of Eq.~(\ref{Hactual}). 

We work with Eq.~(\ref{Hchiral}) instead of Eq.~(\ref{Hactual}) because the former is 
of Richardson-Gaudin\cite{Richardson,Gaudin,Dukelsky} type and therefore integrable;\cite{Richardson02,Skrypnyk09,Sierra09,Sierra10,Ortiz10} e.g., equilibrium properties can be 
extracted exactly via the Bethe ansatz. We can absorb the polar phase into the pseudospins 
$s_{\vex{k}}^- \rightarrow \exp(-i \phi_k) s_{\vex{k}}^-$, and sum spins along
arcs in momentum space:
\begin{align}\label{spinArc}
	\vec{s}_{k}
	\equiv \frac{1}{\pi} \int_0^{\pi} d \phi_k \, \vec{s}_{\vex{k}}.
\end{align}
As a result Eq.~(\ref{Hchiral}) reduces to an effective ``1D'' model
\begin{align}\label{H}
	H
	=
	\sum_{i}
	\e_i
	s_i^z
	-
	G
	\sum_{i}
	\sqrt{\e_i}
	s_i^+
	\sum_{j}
	\sqrt{\e_j}
	s_j^-,
\end{align} 
where $\e_i \equiv k_i^2$.
The Heisenberg equations of motion for the pseudospins are
\begin{align}\label{spinEOM}
\begin{aligned}
	\frac{d}{d t}\langle \vec{s}_i \rangle
	=&\,
	- \langle \vec{B}_i \times \vec{s}_i \rangle,
	\\
	\vec{B}_i
	\equiv&\,
	- \e_i \hat{z} - 2 \sqrt{\e_i} (\Delta_x \hat{x} + \Delta_y \hat{y}),
\end{aligned}
\end{align}
\begin{align}\label{DeltaDef}
	\Delta
	\equiv&\, \Delta_x - i \Delta_y
	\equiv
	- G
	\sum_{i} \sqrt{\e_i} s_i^-.
\end{align}
In the first equation, we take the expectation $\langle \cdots \rangle$ with respect to the initial state.
Due to the infinite-ranged nature of the interactions in the BCS Hamiltonian,
self-consistent mean field theory becomes \emph{exact} in the thermodynamic limit.\cite{AndBCS58,Richardson}
This is because 
$\Delta$ is an extensive variable depending upon all $N$ of the spins in the system, 
and can be replaced by its expectation value in the limit $N \rightarrow \infty$. 
For a global quench, the instantaneous state of the system is described
by a BCS product wavefunction at all times, albeit one parameterized by time-dependent coherence
factors.\cite{Barankov04} This implies that the problem reduces to solving Eq.~(\ref{spinEOM}), 
treating the spins and 
$\Delta$ 
as classical variables.\cite{Barankov04,WarnerLeggett05,YKA05,YuzbashyanAltshuler05}

In Appendix~\ref{Sec: APP--ChiralP}, we demonstrate that the classical dynamics following from a 
$p+ip$ initial pseudospin configuration are in fact \emph{identical} when generated by either
Eq.~(\ref{Hactual}) or Eq.~(\ref{Hchiral}); we therefore expect our predictions hold
for the full quantum dynamics of Eq.~(\ref{Hactual}) as well, in the thermodynamic limit.
The main approximation employed in the present work is not tied to the distinction 
between Eqs.~(\ref{Hactual}) and (\ref{Hchiral}), but rather the neglect of pair-breaking processes.
These are always present, break the integrability of the BCS Hamiltonian, and should ultimately
induce thermalization. We will discuss time scales relevant to pair-breaking in 
Sec.~\ref{Sec: End}.


\begin{figure}
   \includegraphics[width=0.48\textwidth]{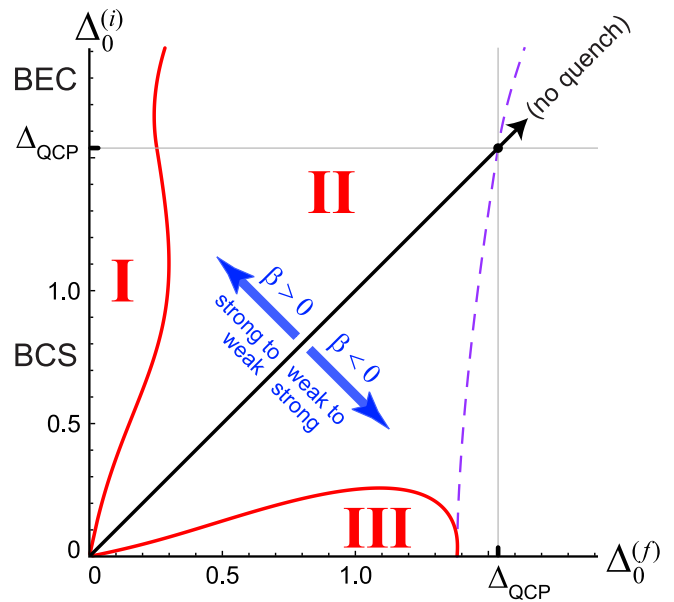}
   \caption{Exact interaction strength quench phase diagram, extracted from the isolated 
	roots of the spectral polynomial.
	The vertical axis measures the initial pairing 
	amplitude 
	$\Di$.
	$\Dqcp$ marks the equilibrium BCS-to-BEC topological quantum phase transition.
	The horizontal axis specifies the post-quench Hamiltonian through
	$\Df$, which is the 
	order parameter
	the system would exhibit in its ground state.
	The diagonal line $\Di = \Df$ is the case of no quench.
	Off-diagonal points to the left (right) describe strong-to-weak (weak-to-strong)
	pairing quenches, wherein the 
	dynamic variable $\Delta(t)$
	evolves away from its initial value.
	Similar to the s-wave case,\cite{Barankov04,amin,simons,WarnerLeggett05,YKA05,YuzbashyanAltshuler05,YuzbashyanAltshuler06,BarankovLevitov06,DzeroYuzbashyan06} 
	the $p+ip$ system exhibits three different dynamical phases defined by the long-time asymptotics ($t \rightarrow \infty$) 
	of 
	$\Delta$.
	In phase {\bf I}, $\Delta(t) \rightarrow 0$ due to dephasing,\cite{BarankovLevitov06,DzeroYuzbashyan06}
	in {\bf II}, $\Delta(t) \rightarrow \Dasy$, a non-zero constant,\cite{YuzbashyanAltshuler05,YuzbashyanAltshuler06} and
	in phase {\bf III}, $\Delta(t)$ shows persistent oscillations.\cite{Barankov04,YuzbashyanAltshuler06,BarankovLevitov06}
	The dashed purple line is the non-equilibrium continuation of the quantum phase transition, in the sense that the
	asymptotic value of the chemical potential $\masy \equiv {\displaystyle{\lim_{t \rightarrow \infty}}} \mu(t)$ 
	vanishes. This leads to a change in the retarded Green's function winding number $W$ (Fig.~\ref{Fig--PhaseDiagW}, below).
	For this plot and all subsequent figures, 
	we choose the Fermi energy $\ef = 2 \pi n = 5.18$ and the energy cutoff $\Lambda = 50 \ef$, so that $\Dqcp = 1.54$.
	}
   \label{Fig--PhaseDiagBasic}
\end{figure}

\subsection{Non-equilibrium phase diagram and asymptotic order parameter dynamics \label{Sec: Results--PD}}

We consider quenches in the model of Eq.~(\ref{Hchiral}) 
[or equivalently, Eq.~(\ref{H})]. The system is initially prepared in the $p+ip$
ground state of the pre-quench Hamiltonian $H_i$, which has interaction strength $\Gi$ and
amplitude $\Di$. At time $t = 0$, the coupling is instantaneously deformed
$\Gi \rightarrow \Gf$. We can label the quench by the initial 
pairing amplitude
$\Di$ and the quench parameter 
\begin{align}\label{betaDef}
	\beta
	\equiv
	2 \pi
	\left(
	\frac{1}{\gf} 
	-
	\frac{1}{\gi}
	\right),
\end{align}
where 
\begin{align}\label{gtoG}
	g \equiv G \ls^2/4
\end{align}
is the interaction strength that remains well-defined in the
thermodynamic limit; $\ls$ is the linear system size.
The case of zero quench has $\beta = 0$; $\beta > 0$ ($\beta < 0$) signifies a quench
towards weaker (stronger) pairing in the post-quench Hamiltonian $H_f$.
Although $g$ carries units of length-squared and is therefore formally irrelevant
in an RG sense, the integrals necessary to compute the quench dynamics are
at most logarithmically divergent in the high energy ultraviolet cutoff $\Lambda$.
These can be evaluated to logarithmic accuracy. 
Parameters $(\Do)^2$, $\mu$, $\beta$, and the fixed particle density $n$ carry the same units;
the latter sets the natural scale.

A quench $\{\Di,\beta\}$ can also be specified via
``quench coordinates'' $\{\Di,\Df\}$, where $\Df$ denotes the pairing 
amplitude
associated to 
the \emph{ground state} of $H_f$. This is not to be confused with the dynamic variable 
$\Delta(t)$, which evolves away from its initial value 
for any $\beta \neq 0$.
Using the BCS 
Eq.~(\ref{BCSEqF}) in Appendix~\ref{Sec: APP--GND}, 
we can express $\beta = \beta(\Di,\Df)$, a function of the pre- and post-quench 
Hamiltonian ground state 
order parameter
amplitudes, with $\beta(\Do,\Do) = 0$.
An explicit formula appears in Eq.~(\ref{betaDiDf}). 

In this work, we employ a generalized Lax construction\cite{YuzbashyanAltshuler05,YKA05} 
to solve the integrable dynamics of the classical pseudospins governed by Eq.~(\ref{H}), given a $p+ip$ paired initial 
state. The key to the solution 
is the so-called ``spectral polynomial'' $\qp(u;\beta)$, defined via a suitable Lax vector norm
(see Sec.~\ref{Sec: Lax} for details).
For a system of $N$ spins, $\qp(u;\beta)$ is a rank $2N$ polynomial in an auxiliary parameter $u$;
it is also a conserved integral of motion for any value of $u$. 
The polynomial coefficients (which are also integrals of motion) are complicated functions of the 
pseudospins $\{\vec{s}_j\}$.
The spectral polynomial encodes all essential aspects of the quench.\cite{YuzbashyanAltshuler05,YKA05,YuzbashyanAltshuler06,BarankovLevitov06}
It is a function of the post-quench coupling strength $\Gf$, and of the pre-quench state;
the coefficients can be evaluated in terms of the spin configuration at $t = 0$.

\begin{figure}[t!]
   \includegraphics[width=0.4\textwidth]{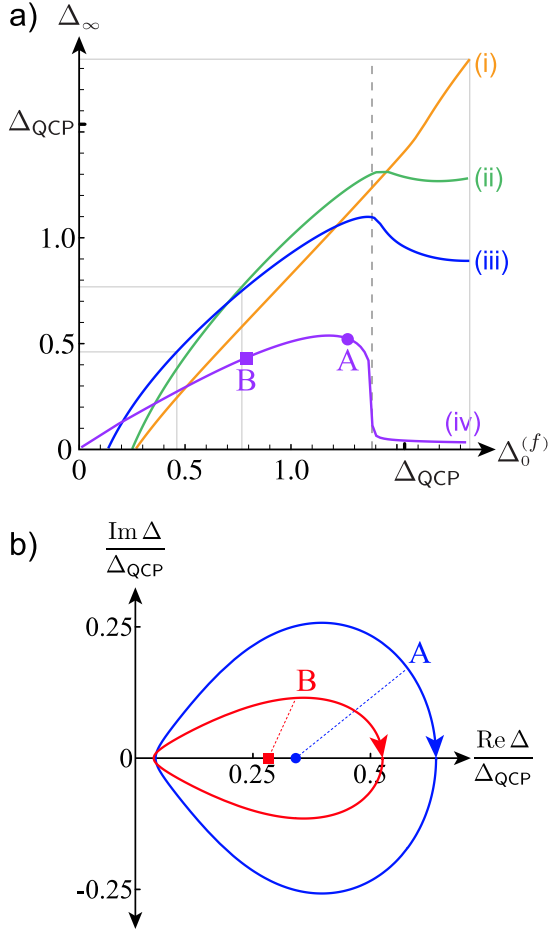}
   \caption{
	Asymptotic values of the non-equilibrium order parameter induced by various quenches.
	In (a), we plot $\lim_{t \rightarrow \infty} \Delta(t) = \Dasy$ as a function of the post-quench 
	ground state amplitude $\Df$, for fixed values of the initial $\Di$. 
	Curves (\tsfb{i})--(\tsfb{iv}) respectively correspond to $\Di / \Dqcp = \{1.2, 0.5, 0.3, 0.00651\}$;
	each gives $\Dasy$ along a horizontal cut across the quench phase diagram in Fig.~\ref{Fig--PhaseDiagBasic}.
	The value of $\Dasy$ is determined by the isolated roots in Eq.~(\ref{IsoRootsPhaseII}). 
	For the quench specified by $\{\Di,\Df\}$, the roots are computed from the conserved spectral	
	polynomial [Eq.~(\ref{qpQuench})].
	The dashed vertical line marks the boundary separating phases {\bf III} and {\bf II} at $\Di = 0$. 
	The portion of curve (\tsfb{iv}) to the left of this line in fact represents quenches in phase {\bf III},
	while the portion to the right resides in phase {\bf II}.
	Instead of asymptoting to a constant, 
	$\Delta(t)$ executes periodic amplitude and phase motion in phase {\bf III}. 
	For a given $\Df$ to the left of the dashed line, 
	the value of the curve (\tsfb{iv}) specifies the average $(|\Delta_{+}| + |\Delta_{-}|)/2$, 
	where $|\Delta_{\pm}|$ denote the turning points of the orbit in the complex $\Delta$-plane. 	
	Phase {\bf III} orbits\cite{footnote--RotFrameOrbit} associated to the 
 	quenches  
	marked A and B appear in (b). 
	}
   \label{Fig--Dasy}
\end{figure}

There is a separation of global versus local dynamics in the BCS quench problem. 
The long-time evolution of the 
order parameter
is determined by the isolated roots\cite{YKA05,YuzbashyanAltshuler05,YuzbashyanAltshuler06} 
of $\qp(u;\beta)$. 
These always appear in pairs and are few in number for a quench.
In a quench with 
$M$ isolated pairs, $\Delta(t \rightarrow \infty)$ is governed by an effective 
$M$-spin problem, with parameters specified by the roots.\cite{YuzbashyanAltshuler06}
Once the asymptotic 
dynamics 
of $\Delta(t)$ 
are determined, more detailed information can be extracted. 
In particular, the pseudospin distribution required to compute winding numbers and Green's functions 
in the limit $t \rightarrow \infty$ follows from the conservation of the spectral polynomial and the 
behavior of $\Delta(t)$.

Our results for the 
order parameter
dynamics are summarized in Fig.~\ref{Fig--PhaseDiagBasic},
which shows the non-equilibrium phase diagram.
The initial pre-quench state is labeled by $\Di$ on the vertical axis; the post-quench
Hamiltonian is identified by $\Df$ (the \emph{ground state} 
pairing amplitude
of $H_f$) 
on the horizontal. The diagonal line $\Di = \Df$ corresponds
to the ground state (no quench), while points to the left (right) of this line indicate strong-to-weak
(weak-to-strong) quenches. Each point in this diagram represents a specific quench.
As in previous studies of the s-wave case,\cite{Barankov04,amin,simons,WarnerLeggett05,YKA05,YuzbashyanAltshuler05,YuzbashyanAltshuler06,BarankovLevitov06,DzeroYuzbashyan06}
we find that the $p+ip$ 
order parameter
exhibits only three different
classes of long-time dynamics, labeled $\{{\bf I},{\bf II},{\bf III}\}$ in Fig.~\ref{Fig--PhaseDiagBasic}. 
For strong-to-weak pairing quenches in phase {\bf I}, 
$\Delta(t)$
decays to zero
due to dephasing; this is the case of zero isolated pairs in $\qp(u;\beta)$.
Phase {\bf I} is an example of gapless superconductivity: 
pair oscillations produce a 
continuous	
frequency spectrum with no isolated frequencies
separated from the continuum.\cite{BarankovLevitov06,DzeroYuzbashyan06} 
The Riemann-Lebesgue lemma then implies that $\Delta(t \rightarrow \infty) = 0$. 
Quenches in phase {\bf II} exhibit $\Delta(t) \rightarrow \Dasy$ as $t \rightarrow \infty$, where
$\Dasy$ is a non-zero constant. In this case, there is a single pair of isolated roots in $\qp(u;\beta)$.
Finally, weak-to-strong quenches in phase {\bf III} induce persistent oscillations in $\Delta(t)$;
here $\qp(u;\beta)$ 
has
two isolated pairs of roots.

\begin{figure}[b]
   \includegraphics[width=0.37\textwidth]{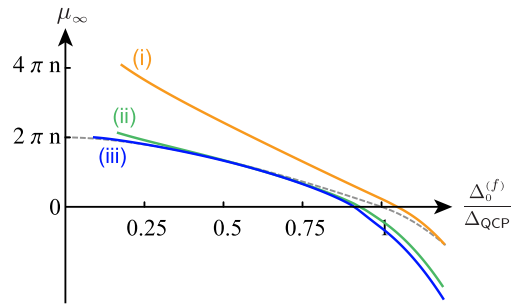}
   \caption{
	Asymptotic value of the non-equilibrium chemical potential  
	$\lim_{t \rightarrow \infty} \mu(t) = \masy$ 
	for quenches in phase {\bf II}, 
	as a function of the post-quench ground state amplitude $\Df$,
	for fixed values of the initial $\Di$.
	Curves (\tsfb{i})--(\tsfb{iii}) correspond to the associated 
	quenches in Fig.~\ref{Fig--Dasy}(a).	
	The dashed line is the ground state curve [Eq.~(\ref{muGND})].
	}
   \label{Fig--masy}
\end{figure}

\begin{figure}
   \includegraphics[width=0.3\textwidth]{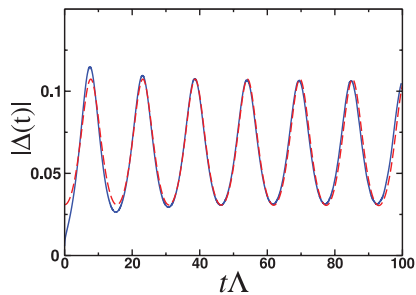}
   \caption{Example of persistent 
	order parameter
	oscillations following a quench.
	The coordinates $\{\Di,\Df\} = \{.00503,0.108\}$ place this quench in phase {\bf III} of Fig.~\ref{Fig--PhaseDiagBasic}. 
	The blue solid curve is the result
	of numerical simulation for classical pseudospins (5024 spins). 
	The red dashed
	curve obtains from the effective two-spin analytical solution, the parameters 
	of which are extracted from the isolated roots of the spectral polynomial $\qp(u;\beta)$ (see text).
	}
   \label{Fig--RegionIIIOsc1}
\end{figure}

A key difference from previous work\cite{Barankov04,WarnerLeggett05,YKA05,YuzbashyanAltshuler05,YuzbashyanAltshuler06,BarankovLevitov06,DzeroYuzbashyan06}
is that the chemical potential $\mu(t)$ is also a dynamical variable 
here.
This occurs because we consider quenches to and from intermediate and strong pairing, wherein
$\mu$ deviates from the Fermi energy even in the ground state [Eq.~(\ref{muGND}) and Fig.~\ref{Fig--MuPlot}].
In phase {\bf II},
the chemical potential asymptotes to a constant 
$\masy$, which is positive (negative) to the left (right) of the 
dashed purple line shown in Fig.~\ref{Fig--PhaseDiagBasic}. This line is 
the
non-equilibrium extension of
the topological quantum phase transition at $\Do = \Dqcp$. As discussed below, the 
asymptotic value ($t \rightarrow \infty$) of the Green's function winding number $W$ [Eq.~(\ref{W})]
associated with Majorana edge modes	
changes across this line. 

In phases {\bf II} and {\bf III} wherein the 
order parameter
remains non-zero, the quantitative description
of the asymptotic dynamics is entirely encoded in the isolated roots of $\qp(u;\beta)$. These solve a particular
transcendental equation in the thermodynamic limit, and can be extracted for any quench. 
Results for the asymptotic order parameter $\Dasy$ and chemical potential $\masy$ amplitudes are
plotted for horizontal cuts across the quench phase diagram in Figs.~\ref{Fig--Dasy}(a) and \ref{Fig--masy}.
Persistent 
oscillations in phase {\bf III} are depicted in Figs.~\ref{Fig--Dasy}(b) and \ref{Fig--RegionIIIOsc1}.
In the latter, the result 
obtained from the isolated roots is compared to a direct simulation of 5024 coupled spins.


\subsection{One quench, two winding numbers}

Our main purpose is to characterize the dynamics of the system topology
following a global quench. We compute the winding numbers $Q$ and
$W$ respectively defined by Eqs.~(\ref{Q}) and (\ref{W}) for quenches in the dynamical 
phases 
{\bf I}--{\bf III} of Fig.~\ref{Fig--PhaseDiagBasic}.

\begin{figure}[b]
   \includegraphics[width=0.35\textwidth]{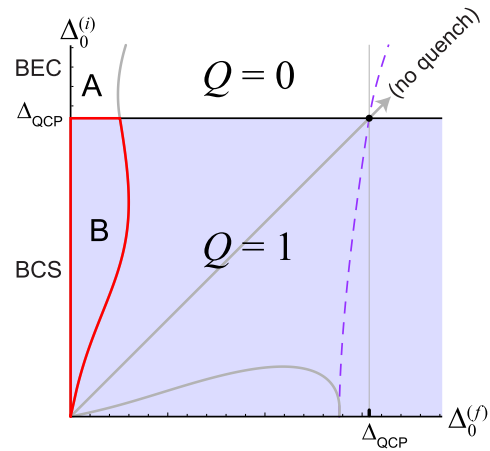}
   \caption{Quench phase diagram: Pseudospin winding number $Q$.
	We find that $Q$ is unchanged from its initial value,
	such that $Q = 1$ ($Q = 0$) for an initial BCS $\Di < \Dqcp$ 
	(BEC $\Di > \Dqcp$) state.
	The highlighted region {\bf B} is a ``topological gapless'' phase. 
	Throughout phase {\bf I} of Fig.~\ref{Fig--PhaseDiagBasic}, 
	$\Delta(t)$ vanishes as $t \rightarrow \infty$. 
	In subregion {\bf B}, the pseudospin texture nevertheless retains a non-zero winding
	$Q = 1$, Fig.~\ref{Fig--Gapless-B} and Eq.~(\ref{GaplessPrec}).
	The state can be visualized as a skyrmion texture as in	
	Fig.~\ref{Fig--Skyrm}(a), but now the tilted pseudospins \emph{precess} 
	at different frequencies about $\hat{z}$ in spin space---see Eq.~(\ref{GaplessPrec}). 
	The notion of smooth topology for the evolving texture remains 
	well-defined for times up to the 
	inverse level spacing.
	}
   \label{Fig--PhaseDiagQ}
\end{figure}

In the initial BCS or BEC ground state, $Q = W$.
We find that the pseudospin winding number $Q$ does not change following a quench,
as indicated in Fig.~\ref{Fig--PhaseDiagQ}. 
By contrast, the retarded Green's function winding $W$
undergoes a dynamical transition for a quench across the quantum critical point.
We argue below that $W$ determines the presence or absence of 
chiral Majorana edge modes in the post-quench asymptotic state
for quenches in phase {\bf II}. 
A quench in which $W \neq Q$ as $t \rightarrow \infty$
incurs a non-equilibrium topological quantum phase transition.

\begin{figure}[t]
   \includegraphics[width=0.35\textwidth]{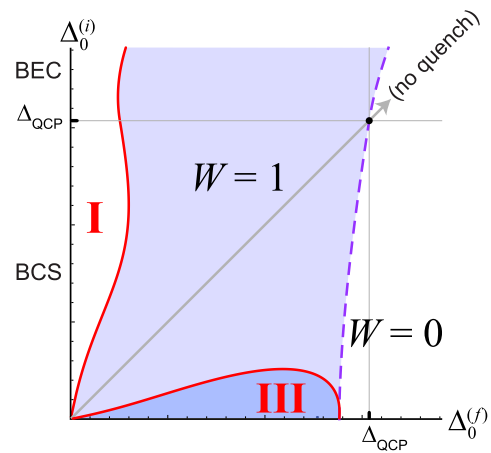}
   \caption{Quench phase diagram: Retarded Green's function winding number $W$ and Majorana edge modes.
	In phase {\bf II} of Fig.~\ref{Fig--PhaseDiagBasic}, 
	$\{\Delta(t),\mu(t)\} \rightarrow \{\Dasy,\masy\}$ as $t \rightarrow \infty$,
	with $\Dasy$ a non-zero constant.
	Along the dashed purple line, $\masy = 0$; this 
	is 
	a non-equilibrium extension
	of the ground state quantum phase transition;
	see also Fig.~\ref{Fig--masy}.
	To the left (right) of this line, $\masy > 0$ ($\masy < 0$), leading to $W = 1$ ($W = 0$).
	$W$ is ill-defined in the gapless phase {\bf I}.
	Phase {\bf III}, wherein $\Delta(t)$ exhibits persistent amplitude and phase oscillations
	[Figs.~\ref{Fig--Dasy}(b) and \ref{Fig--RegionIIIOsc1}]
	is topologically non-trivial.\cite{PwaveLett}
	Both phase {\bf III} and the $W = 1$ region of phase {\bf II} support 
	gapless Majorana edge modes.
	The former is confirmed by the Floquet analysis in Ref.~\onlinecite{PwaveLett},
	while the latter is established here via Eq.~(\ref{Gasy}) and the surrounding discussion.
	}
   \label{Fig--PhaseDiagW}
\end{figure}

In phase {\bf II} of Fig.~\ref{Fig--PhaseDiagBasic} wherein $\{\Delta(t),\mu(t)\} \rightarrow \{\Dasy,\masy\}$
as $t \rightarrow \infty$, $W = 1$ ($W = 0$) when $\masy > 0$ ($\masy < 0$). These regions are indicated in 
Fig.~\ref{Fig--PhaseDiagW}. The dashed purple line has $\masy = 0$, and is the 
extension of the topological quantum critical point into the 
non-equilibrium phase diagram. 

\begin{figure}[b]
   \includegraphics[width=0.35\textwidth]{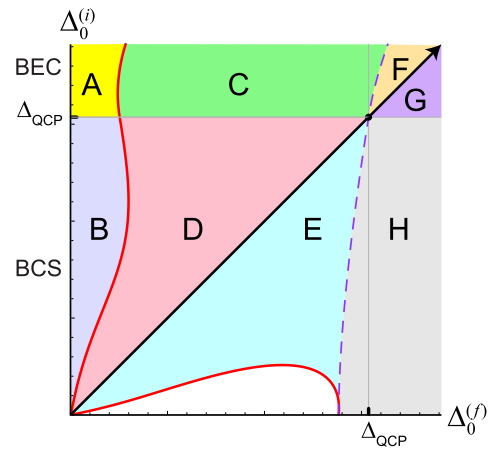}
   \caption{Sectioned phase diagram.
	Regions {\bf C} and {\bf H} include strong-to-weak and weak-to-strong quenches 
	across the quantum critical point at $\Do = \Dqcp$. 
	}
   \label{Fig--PhaseDiagSec}
\end{figure}

Phase {\bf II} quenches in which $W$ changes relative to $Q$ occur in 
two regions. Strong-to-weak pairing quenches across the critical point 
($\Di > \Dqcp$, $\Df$ to the left of the $\masy = 0$ line) 
have $Q = 0$ and $W = 1$, region {\bf C} in Fig.~\ref{Fig--PhaseDiagSec}.
Weak-to-strong quenches across the non-equilibrium quantum phase
boundary ($\Di < \Dqcp$, $\Df$ to the right of the $\masy = 0$ line)
have $Q = 1$ and $W = 0$, region {\bf H} in Fig.~\ref{Fig--PhaseDiagSec}.
Our methods allow access to the asymptotic behavior; we do not 
compute the transient kinetics of the topological transition wherein $W$ changes. 

We now discuss implications specific to the particular winding numbers.

\subsubsection{Non-equilibrium gapless topological phase \label{Sec: Topogapless}}

The conservation of the pseudospin winding number $Q$ [Eq.~(\ref{Q})] is simple to understand.
Under the dynamics induced by Eq.~(\ref{Hchiral}), spins along equal-radius arcs in momentum space
evolve collectively. For a $p+ip$ initial state, the relative canting of spins with equal $k$ is 
determined by the polar phase $\phi_k$, and this does not change; the spin texture 
is chiral at any time $t \geq 0$.
The pseudospins can be parameterized as in Eq.~(\ref{PseudoTex}), with $\varrho(k)$ and $\Theta(k)$ now time-dependent
parameters. 
The effective dynamics are captured by the ``1D'' model in Eq.~(\ref{H}). 
The spin at zero energy is stationary because it 
is decoupled from $\Delta$, see Eq.~(\ref{spinEOM}). By continuity, low energy spins remain close to the 
zero energy spin over a time interval of order the inverse level spacing, beyond which the notion of smooth 
topology becomes meaningless. Up until this time, $Q$ is conserved.  

This has interesting implications in phase {\bf I} of Fig.~\ref{Fig--PhaseDiagBasic}, wherein
$\Delta(t)$ decays to zero. At sufficiently long times, the effective magnetic field acting
upon spin $\vec{s}_i$ reduces to $\vec{B}_i = - \e_i \hat{z}$. However, the gapless phase is \emph{not}
a Fermi liquid ground state, which would have spins aligned along the field,\cite{footnote--NotAlignGND}
nor can it be understood as a finite temperature normal fluid.
Instead, phase {\bf I} is a quench-induced state of gapless superconductivity with a non-zero superfluid density.\cite{DzeroYuzbashyan06}
The spin configuration can be parameterized as
\begin{align}\label{GaplessPrec}
	\vec{s}_i(t)
	=
	\ts{\frac{1}{2}}
	\sqrt{1 - \gamma_i^2}
	\left\{
	\begin{aligned}
	&\!
	\cos(\e_i t + \Theta_i) \hat{x}
	\\
	&\!
	+
	\sin(\e_i t + \Theta_i) \hat{y}
	\end{aligned}
	\right\}
	+
	\ts{\frac{1}{2}}
	\gamma_i
	\hat{z},
\end{align}
where $\Theta_i$ is some constant phase.
The precession frequency of a spin at radius $k_i$ is $\e_i = k_i^2$,
twice the bare energy. The parameter $\gamma_i$ gives the 
z-projection
of the 
$\mathit{i}^{\mathit{th}}$
spin
in the $t \rightarrow \infty$ limit.
This is the ``distribution function'' for the Anderson pseudospins,
equivalent to the fermion mode occupation minus one,	
which characterizes the out-of-equilibrium state.
The zero temperature Fermi liquid would have $\gamma_i = \sgn(\e_i - 2 \ef)$,
with $\ef$ the Fermi energy. For the quench, we compute $\gamma_i$ exactly in the 
thermodynamic limit using the conservation of the Lax vector norm.

\begin{figure}[t]
   \includegraphics[width=0.325\textwidth]{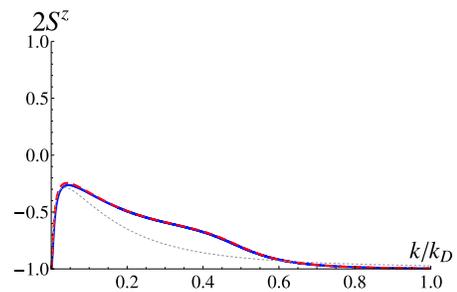}
   \caption{
	Asymptotic (infinite time) pseudospin distribution function 
	for a representative quench
	in the gapless region {\bf A} of Fig.~\ref{Fig--PhaseDiagQ}.
	The solid blue curve is the result of numerical simulation for
	5024 classical pseudospins; the red dashed curve is the analytical
	solution obtained from the Lax construction.
	The dotted gray line gives the initial ground state distribution.
	Although the initial and asymptotic distributions share the
	same winding $Q = 0$, in the latter case 
	$\Delta(t)$
	has decayed to
	zero, due to the dephasing of precessing pseudospins
	[Eqs.~(\ref{GaplessPrec}) and (\ref{DDecayG})]. 
	By contrast, in the ground state each pseudospin is aligned along
	its magnetic field.
	The quench coordinates are 
	$\{\Di,\Df\} = \{1.56, 0.00211\}$, and
	$k_D \equiv 4 \sqrt{4 \pi n} = 12.9$.
	}
   \label{Fig--Gapless-A}
\end{figure}

\begin{figure}[b]
   \includegraphics[width=0.325\textwidth]{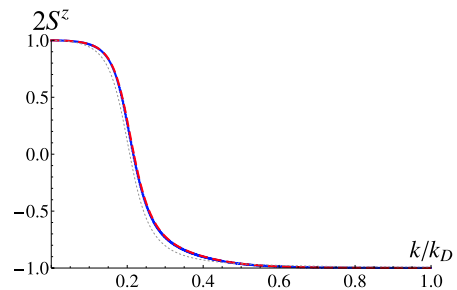}
   \caption{
	The same as Fig.~\ref{Fig--Gapless-A}, but for a point in the gapless 
	region {\bf B} of Fig~\ref{Fig--PhaseDiagQ}.
	In this case $Q = 1$, and the decay of 
	$\Delta(t)$
	to zero implies that the
	fluctuating region {\bf B} is a ``topological gapless'' phase.
	The quench coordinates are 
	$\{\Di,\Df\} = \{0.750, 0.00224\}$.
	}
   \label{Fig--Gapless-B}
\end{figure}

In Figs.~\ref{Fig--Gapless-A} and \ref{Fig--Gapless-B}, we plot $2 s^z_i = \gamma_i$ for
representative quenches in regions {\bf A} and {\bf B} of 
Fig.~\ref{Fig--PhaseDiagQ}. The pseudospin distributions resemble
those of the initial, pre-quench ground state with 
pairing amplitude
$\Di$, and the winding $Q$ 
is the same. Nevertheless, the post-quench state is gapless, due to dephasing of the spins. 
In particular, a quench in region {\bf B} induces a ``gapless topological'' state 
with $Q = 1$ and $\Delta(t \rightarrow \infty) = 0$. The state can be visualized 
as an undulating (time-evolving) variant of the skyrmion texture shown in Fig.~\ref{Fig--Skyrm}(a),
where the pseudospin $\vec{s}_{\vex{k}}$ precesses about $\hat{z}$ at frequency $k^2$ [Eq.~(\ref{GaplessPrec})].

Knowledge of $\gamma_i$ allows the self-consistent determination of $\Delta(t)$.
For an initial state not at the quantum critical point $\Di \neq \Dqcp$, we find that
\begin{align}\label{DDecayG}
	\Delta(t \rightarrow \infty)
	\propto		
	\frac{c_1}{\Lambda}
	\left(\frac{1}{t}\right)
	\exp(-2 i \Lambda t)
	+
	c_2
	\frac{1}{t^2},
\end{align}
where $c_{1,2}$ are constants and $\Lambda$ is the high-energy cutoff appearing
in (e.g.) Eq.~(\ref{muGND}).
Ignoring the high-frequency, cutoff-dependent piece, the dominant decay is $1/t^2$.
By contrast, for $\Di = \Dqcp$ the cutoff-independent decay law is slower:
\begin{align}\label{DDecayQCP}
	\Delta(t \rightarrow \infty)
	\propto		
	\frac{\tilde{c}_1}{\Lambda}
	\left(\frac{1}{t}\right)
	\exp(-2 i \Lambda t)
	+
	\tilde{c}_2
	\frac{1}{t^{3/2}}.
\end{align}


\subsubsection{Asymptotic Bogoliubov-de Gennes spectrum and edge states \label{Sec: QuenchEdge}}

The retarded Green's function winding $W$ in Eq.~(\ref{W}) is well-defined as 
$t \rightarrow \infty$ 
in	
phase {\bf II} of Fig.~\ref{Fig--PhaseDiagBasic}, 
whereupon	
$\{\Delta(t),\mu(t)\} \rightarrow \{\Dasy,\masy\}$.
As discussed above, $W$ changes from its initial value $Q$ for quenches across the topological quantum critical point,
regions {\bf C} and {\bf H} in Fig.~\ref{Fig--PhaseDiagSec}.
This follows from solving the Bogoliubov-de Gennes equation in the asymptotic steady state:
\begin{align}\label{Gasy}
	i \frac{d }{d t}
	\G_{\vex{k}}(t,t')
	=&\,
	\begin{bmatrix}
	- \frac{k^2}{2} + \masy & (k_x + i k_y)\Dasy \\ 
	(k_x - i k_y)\Dasy & \frac{k^2}{2} - \masy 
	\end{bmatrix}
	\G_{\vex{k}}(t,t'),
\end{align}
subject to the initial condition
\begin{align}
	\label{GasyIC}
	\lim_{
	\delta t 
	\rightarrow 0^+} \G_{\vex{k}}(t+ \delta_t,t) = -i
	\begin{bmatrix}
	1 & 0 \\
	0 & 1
	\end{bmatrix}.
\end{align}
Technically it is the magnitude 
$|\Delta(t)|$
that asymptotes to a finite constant in phase {\bf II}:
Using
the definition in Eq.~(\ref{DeltaDef}), the phase of the 
order parameter
winds according to 
$\Delta(t) \rightarrow \Dasy \exp(-2 i \masy t)$. 
In Eq.~(\ref{Gasy}) and all following equations in this section, 
we work in the rotating frame that eliminates this phase.\cite{footnote--NotAlignGND}
The function $\G_{\vex{k}}(t,t')$ encodes only the asymptotic pairing 
amplitude
and chemical potential, not the non-equilibrium 
spin distribution function.
The solution to Eq.~(\ref{Gasy}) is identical to that in the BCS or BEC ground state,
but with $\Dasy$ and $\masy$ determined by the quench.
Eq.~(\ref{W})
then implies that $W = 1$ ($W = 0$) for $\masy > 0$ ($\masy < 0$).
This is specific to the retarded function; other Green's functions 
(e.g., Keldysh) do depend upon the asymptotic pseudospin configuration.

The question of Majorana edge modes in the spectrum of a system with a boundary
is determined by solving the effective Bogoliubov-de Gennes mean field Hamiltonian $H_{\msf{BdG}}$
in the appropriate geometry. For the quench, we have
\begin{align}
	H_{\msf{BdG}}(t) 
	=
	\sum_{\vex{k}}'
	\left\{
	[k^2 - 2 \mu(t)]
	s_{\vex{k}}^z
	+
	\kc
	\Delta(t)
	s_{\vex{k}}^+
	+
	\kc^{*}
	\Delta^{*}(t)
	s_{\vex{k}}^-
	\right\}\!,
\end{align}
where $\kc \equiv k_x - i k_y$.
Since $H_{\msf{BdG}}$ and $\G_{\vex{k}}(t,t')$ encode the same information
as $t \rightarrow \infty$, one expects edge modes in the asymptotic spectrum when $W = 1$.

We conclude that Majorana edge modes appear when $W \neq 0$ in the post-quench state.
This implies that the edge spectrum can change following a quench across the critical point.
In region {\bf C} of Fig.~\ref{Fig--PhaseDiagSec}, the initially trivial BEC state develops
edge 
modes
in the $H_{\msf{BdG}}$ spectrum, while the 
modes present in the initial
BCS state disappear from the spectrum in region {\bf H}. 
Using a Floquet analysis,\cite{Floquet-1,Floquet-2,Floquet-3,Floquet-4} we have established that phase {\bf III} 
also hosts gapless Majorana edge modes. Details appear elsewhere.\cite{PwaveLett}

Within the bulk integrable theory, 
we cannot determine the \emph{occupation} of edge states following a quench. This is because a spatial edge breaks the integrability.
In future work we will investigate possible experimental signatures of the edge states
following a quench, such as a quantized dissipationless energy current.\cite{KaneFisher,ReadGreen2000,Capelli02} 
The characterization of Majorana zero modes following a quench in a 1D topological superconductor has 
been studied numerically,\cite{Perfetto2013} for a parameter change in a non-interacting (non-self-consistent) 
Bogoliubov-de Gennes model.


\subsection{A new $\mathbb{Z}_2$: Parity of the non-equilibrium Cooper pair distribution \label{Sec: CPRF}}

We have established above that the two different formulations of the bulk topological invariant,
although equivalent in equilibrium, can differ following a quench, and we have discussed implications
for the presence or absence of chiral Majorana modes at the edge of the sample. 
Are there any experimentally-accessible bulk signatures of the topological transition that occur when $W \neq Q$? 
Here we discuss one possibility in the parity of zeroes of the Cooper pair distribution function,
defined below. 

For a quench in which the 
order parameter
asymptotes to a non-zero constant (phase {\bf II} of Fig.~\ref{Fig--PhaseDiagBasic}),
the pseudospins precess around the ``effective ground state field''
\begin{align}\label{GndField}
	\vec{B}_i = - (\e_i - 2 \masy) \hat{z} - 2 \sqrt{\e_i} \Dasy \hat{x}.
\end{align}
The solution is
\begin{align}\label{GappedPrec}
	\vec{s}_{i}(t)
	=
	{\ts{\frac{1}{2}}}
	\sqrt{1 - \gamma_i^2}
	\left\{
	\begin{aligned}
	&\!
	\cos\left[2 \Easy(\e_i) t + 
	\Theta_i
	\right] \hat{B}_i \times \hat{y}
	\\&\!
	+
	\sin\left[2 \Easy(\e_i) t + 
	\Theta_i
	\right] \hat{y}
	\end{aligned}
	\right\}
	-
	\frac{\gamma_i}{2}
	\hat{B}_i,
\end{align}
where 
$\hat{B}_i \equiv \vec{B}_i / |\vec{B}_i|$,	
$\Easy(\e) \equiv E(\e;\Dasy,\masy)$, and 
\begin{align}\label{QPNRG}
	E(\e;\Do,\mu)
	=	
	\sqrt{\left(\frac{\e}{2} - \mu\right)^2 + \e \Do^2}.
\end{align}
In Eq.~(\ref{GappedPrec}), $\gamma_i$ is the ``Cooper pair distribution,''
which measures the projection of the pseudospin onto $-\vec{B}_i$. 
[Note that this is different from the definition employed in the
gapless case, Eq.~(\ref{GaplessPrec}).]
In the ground state, $\gamma_i = -1$ for all pseudospins,
while the configuration with $\gamma_i = +1$ for all $i$ is a metastable
negative temperature state.
We refer to a spin with $\gamma_i = -1$ as a ground state pair,
while $\gamma_i = +1$ is an excited pair.\cite{DzeroYuzbashyan07}

\begin{figure}
   \includegraphics[width=0.35\textwidth]{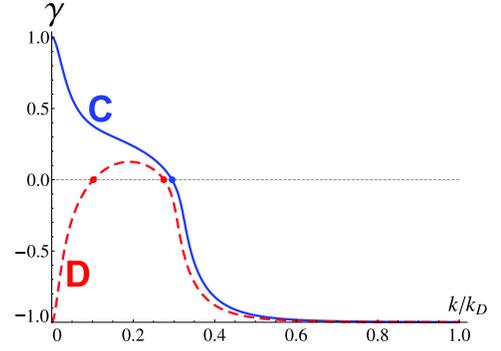}
   \caption{
	Non-equilibrium winding number: Parity of zeroes in the 
	Cooper pair distribution function of phase {\bf II},
	for representative quenches in regions {\bf C} and {\bf D} 
	[Fig.~\ref{Fig--PhaseDiagSec}].
	The Cooper pair distribution function
	is the pseudospin projection 
	$\gamma(k)$ 
	at each
	momentum
	$k$ onto the effective field 
	$-\vec{B}(\e = k^2)$.
	The latter
	encodes the asymptotic global parameters 
	$\{\Dasy,\masy\}$, see Eqs.~(\ref{GndField}) and (\ref{GappedPrec}).
	Distributions {\bf C} (blue solid line) and {\bf D} (red dashed line) belong to 
	quenches in 	
	the corresponding
	regions 
	highlighted		
	in Fig.~\ref{Fig--PhaseDiagSec}. 
	Whenever $Q \neq W$ ($Q = W$), $\gamma(k)$ exhibits an odd (even) number of zeroes,
	illustrated here by curve {\bf C} ({\bf D}); c.f.\ Figs.~\ref{Fig--PhaseDiagQ} and \ref{Fig--PhaseDiagW}.
	The parity of these zeroes is a $\mathbb{Z}_2$-valued quantum number that can 
	in principle	
	be extracted from an RF spectroscopic measurement in a cold atomic realization of the quench.
	The quench coordinates for curves C and D are  
	$\{\Di,\Df\} = \{1.65,0.359\}$
	and 
	$\{1.47, 0.365\}$, 
	respectively.
 	Both curves were obtained from the analytical solution.
	}
   \label{Fig--ZeroParity1}
\end{figure}

For the quench, we compute $\gamma_i$ exactly in the limit $t \rightarrow \infty$ 
using the conservation of the Lax vector norm.  
We find that $\gamma_i$ exhibits an odd number of zeroes whenever 
$W \neq Q$, whereas the number of zeroes is even (and typically zero) 
when $W = Q$. Two examples are shown in Fig.~\ref{Fig--ZeroParity1}:
The first is a quench across the quantum critical point from strong-to-weak
pairing, region {\bf C} in Fig.~\ref{Fig--PhaseDiagSec} with $W = 1$ and $Q = 0$.
The second is a strong-to-weak quench within the BCS phase, region {\bf D}
in Fig.~\ref{Fig--PhaseDiagSec} with $W = Q = 1$. 

The presence of an odd number of zeroes in $\gamma_i$ is required by
the conservation of the pseudospin winding $Q$. The effective field in Eq.~(\ref{GndField})
``winds'' when $W = 1$ ($\masy >0$). When $W$ obtained in the asymptotic steady state
differs from its initial, pre-quench value, $\gamma_i$ must also ``wind'' so that
the pseudospin index $Q$ is conserved. 
Thus the parity of the number of zeroes in $\gamma_i$ constitutes a new $\mathbb{Z}_2$-valued
index that encodes the retarded Green's function invariant $W$, which can change following 
a quench.

The Cooper pair distribution appears in the amplitude for photon absorption or emission
via RF spectroscopy.\cite{DzeroYuzbashyan07} 
In a cold atomic realization, absorption of an RF photon can destroy an 
Anderson pseudospin by breaking a Cooper pair. The photon is absorbed by one partner,
which is subsequently excited to a different internal state, denoted below as ``3.'' 
An atom in state 3 does not participate in pairing.

The RF-induced tunneling Hamiltonian is
\begin{align}
	H_T = 
	T
	\sum_{\vex{k}}
	\left[
	e^{i \omega_L t} c^\dagger_{\vex{k}} d_{\vex{k}} 
	+
	e^{-i \omega_L t} d^\dagger_{\vex{k}} c_{\vex{k}} 
	\right].
\end{align}
Here $d_{\vex{k}}$ annihilates a state 3 atom with momentum $\vex{k}$
and $\omega_L > 0$ is the frequency of the RF source. 
We denote the total number of state 3 atoms by
\[
	N_d 
	\equiv
	\sum_{\vex{k}}
	d^\dagger_{\vex{k}} d_{\vex{k}}.
\]
In the asymptotic steady-state following a quench in phase {\bf II}, a straight-forward 
linear response calculation gives the RF current
\begin{align}
	\label{Current}
	\left\langle \frac{d}{d t} N_d(t) \right\rangle
	=&\,
	2 \pi \dos T^2
	\frac{\e_\omega \Dasy^2}{|\omega||\omega + \Dasy^2|}
	\theta(\e_\omega)
	A(\omega),
\end{align}
where
\begin{align}\label{A}
	A(\omega)
	=&
	\left\{
	\begin{aligned}
	&
	[1 - \gamma(\e_\omega)]
	[1 - n^{\pup{d}}(\e_\omega)]
	\\&
	-
	[1 + \gamma(\e_\omega)]
	n^{\pup{d}}(\e_\omega)
	\end{aligned}
	\right\}
	\,
	\theta(\omega) 
	\nonumber\\
	&
	+
	\,
	\left\{
	\begin{aligned}
	&
	[1+\gamma(\e_\omega)]
	[1 - n^{\pup{d}}(\e_\omega)]
	\\&
	-
	[1 - \gamma(\e_\omega)]
	n^{\pup{d}}(\e_\omega)
	\end{aligned}
	\right\}
	\,
	\theta(-\omega).
\end{align}
In these equations, $\theta(\e)$ denotes the Heaviside unit step function,
and $n^{\pup{d}}(\e = k^2)$ is the initial occupation of atoms in state 3, 
equal to $\langle d^\dagger_{\vex{k}} d_{\vex{k}} \rangle_0$. 
The frequency $\omega$ is defined by
\[
	\omega \equiv \omega_L - \eRF,
\]
where $\eRF > 0$ denotes the atomic transition energy between states 3 and 2. 
We ignore high frequency processes that involve counter-rotating terms with 
$\omega_L \rightarrow -\omega_L$.
In Eq.~(\ref{Current}), $\dos$ denotes the bare 
density of states [Eq.~(\ref{nuDef})] and the mode energy 
$\e_\omega/2$ is defined below in Eq.~(\ref{eomega}).

Eq.~(\ref{A}) follows from simple kinematics.
The first term proportional to $[1 - n^{\pup{d}}]$ 
describes the process wherein a photon with energy 
$\omega_L > \eRF$ 
is absorbed by a ground state Cooper pair with initial energy 
$-E(\e_\omega)$, exciting one partner to state 3 with energy 
$\e_\omega/2 + \eRF - \masy$. The remaining unpaired fermion 
carries energy zero, since $s_{\vex{k}}^a c^\dagger_{\vex{k}} \ket{0} = s_{\vex{k}}^a c^\dagger_{-\vex{k}} \ket{0} = 0$,
where modes $\{\vex{k},-\vex{k}\}$ are vacant in $\ket{0}$. 
The conservation of energy gives 
\[
	\omega - \Easy(\e_\omega) = \frac{\e_\omega}{2} - \masy,	
\]
which has the unique solution
\begin{align}\label{eomega}
	\e_\omega = \frac{\omega(\omega + 2 \masy)}{\Dasy^2 + \omega}.
\end{align}
The second term in Eq.~(\ref{A}) proportional to $n^{\pup{d}}$ 
is the inverse stimulated emission process.
The third term describes the destruction of an excited state Cooper pair
due to a photon absorption with $\omega_L < \eRF$, 
again creating a state 3 atom with energy $\e_\omega/2 + \eRF - \masy$ and 
an unpaired particle with zero energy.
Energy balance is
\[
	\omega + \Easy(\e_\omega) = \frac{\e_\omega}{2} - \masy,
\]
with $\e_\omega$ again given by Eq.~(\ref{eomega}) and $\omega < 0$.
The fourth term is the inverse emission process. 
The factor $\theta(\e_\omega)$ in Eq.~(\ref{Current}) disallows
unphysical processes requiring negative mode energies.

Conceptually, the simplest situation has an initially empty state 3 band;
then $n^{\pup{d}}(\e) = 0$ for all $\e$. Eq.~(\ref{A}) 
implies that the Cooper pair distribution function $\gamma(\e)$ can
in principle be extracted from the RF spectroscopy current.
However, this result ignores complications involving ``off-diagonal'' 
processes\cite{DzeroYuzbashyan07} that can become important for transitions involving 
states far from the Fermi energy. We defer a full treatment to future work.

\section{Quench phase diagram \label{Sec: QPD}}

In this section, we derive the quench phase diagram in Fig.~\ref{Fig--PhaseDiagBasic}.
Ground state properties of the model
are reviewed in Appendix~\ref{Sec: APP--GND}.


\subsection{Lax construction, spectral polynomial, separation variables \label{Sec: Lax}}

Our starting point is the ``1D'' Hamiltonian in Eq.~(\ref{H}).
This can be derived from the 2D chiral p-wave model\cite{Richardson02,Sierra09} in Eq.~(\ref{Hchiral})
by a canonical rescaling of the pseudospins,
\begin{align}\label{PolarRescale}
	s_{\vex{k}}^- \rightarrow \exp(-i \phi_k) \, s_{\vex{k}}^-,
	\;\;
	s_{\vex{k}}^+ \rightarrow \exp(i \phi_k) \, s_{\vex{k}}^+,
\end{align}
where $\phi_k$ is the polar angle of $\vex{k}$.
Applying Eq.~(\ref{PolarRescale}) to $H$ in Eq.~(\ref{Hchiral}) eliminates the
phases of the complex momenta appearing in the pairing term. 
Pseudospins with the same momentum radius evolve collectively. 
For each $k$, we sum spins along the arc in Eq.~(\ref{spinArc}) 
to obtain a single radial pseudospin $\vec{s}_k$. 
The Hamiltonian reduces to Eq.~(\ref{H}), where $\e_i \equiv k_i^2$ and
$\vec{s}_i \equiv \vec{s}_{k_i}$. 
At any time $t$ following a quench, the full 2D spin configuration is easily reconstructed.

In what follows, we switch frequently between discrete and continuum
formulations of the problem. The connection is given by
\begin{align}\label{ContDisc}
	\sum_{i} \Leftrightarrow \dos \inte,
\end{align}
where 
\begin{align}\label{nuDef}
	\dos \equiv \frac{\ls^2}{8 \pi}
\end{align}
is the (bare) density of states and $\ls$ denotes the linear system size.
On the right-hand side of Eq.~(\ref{ContDisc}), $\Lambda$ is the high-energy cutoff;
the chemical potential is incorporated here as a convenience, see Appendix~\ref{Sec: APP--GND}. 
Using these conventions, all spins have $(\vec{s}_i)^2 = 1/4$.

Although the model in Eq.~(\ref{H}) is classically integrable as we demonstrate below,
the spin equations of motion in Eq.~(\ref{spinEOM}) are not directly useful. 
Instead, we introduce a new Lax vector construction inspired by the s-wave case\cite{Richardson,Gaudin,YKA05,YuzbashyanAltshuler05,YuzbashyanAltshuler06} 
and the Bethe ansatz formulation\cite{Sierra09,Sierra10,Ortiz10} of the p-wave model. 

For a system of $N$ pseudospins, we define the Lax vector components 
\begin{align}\label{LaxDef}
\begin{aligned}
	L^{+}(u)
	\equiv&\,
	\sum_{i = 1}^N
	\frac{\sqrt{\e_i} s_i^{+}}{\e_i - u},
	\\
	L^{-}(u)
	\equiv&\,
	\sum_{i = 1}^N
	\frac{\sqrt{\e_i} s_i^{-}}{\e_i - u},
	\\
	L^{z}(u)
	\equiv&\,
	\sum_{i = 1}^N
	\frac{\e_i s_i^{z}}{\e_i - u} + \frac{1}{2 G},
\end{aligned}
\end{align}
where $G$ is the interaction strength in Eq.~(\ref{H}).
In these equations, $u$ denotes an arbitrary complex-valued
parameter.
We also introduce a Lax vector norm:
\begin{align}\label{L2Def}
	L_{2}(u) \equiv u L^+(u) L^-(u) + [L^z(u)]^2.
\end{align}
Unlike the s-wave case,\cite{Richardson,Gaudin,YKA05,YuzbashyanAltshuler05,YuzbashyanAltshuler06} 
the norm is not Euclidean. This is a key distinction that produces a different structure for 
isolated roots of the spectral polynomial, defined below.

Employing canonical Poisson bracket relations for the spins 
\begin{equation}\label{SpinPoisson}
	\{s_i^a,s_j^b\} = \delta_{i j} \epsilon^{a b c} s_j^c,	
\end{equation}
it is easy to show that
\begin{align}\label{LaxPB}
\begin{aligned}
	\{L^{+}(u),L^{-}(v)\}
	=&
	2 \left[\frac{L^z(u) - L^z(v)}{u - v}\right],
	\\
	\{L^{z}(u),L^{+}(v)\}
	=&
	\frac{u L^{+}(u) - v L^{+}(v)}{u - v},
	\\
	\{L^{z}(u),L^{-}(v)\}
	=&
	-\frac{u L^{-}(u) - v L^{-}(v)}{u - v}.
\end{aligned}
\end{align}
These in turn imply that
\begin{align}\label{LaxNormPB}
	\{L_2(u),L_2(v)\} = 0.
\end{align}

The Lax norm $L_2(u)$ is a generator for integrals of motion.
Explicitly,
\begin{align}\label{L2Explicit}
	L_2(u)
	=&\,
	\sum_{i}
	\frac{H_i}{u - \e_i}
	+
	\sum_{i}
	\frac{
	\e_i^2
	}{4(u - \e_i)^2}
	+ 
	\frac{1}{4 G^2},
\end{align}
where $H_i$ denotes a central-spin type Hamiltonian,
\begin{align}
	H_i
	=&\,
	-
	\frac{1}{G}
	\e_i
	s^z_{i}
	+
	\e_i
	s^+_{i} s^-_{i}
	\nonumber\\
	&\,
	+
	\sum_{j \neq i}
	\left\{
	\frac{
	\e_i \sqrt{\e_i \e_j} \left[s^+_{i} s^-_{j} + s^+_{j} s^-_{i} \right]
	+
	2 \e_i \e_j
	s^z_{i} s^z_{j}}{(\e_i - \e_j)}
	\right\}.
\end{align}
There are $N$ independent $H_i$'s in a system of $N$ spins. 
Because Eq.~(\ref{LaxNormPB}) holds for generic $u$ and $v$, it implies
that the $H_i$'s are mutually conserved:
\begin{align}\label{ConsHis}
	\{H_i,H_j\} = 0.
\end{align}
The BCS Hamiltonian in Eq.~(\ref{H}) is given by the sum
\begin{align}
	H 
	=
	-G
	\sum_{i}
	H_{i}.
\end{align}
For the spin dynamics generated by $H$  
[Eq.~(\ref{spinEOM})], the Lax components evolve according to 
\begin{align}
	\label{LaxEOMDet}
	\frac{d\vec{L}(u)}{d t}
	=
	\det
	\begin{bmatrix}
	\hat{x} & \hat{y} & u \hat{z} \\
	2 \Delta^x & 2 \Delta^y & u \\
	L^x(u) & L^y(u) & L^z(u)
	\end{bmatrix},
\end{align}
where $\vec{L} \equiv \hat{x} L^x + \hat{y} L^y + \hat{z} L^z$ and 
$L^{\pm} = L^x \pm i L^y$.

We define the spectral polynomial 
\begin{align}\label{qpDef}
	\qp(u)
	\equiv
	G^2
	\prod_{j = 1}^{N}
	(u - \e_j)^2
	L_2(u).
\end{align}	
This is a polynomial of degree $2N$ in $u$, with coefficients that
depend upon (a) the coupling strength $G$ and (b) the pseudospin configuration $\{\vec{s}_i\}$.
Eq.~(\ref{LaxEOMDet}) implies that $L_2(u)$ and $\qp(u)$ are integrals of motion.

For a quench, the roots of $\qp(u)$ provide the key to determine the long-time asymptotic dynamics.
Part of the story involves trading the spins for a more convenient set of coordinates.
From Eq.~(\ref{LaxDef}), we write
\begin{align}\label{SepVars}
	L^{-}(u)
	=
	\frac{\Delta}{G}
	\frac{\prod_{\beta=1}^{N-1}(u - u_\beta)}{\prod_{j = 1}^{N}(u - \e_j)}.
\end{align}
In this equation, we have formed a common denominator. The numerator is a polynomial in $u$ of degree $N - 1$
with zeroes $\{u_{\beta}\}$, which we term 
\emph{separation variables}.\cite{sklyanin,vadim,YuzbashyanAltshuler05}		
Each $u_\alpha$  is a complicated function of all $N$ $\{s_i^{-}\}$, the precise form of which we will not need.
The separation variables satisfy the Poisson bracket relations $\{u_{\alpha},u_{\beta}\} = 0$.
The prefactor in Eq.~(\ref{SepVars}) follows by expanding the numerator and matching the coefficient of $u^{N-1}$
with Eq.~(\ref{LaxDef}), using Eq.~(\ref{DeltaDef}).

The BCS evolution of the Lax vector in Eq.~(\ref{LaxEOMDet}) implies that 
\[
	\frac{d L^{-}(u)}{d t} 
	=
	-
	i
	\left[
	u
	L^{-}(u) 
	-
	2
	\Delta L^{z}(u)
	\right]
	+
	\frac{\partial L^{-}(u)}{\partial u}
	\left(\frac{d u}{d t}\right),
\]
allowing for a time-dependent parameter $u$. Evaluating this equation for a separation variable gives
\begin{align}
	0
	=&\,
	2 i
	\Delta L^{z}(u_\alpha)
	+
	\frac{\Delta}{G}
	\frac{\prod_{\beta \neq \alpha}(u_\alpha - u_\beta)}{\prod_{j = 1}^{N}(u_\alpha - \e_j)}
	\left(\frac{d u_\alpha}{d t}\right).
\end{align}
Using Eqs.~(\ref{L2Def}) and (\ref{qpDef}), we obtain the equations of motion
\begin{align}\label{SepVarEOM}
	\frac{d u_\alpha}{d t}
	=
	-2 i
	\frac{\sqrt{\qp(u_\alpha)}}{\prod_{\beta \neq \alpha}(u_\alpha - u_\beta)}.	
\end{align}
The spins have been entirely eliminated in favor of coupled equations for the
separation variables. 
One can also derive the following equation of motion for 
$\Delta(t)$,
employing ~Eqs.~(\ref{spinEOM}), (\ref{DeltaDef}), and (\ref{SepVars}): 
\begin{align}\label{GapEOM}
	\frac{d \Delta}{d t}
	=&\, 
	i
	\Delta
\left[
	\sum_{\beta = 1}^{N-1}
	u_\beta
	-
	\sum_{j} \e_j 
	- 
	2 G
	H
	- 
	2
	|\Delta|^2
\right].
\end{align}

Separation variables are in general complex-valued; to solve the
equations of motion in (\ref{SepVarEOM}), one has to choose a proper branch of 
$\sqrt{\qp(u)}$ in the plane of complex $u$. 
This can be done by connecting pairs of roots of the polynomial $\qp(u)$ with branch cuts. 
Separation variables cannot cross these cuts in their motion. Note that the number of separation 
variables ($N-1$) is one less than the number of the branch cuts ($N$).


\subsection{Ground state roots and spectral transitions \label{Sec: GNDRoots}}

In the BCS or BEC
ground state, the pseudospins satisfy
\begin{gather}\label{spinGND}
	s^{-}_i
	=
	-
	\frac{
	\sqrt{\e_i} \Do
	}{2 E(\e_i;\Do,\mu)},	
	\;\;
	s^z_i
	=
	-
	\frac{
	(\e_i - 2 \mu) 
	}{4 E(\e_i;\Do,\mu)},
\end{gather}
where $E(\e;\Do,\mu)$ is the quasiparticle energy, defined by Eq.~(\ref{QPNRG}).
We take $\Do$ real and positive without loss of generality.
Spin $\vec{s}_i$ lies along the field $\vec{B}_i + 2 \mu \hat{z}$, which incorporates $\mu$ 
so as to fix the total density $n$.
The 
pairing amplitude
in Eq.~(\ref{DeltaDef}) solves the BCS equation
\begin{align}\label{BCSEqGap} 
	\frac{1}{G} = \sum_{i} \frac{\e_i}{2 E(\e_i;\Do,\mu)}.
\end{align}
Eq.~(\ref{muGND}) determines $\mu$ in terms of the density $n$ and 
$\Do$.
Eq.~(\ref{BCSEqF}) relates these to the interaction strength, see Appendix~\ref{Sec: BCSEq} for details.

We evaluate the ground state spectral polynomial [Eq.~(\ref{qpDef})] by combining Eqs.~(\ref{spinGND}),
(\ref{LaxDef}), and (\ref{L2Def}). The result is
\bsub\label{qpGND}
\begin{align}
	\qp(u) 
	=&\,
	\frac{G^2}{4}
	(u - \uoqa{+})
	(u - \uoqa{-})
	\left[P_{N-1}(u)\right]^2,
	\\
	P_{N-1}(u)
	\equiv&\,
	\prod_{j = 1}^{N}(u - \e_j)
	\,
	F(u;\Do,\mu),
\end{align}
\begin{align}
	F(u;\Do,\mu)
	\equiv&\,
	\sum_{i = 1}^{N}
	\frac{\e_i}{2 (u - \e_i) E(\e_i;\Do,\mu)}.
	\label{FDef}
\end{align}
\esub
$P_{N-1}(u)$ denotes a polynomial of degree $N-1$ in $u$. 
The zeroes of $F(u)$ fall between adjacent mode energies $\{\e_j,\e_{j+1}\}$,
the latter non-negative and non-degenerate.
The $N-1$ distinct, positive real zeroes of $F(u)$ are the roots of $P_{N-1}(u)$.
Each of these is a doubly-degenerate root of $\qp(u)$. 
The remaining two roots $\uoqpm$ solve $E(u;\Do,\mu) = 0$ [Eq.~(\ref{QPNRG})],
\begin{align}\label{IsoRootsGND}
	\uoqpm
	=
	2 \left[\mu - (\Do)^2 \pm \Do \sqrt{(\Do)^2 - 2 \mu}\right].
\end{align}
 
We first consider the $2(N-1)$ positive real roots of $\qp(u)$.
Eq.~(\ref{L2Def}) implies that each such root $\uoq$ also satisfies
$L^{-}(\uoq) = 0$, i.e.\ $\uoq$ is a separation variable [Eq.~(\ref{SepVars})].
Eq.~(\ref{SepVarEOM}) shows that this is a stationary solution to the equations of motion.
Thus, in the BCS ground state, the $N-1$ separation variables $\{u_\alpha\}$ are locked to 
the $N-1$ distinct, positive real roots of $\qp(u)$.

The single pair of isolated roots in Eq.~(\ref{IsoRootsGND}) encodes key macroscopic 
features of the superfluid state: the 
order parameter
$\Do$ and the chemical potential $\mu$. 
These are constrained by Eq.~(\ref{muGND}) for a fixed particle density $n$, Fig.~\ref{Fig--MuPlot}. 
Roots $\uoqpm$ take values away from the positive real axis for all $\Do > 0$
[except at the critical point $\Do = \Dqcp$, Eq.~(\ref{IsoRootsQCP}) below], and
cannot serve as stationary solutions for separation variables. 
Different from the s-wave case, the isolated roots for p-wave can be complex or negative
real, due to the non-Euclidean norm in Eq.~(\ref{L2Def}).

\begin{figure}[b]
   \includegraphics[width=0.35\textwidth]{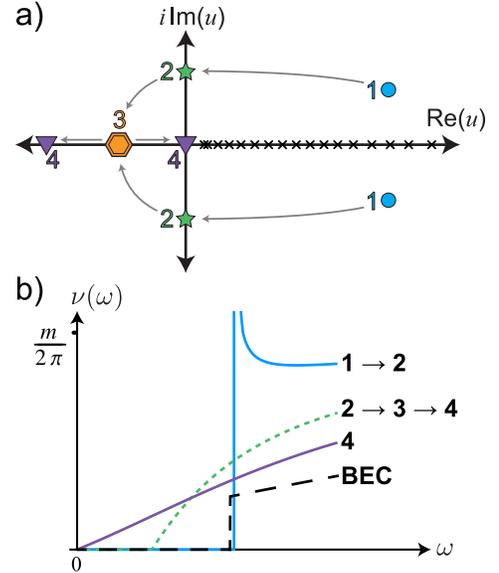}
   \caption{
	Isolated roots and ground state spectral transitions.
	Subfigure (a) shows the positions of the isolated root pair $\uoqpm$ 
	[Eq.~(\ref{IsoRootsGND})] in the ground state spectral polynomial, for increasing values of $\Do$.
	Particular root configurations are labeled (1)--(4).
	For $\Do < \Dcoh$ (1), the isolated roots are a conjugate pair with positive real part.
	At $\Do = \Dcoh$ (2), the roots become purely imaginary. 
	At $\Do = \Dmr$ 
	(``Moore-Read,'' see text)		
	(3), the roots become negative real and degenerate.
	For $\Dmr < \Do < \Dqcp$, the roots split and travel along the negative real axis.
	At $\Do = \Dqcp$ (4), the retreating root hits zero.
	The thresholds $\{\Dcoh,\Dmr\}$ lie within the BCS phase ($\mu > 0$), while 
	$\Dqcp$ marks the topological transition. All are defined explicitly in Appendix~\ref{Sec: TDOS},
	Eqs.~(\ref{DeltaVals}) and (\ref{DDefs}).
	For $\Do > \Dqcp$ (BEC), both roots are again non-degenerate and negative real.
	The zero temperature tunneling density of states $\nu(\omega)$ is shown for the corresponding
	root positions in subfigure (b). The weak pairing coherence peak 
	visible in the trace $(1 \rightarrow 2)$ 
	disappears for
	$\Do \geq \Dcoh$. The difference between the ``soft'' and ``hard'' gaps 
	in the BCS ($2 \rightarrow 3 \rightarrow 4$) and BEC regimes is a coherence
	factor effect. The spectrum is gapless at $\Do = \Dqcp$ (4).
	}
   \label{Fig--Roots and TDOS}
\end{figure}

For the p-wave model, the pattern of isolated roots is tied to the strength of 
the pairing as measured by $\Do$. Three special values $\Dcoh < \Dmr < \Dqcp$ separate 
four domains. These are implicitly defined through Eq.~(\ref{DeltaVals}) in Appendix~\ref{Sec: TDOS}, 
which specifies the relation of each to the chemical potential.
All three 
pairing amplitudes
$\{\Dcoh,\Dmr,\Dqcp\}$ are of order $\sqrt{4 \pi n / \log(\Lambda/ 2 \pi n)}$, 
and differ by terms of size $\sqrt{n}[\log(\Lambda/ 2 \pi n)]^{-3/2}$; explicit values are transcribed in Eq.~(\ref{DDefs}).
In the following, we describe the four
pairing domains in terms of the isolated roots,
and via spectral features detectable in the tunneling density of states (TDoS).
Both are depicted in Fig.~\ref{Fig--Roots and TDOS}.
 
For the weakest coupling  
strengths such that $\Do < \Dcoh$,
the minimum of the quasiparticle spectrum $E(\e;\Do,\mu)$ occurs at a non-zero mode energy 
$\esp \equiv 2(\mu - \Do^2)$. The quasiparticle gap is
\begin{align}\label{EminDef--WP}
	\Emin 
	\equiv&\,
	E(\esp;\Do,\mu)
	=
	\Do \sqrt{2 \mu - \Do^2}, 
	\;\; \Do < \Dcoh.
\end{align}
In this regime, the TDoS $\nu(\omega)$ possesses a coherence peak at the gap edge,
Eq.~(\ref{TDOS}) in Appendix~\ref{Sec: TDOS} and Fig.~\ref{Fig--Roots and TDOS}(b). 
The corresponding isolated roots form a complex conjugate pair with positive real part. 
At $\Do = \Dcoh$ [where $\mu = (\Dcoh)^2$, Eq.~(\ref{DeltaVals})], 
the TDoS coherence peak and the real part of $\uoqpm$ both vanish. 
For stronger pairing, the bulk quasiparticle gap is determined by the chemical potential,
and resides at $\e = 0$:
\begin{align}\label{EminDef--SP}
	\Emin 
	=&\,
	E(0;\Do,\mu)
	=
	|\mu|, 
	\;\; \Do \geq \Dcoh.
\end{align}

Increasing the interaction strength (yet remaining within the topologically non-trivial BCS phase), 
the roots move into the left-hand complex plane. At $\Do = \Dmr$ [$\mu = (\Dmr)^2/2$], the roots collapse to a
degenerate value on the negative real axis. This is a point (for fixed density) on the ``Moore-Read'' 
line discussed in Ref.~\onlinecite{Sierra09}.
Although gapped, the spectrum $E(\e;\Dmr,\mu)$ exhibits zero curvature (Appendix~\ref{Sec: TDOS}). As $\Do$ is further increased
towards the topological phase transition at $\Dqcp$, the roots split along the negative real axis;
$\uoqa{+}$ ($\uoqa{-}$) becomes less (more) negative. 
At the critical point $\Do = \Dqcp$ and $\mu = 0$, the quasiparticle gap vanishes as indicated by the TDoS in Fig.~\ref{Fig--Roots and TDOS}(b).
The roots are 
\begin{align}\label{IsoRootsQCP}
	\uoqa{+} = 0, 
	\;\; 
	\uoqa{-} = - 4 (\Dqcp)^2.
\end{align}
Entering the BEC phase with $\Do > \Dqcp$ (and $\mu < 0$),
$\uoqa{+}$ retreats back along the negative axis. Throughout the BEC phase, the TDoS is 
gapped, and $\uoqpm$ remain non-degenerate and negative real.

Thus, the three thresholds $\Do = \{\Dcoh,\Dmr,\Dqcp\}$ correspond to three special configurations of the
isolated roots in the $p + i p $ ground state, marked (2), (3), and (4) in Fig.~\ref{Fig--Roots and TDOS}(a).
Non-equilibrium extensions of all three appear in the quench phase diagram as the lines marked 
$\beta_{\msf{coh}}$, $\beta_{\msf{MR}}$, and $\beta_{\msf{QCP}}$ in Fig.~\ref{Fig--PhaseDiagFull}, discussed below.


\subsection{
Roots of the spectral polynomial and the asymptotic behavior	
\label{Sec: LaxRed}}

An instantaneous quench of the BCS coupling strength sends $\Gi \rightarrow \Gf$ in Eq.~(\ref{H}).
The initial condition is taken as the (BCS or BEC) $p + i p$ ground state of the pre-quench Hamiltonian.
Following the quench, the spins evolve according to Eq.~(\ref{spinEOM}), and $\Delta(t)$ is self-consistently
determined by (\ref{DeltaDef}), with $G = \Gf$. 

We label the strength of the quench by $\beta$, defined in terms of 
$\{\Gi,\Gf\}$ via Eqs.~(\ref{betaDef}) and (\ref{gtoG}). 
We denote the corresponding spectral polynomial as $\qp(u;\beta)$ [Eq.~(\ref{qpDef})]. This is a function of the instantaneous spin state 
$\{\vec{s}_i(t)\}$. Because it is an integral of motion, we can evaluate $\qp(u;\beta)$ at $t = 0$ in terms of the pre-quench
ground state spin configuration in Eq.~(\ref{spinGND}), wherein the initial pairing amplitude $\Di$ 
is related to $\Gi$ via the BCS Eq.~(\ref{BCSEqGap}). 

In the ground state, all but two of the $2N$ roots of $\qp(u;0)$ reside along the positive
real axis; the remaining isolated 
roots $\uoqpm$ in Eq.~(\ref{IsoRootsGND}) are determined by 
the pairing amplitude $\Do$ and the chemical potential $\mu$. 
Various equilibrium spectral transitions (including the topological BCS-BEC transition) are encoded
in the isolated root positions, see Fig.~\ref{Fig--Roots and TDOS}.

Because a quench is a violent perturbation to the many-pair superfluid, one expects to find 
a different pattern of roots in $\qp(u;\beta)$ for any $\beta \neq 0$. In particular, for a \emph{finite} number of 
spins, all of the real, doubly-degenerate, positive ground state roots split into complex 
conjugate pairs for an arbitrarily weak quench. However, the splitting for most roots turns 
out to be small, of the order of the level spacing. In fact, when $\qp(u;\beta)$ is evaluated 
at points $\{\Di,\Df\}$ throughout the quench phase diagram in Fig.~\ref{Fig--PhaseDiagBasic}, 
for even a modest number (e.g.~100) of spins, one finds that all but a few roots always cluster 
around the positive real axis, even for ``large'' quenches. As $\beta$ carries units of density, 
a large quench has $|\beta| \gg n$.

For all quenches depicted in Fig.~\ref{Fig--PhaseDiagBasic}, including those across the topological 
quantum phase transition (e.g., $\Di > \Dqcp$ and $\Df < \Dqcp$), we find that $\qp(u;\beta)$ exhibits zero, one, 
or two \emph{isolated} pairs of roots. An isolated pair is well-separated from the positive real $u$ axis, as is the case
for the ground state pair $\uoqpm$ in Eq.~(\ref{IsoRootsGND}). 
The following picture therefore emerges, identical to the 
s-wave\cite{YKA05,YuzbashyanAltshuler05,YuzbashyanAltshuler06,BarankovLevitov06,DzeroYuzbashyan06} 
case: The pattern of roots in $\qp(u;\beta)$ for a quench  
is similar to that of the ground state, except that the number $M$ of isolated root pairs can change. 
There is also a small splitting of the remaining $2(N-M)$ roots away from the positive real $u$-axis. 

The importance of the spectral polynomial roots can be appreciated from the following argument. 
Suppose $\qp(u)$ has a positive real zero $\uoq$, i.e.\ ${\cal Q}_{2N}(\uoq)=0$. 
Zeroes of the spectral 
polynomial coincide with the zeros of $L_2(u)$ and because by definition 
this quantity is 
nonnegative	
when $u>0$, any real positive root of $L_2(u)$ must also be a double root. 
Further, since both terms in 
Eq.~(\ref{L2Def}) are nonnegative for $u>0$, it follows that $\uoq$ must be a root of 
both $L^z(u)$ and $L^-(u)$. 
But the roots of $L^-(u)$ are defined to be the separation variables. This implies that 
one of the separation variables must coincide with the positive real root: 
$u_\beta(t) =\uoq$. It is then ``frozen'' in time; this is consistent with the equations of motion 
because both sides of Eq.~(\ref{SepVarEOM}) vanish for $u_\beta(t) = \uoq = \mbox{const.}$

Note that $u_\beta = \uoq$ also drops out from the equations of motion for the remaining separation 
variables. Indeed, we have 
$\qp(u_\alpha) = \left(u_\alpha - \uoq\right)^2 {\cal Q}_{2N-2}(u_\alpha)$ 
because $\uoq$ is a double root of $\qp(u)$; the factor of $(u_\alpha - \uoq)$ 
in the numerator of Eq.~(\ref{SepVarEOM}) cancels $(u_\alpha - u_\beta)$ in the denominator. Note also 
that the order of the spectral polynomial drops by two. It turns out that this kind of reduction occurs 
for our quench initial conditions in the continuum, $N \to \infty$, limit. As a result the order of 
the spectral polynomial drops dramatically to either 0, 2 or 4, and the resulting equations of motion 
can be explicitly solved.

Suppose there are $N-M$ real double roots $\{\uoqa{\beta}\}$. 
The spectral polynomial in Eq.~(\ref{qpDef}) then reduces to
\begin{equation}\label{qpResTor} 
	\qp(u) = \prod_{\beta=1}^{N-M} \left( u -  \uoqa{\beta} \right)^2 {\cal Q}_{2M}(u),
\end{equation}
where ${\cal Q}_{2M}(u)$ is a polynomial of order $2M$ whose roots are isolated. 
Now $N-M$ separation variables are equal to the roots, and the remaining nontrivial $M-1$ variables $u_\alpha$ satisfy,
as a consequence of Eq.~(\ref{SepVarEOM}),
\begin{equation}   
	\frac{d u_\alpha}{d t}  
	= 
	- 2i \frac{ \sqrt{ {\cal Q}_{2M}(u_\alpha) } 
	}{ 
	\prod_{\beta \not = \alpha}\left( u_\alpha- u_\beta \right)}.
\end{equation}
These are the reduced equations of motion for the remaining $M-1$ separation variables, which have the same form 
as the original equations of motion for $N-1$ variables. 

It is possible to reduce the number of degrees of freedom for this problem 
by using an explicit
(``Lax reduction'')
procedure, whose outcome allows one to find the asymptotic behavior of the 
order parameter. We introduce $M$ collective spin variables 
$\vec \sigma_r$, where $r = 1, 2, \dots, M$. They satisfy the same Poisson bracket relations as the 
original spins, Eq.~(\ref{SpinPoisson}). 
The collective spins have their own Lax vector $\vec L_{\sigma}$ defined analogously to 
Eq.~(\ref{LaxDef}) as
\begin{align}\label{LaxRedDef}
\begin{aligned}
	L_\sigma^{\pm}(u)
	\equiv&\,
	\sum_{r = 1}^M
	\frac{
	\sqrt{\chi_r} \,
	\sigma^{\pm}_{r}}{\chi_r - u},
	\\
	L_\sigma^{z}(u)
	\equiv&\,
	\sum_{r = 1}^M
	\frac{
	\chi_r
	\sigma^z_{r}}{\chi_r - u}
	+ 
	\frac{1}{2 G}.
\end{aligned}
\end{align}
Here the parameters $\{\chi_r\}$ are chosen in such a way that 
\begin{align}\label{LaxRed}
	\vec{L}(u)
	=
	\mathcal{A}(u)
	\,
	\vec{L}_\sigma(u),
\end{align}
where
\begin{align}\label{ALaxRed}
	\mathcal{A}(u)
	\equiv
	1
	+
	\sum_{j = 1}^{N}
	\frac{d_j}{\e_j - u}.
\end{align}
Matching the residues of the poles with Eq.~(\ref{LaxDef}), we require that
\bsub
\begin{gather}
	\label{chiDef}
	\sum_{j = 1}^{N}
	\frac{d_j}{\e_j - \chi_r}
	=
	-1,
	\;\;
	r \in \{1,\dots,M\},
	\\
\begin{gathered}[t]
	s^{\pm}_{i}
	=
	\frac{d_i}{\sqrt{\e_i}}
	L_\sigma^{\pm}(\e_i),
	\;\;
	s^z_{i}
	=
	\frac{d_i}{\e_i}
	L_\sigma^{z}(\e_i),
	\\
	i \in \{1,\ldots,N\}.
\end{gathered}
	\label{Redspin}
\end{gather}
\esub
The first Eq.~(\ref{chiDef}) constrains the parameters $\{\chi_r\}$, while
Eq.~(\ref{Redspin}) determines the coefficients
\begin{align}\label{diDef}
	d_i
	=&\,
	-
	\frac{\e_i \zeta_i
	}{
	2 \sqrt{
	L_2^{(\sigma)}(\e_i)
	}
	},	
\end{align}
where $L_2^{(\sigma)}(u) \equiv u L_\sigma^{+} L_\sigma^{-}(u) + (L_\sigma^{z})^2(u)$
and $\zeta_i \in \pm 1$. To obtain Eq.~(\ref{diDef}) we have used the fact that
$\vec{s}_i^2 = 1/4$.
The spectral polynomial indeed takes the form in Eq.~(\ref{qpResTor}).

The effective spin variables $\{\vec{\sigma}_r\}$ evolve according to the same 
Hamiltonian in Eq.~(\ref{H}) except with energies $\{\chi_r\}$ 
and with $M$ spins instead of the original $N$. 
The order parameter is expressed in terms of $\{\vec{\sigma}_r\}$ 
in the same way as $\vec{s}_i$ in Eq.~(\ref{DeltaDef}), that is
\begin{align}
	\Delta
	=
	-
	G
	\sum_{r = 1}^M
	\sqrt{\chi_r} 
	\sigma^{-}_{r}
	\sum_{i = 1}^N 
	\frac{d_i}{\chi_r - \e_i}
	=
	-
	G
	\sum_{r = 1}^M
	\sqrt{\chi_r} 
	\sigma^{-}_{r}.
\end{align}

As discussed above, the roots of the spectral polynomial for a quench $\qp(u;\beta)$ fall into two classes.
This equation can be studied numerically or analytically for some finite large value of $N$.	
Such a study 
reveals	
that most of the roots come in complex conjugate pairs that lie close to the real axis. 
Their imaginary parts scale as $1/N$ for large $N$. For each such root pair there is a separation variable 
that remains close to it (at a distance of order $1/N$) at all times. We call these variables continuum 
separation variables and the respective zeroes of $\sqrt{L_2(u)}$ continuum roots.
[Recall that $L_2(u)$ is proportional to $\qp(u;\beta)$ through Eq.~(\ref{qpDef}).]
In the thermodynamic limit $N \rightarrow \infty$, the continuum roots of $\sqrt{L_2}$ merge with its 
poles into a cut on the real axis. However, several of the zeros, which we can call the isolated roots, remain 
far from each other even in the thermodynamic limit. 

The contribution of the continuum separation variables to the equations of motion (\ref{SepVarEOM}) 
for the isolated ones as well as to Eq.~(\ref{GapEOM}) vanishes as $t\to\infty$. 
This can be shown explicitly assuming the joined Fourier spectrum of the continuum
separation variables is continuous. Then, for example, 
\[	
	\sum_\beta u_\beta \to \int A(\omega) e^{i\omega t} d\omega\to 0 
\] 
as $t\to\infty$.
Here the summation is over 
continuum separation variables only. Thus at large times $\Delta(t)$ and the isolated separation variables are given by an effective 
$M$-spin solution, as outlined above.	

With this information it is straightforward to construct the large time asymptotic solutions of the 
equations of motion. For example, in phase {\bf II} the spectral polynomial has one pair of isolated roots. 
That means that the reduced problem has only one collective spin ($M = 1$), and the 
order parameter behaves as $\Delta(t) = \Dasy \exp(-2 i \masy t)$, 
as follows from the solution of the equations of motion for just one spin. In turn, 
it is possible to relate $\Delta_\infty$ and $\mu_\infty$ to the position of that isolated root, by 
calculating the spectral polynomial for this single-spin problem. 
Phases {\bf I} and {\bf III} respectively correspond to zero and two-spin problems 
associated to zero and two pairs of isolated roots.   
The precise relations between the isolated roots and the order parameter dynamics in 
phases {\bf II} and {\bf III} are determined in Sec.~\ref{Sec: GapDyn}.


\subsection{Spectral polynomial and isolated roots for a quench}

The spectral polynomial is defined by Eqs.~(\ref{qpDef}), (\ref{L2Def}), and (\ref{LaxDef}), where explicit factors 
of the coupling $G$ take the post-quench value $G_f$.
We evaluate this in terms of the initial pre-quench state in Eq.~(\ref{spinGND})
with $\{\Do,\mu\} = \{\Di,\mui\}$, where the chemical potential $\mui$ 
is that associated to the initial pre-quench order parameter $\Di$ via Eq.~(\ref{muGND}).
The result is
\bsub\label{qpQuench}
\begin{align}
	\qp(u;\beta)
	=&\,
	G_f^2
	\prod_{j = 1}^{N}
	(u - \e_j)^2
	L_2(u;\beta),
	\\
	L_2(u;\beta)
	=&\, 
	\left(\frac{\dos}{2}\right)^2
	\Big\{
	\left[ \Ei(u) \ffi(u) \right]^2
	\nonumber\\
	&\,
	+
	\beta
	\left(u - 2 \mui \right)
	\ffi(u)
	+
	\beta^2
	\Big\},
	\label{L2quench}
\end{align}
\esub
where the ``quench parameter'' $\beta$ was defined in Eq.~(\ref{betaDef}). 
In Eq.~(\ref{qpQuench}), we have introduced the following quantities that
characterize the pre-quench state:
\begin{align}\label{qpQuenchDefs}
\begin{aligned}
	\Ei(u) =&\, E(u;\Di,\mui), 
	\\
	\ffi(u) =&\, \frac{2}{\dos} F(u;\Di,\mui),
\end{aligned}
\end{align}
where 
$F$ denotes the function appearing in the ground state spectral polynomial [Eq.~(\ref{FDef})], 
and $\dos$ is the density of states in Eq.~(\ref{nuDef}).

In the remainder of this section,
we demonstrate how to extract isolated roots from Eq.~(\ref{qpQuench}) in the thermodynamic limit,
and we establish the boundaries of the phase diagram in Fig.~\ref{Fig--PhaseDiagBasic}.
We also determine the non-equilibrium extensions of the special 
pairing
amplitudes
$\{\Dcoh,\Dmr,\Dqcp\}$ 
discussed in Sec.~\ref{Sec: GNDRoots}.

\begin{figure}
   \includegraphics[width=0.48\textwidth]{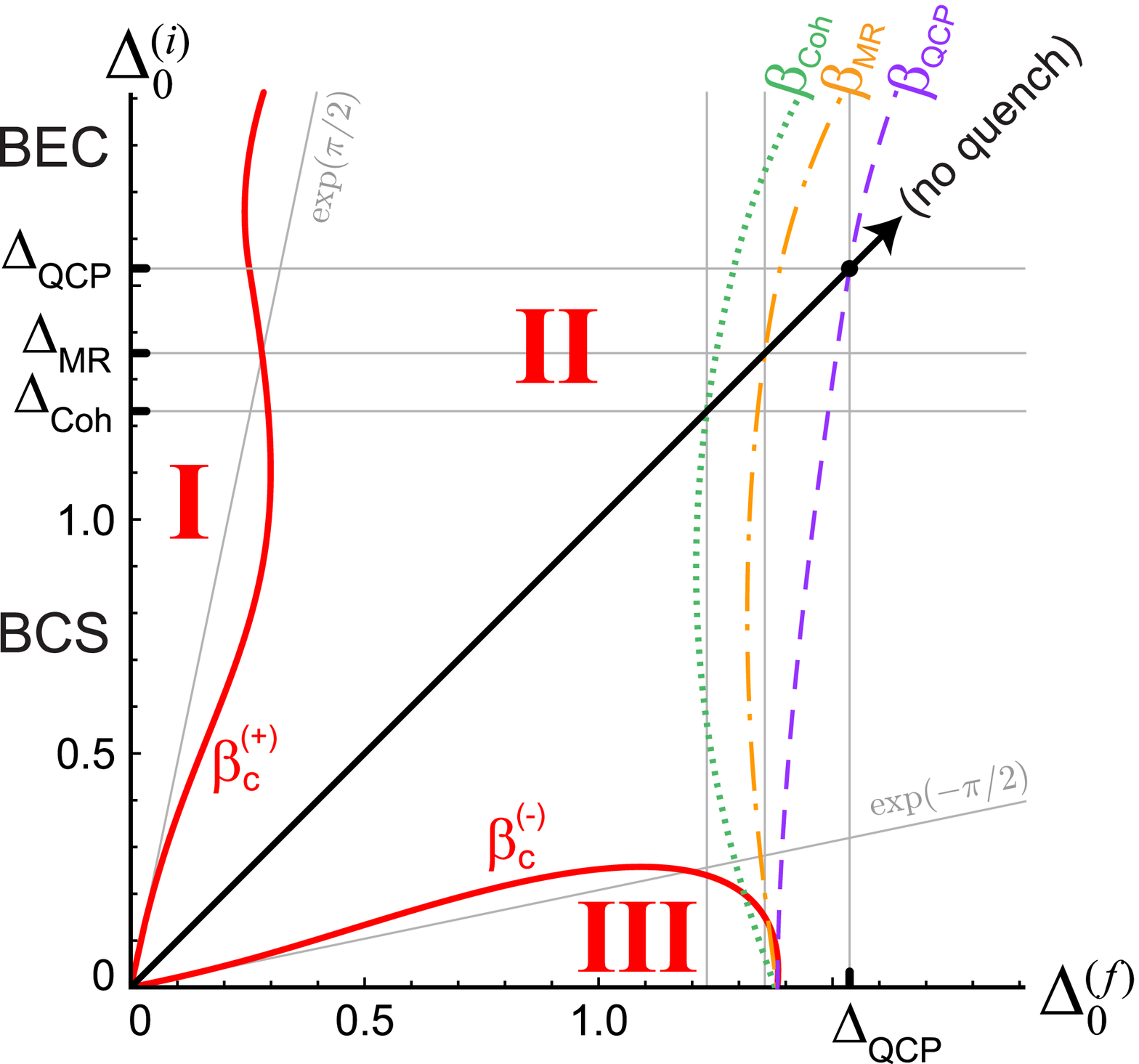}
   \caption{
	Detailed quench phase diagram.
	The dynamical phase boundaries $\bcpm$ were obtained from a numerical solution to 
	Eq.~(\ref{IsoRootsThresh}), using Eq.~(\ref{betaDiDf}).
	The 
	pairing amplitude
	values $\Dcoh$, $\Dmr$, and $\Dqcp$ mark spectral transitions in the equilibrium ground state,
	as described in Sec.~\ref{Sec: GNDRoots} and illustrated in Fig.~\ref{Fig--Roots and TDOS}.
	Explicit values appear in Eqs.~(\ref{DeltaVals}) and (\ref{DDefs}).
	The lines marked $\bcoh$, $\bmr$, and $\bqcp$ are the non-equilibrium extensions of these
	ground state spectral transitions, as discussed in Sec.~\ref{Sec: NEQMTrans}.
	}
   \label{Fig--PhaseDiagFull}
\end{figure}

The roots of $\qp(u;\beta)$ in Eq.~(\ref{qpQuench}) satisfy
\bsub\label{IsoRootsQuench}
\begin{align}
	\ffi(u)
	=&\,
	-
	\frac{\beta}{\Eot(u)},
	\\
	\Eot(u)
	\equiv&\,
	\left(\frac{u}{2} - \mui\right)
	\mp i \Di u^{1/2},
	\label{EotDef}
\end{align}
\esub
where we have solved the quadratic equation for $\ffi$. 
In the thermodynamic limit, the left-hand side of this equation becomes
[via Eq.~(\ref{FDef})] 
\begin{align}\label{ffiint}
	\ffi(u)
	=&\,
	\intei
	\frac{\e}{(u - \e) \Ei(\e)}.
\end{align}
To logarithmic accuracy in the cutoff $\Lambda$,
\begin{widetext}
\begin{align}\label{ffiEval}
	\ffi(u;\zeta)
	=&\,
	-2 \log\left[\frac{2 \Lambda}{(\Di)^2 + 2 |\mui| \, \theta(-\mui)}\right]
	+
	\frac{u}{\Ei(u)}
	\log\left\{
	\frac{\zeta u\left[u+2(\Di)^2 -2 \mui + 2 \Ei(u)\right]}{2\left[u(\Di)^2 -u \mui + 2 (\mui)^2 +2 |\mui| \, \Ei(u) \right]}
	\right\}.
\end{align}
\end{widetext}
The parameter $\zeta$ determines the branch cut in the complex $u$-plane;
taking the principal branch for $\log(z)$, the cut lies along the positive
(negative) $u$-axis for $\zeta = -1$ ($\zeta = 1$).

In the thermodynamic limit, isolated roots are solutions to Eqs.~(\ref{IsoRootsQuench}) and (\ref{ffiEval}) for 
$u$ away from the positive real axis (so that we should take $\zeta = -1$). A given quench is defined by the initial 
pairing amplitude
$\Di$ and the
quench parameter $\beta$ [Eq.~(\ref{betaDef})]. Alternatively, one can specify coordinates $\{\Di,\Df\}$ in the quench
phase diagram, Fig.~\ref{Fig--PhaseDiagBasic}. $\beta$ is determined through the ground state BCS equation for 
the initial and final coupling strengths through Eq.~(\ref{BCSEqF}), leading to
\begin{align}\label{betaDiDf}
	\beta
	=&\,
	2 \muf \log\left[\frac{2 \Lambda e}{(\Df)^2 + 2 |\muf| \theta(-\muf)}\right]
	\nonumber\\&\,
	-
	2 \mui \log\left[\frac{2 \Lambda e}{(\Di)^2 + 2 |\mui| \theta(-\mui)}\right].
\end{align}


\subsection{Threshold roots: Dynamical phase boundaries}

A quench located within the 
dynamical phases marked {\bf I}, {\bf II}, or {\bf III} in Fig.~\ref{Fig--PhaseDiagBasic}
respectively exhibits zero, one, or two isolated pairs of spectral polynomial roots. 
To determine the boundaries of these regions, we look for \emph{threshold} conditions, wherein a complex conjugate pair
first separates from or merges with the positive real axis.
To that end, we write
\[
	u \rightarrow u \pm i \sgn(\beta) \eta,
\]
where $\eta$ denotes a positive infinitesimal, and we take $u$ real and positive on the right-hand side of this equation.
The real and imaginary parts of Eq.~(\ref{IsoRootsQuench}) then imply that 
\bsub\label{IsoRootsThresh}
\begin{align}
	\ffi(u;1) 
	=&\,
	-
	\sgn(\beta)	
	\frac{\pi\sqrt{u}}{\Di}
	\frac{\left(\frac{u}{2} - \mui\right)}{\Ei(u)},
	\\
	|\beta|	
	=&\,
	\frac{\pi\sqrt{u}}{\Di} \Ei(u).
\end{align}
\esub
For the generic quench, two pieces of information such as $\{\Di,\beta\}$ 
must be specified to determine the isolated roots. In Eq.~(\ref{IsoRootsThresh}), $|\beta|$ is no longer a free parameter,
as we have constrained the imaginary part of $u$ to be infinitesimal. For a given $\Di$
and $\sgn(\beta)$, the real positive roots of 
Eq.~(\ref{IsoRootsThresh}) determine $|\beta|$ and $\Df$ as functions of $\Di$, leading to one-parameter curves
in the phase diagram shown in Fig.~\ref{Fig--PhaseDiagBasic}.

The phase boundaries are labeled $\bcpm$ in Fig.~\ref{Fig--PhaseDiagFull},
which depicts a more detailed version of the quench phase diagram. 
These curves were obtained through the numerical solution of Eq.~(\ref{IsoRootsThresh}),
using Eq.~(\ref{betaDiDf}) to determine $\Df$. In the following, we derive
analytical results for quenches from a weakly-paired initial BCS state.


\subsubsection{Threshold roots at weak initial and final pairing, $\{\Di,\Df\} \ll \Dqcp$ \label{Sec: ThreshWeak}}

For weak initial pairing defined as $\Di \ll \Dqcp$,
the threshold root
\begin{align}\label{uc1Def}
	\uc{1}
	\equiv
	2 \mui \simeq 4 \pi n
	+
	\ord{\Di}^2
\end{align}
solves Eq.~(\ref{IsoRootsThresh}) for both $\sgn(\beta) = \pm 1$.
The corresponding quench parameters are 
\begin{align}\label{betapm}
	\bcpm
	=
	\pm (2 \pi)^2 n.
\end{align}
Eq.~(\ref{betaDiDf}) reduces to $\Di/\Df \simeq \exp(\beta/8 \pi n)$ for $\{\Di,\Df\} \ll \Dqcp$. 
Phase boundaries corresponding to Eq.~(\ref{betapm}) are given by the lines
\begin{align}\label{QPDpm}
	\Di = e^{\pm \pi/2} \Df.
\end{align}
These are plotted in Fig.~\ref{Fig--PhaseDiagFull}.
The weak-pairing thresholds in Eq.~(\ref{QPDpm}) correspond to the straight-line portions of the curves
marked $\bcpm$ near the origin $\{\Di,\Df\} = \{0,0\}$ 
in Fig.~\ref{Fig--PhaseDiagFull}.

We can go further and determine the roots throughout the weak-pairing BCS-to-BCS
region of the quench phase diagram. In phases {\bf II} and {\bf III} with $\{\Di,\Df\} \ll \Dqcp$, 
an isolated root can be parameterized as 
\begin{align}\label{deltaDef}
	u \simeq \uc{1} + 2 i \delta.
\end{align}
Employing the branch of Eq.~(\ref{IsoRootsQuench})
with the minus sign [$\Eone(u)$] 	
and using $\zeta = -1$ in Eq.~(\ref{ffiEval}),
a single complex conjugate isolated pair obtains for quenches satisfying
$\beta < \bcp$. 
For $\beta \geq \bcp$, there are no isolated roots (phase {\bf I} in Fig.~\ref{Fig--PhaseDiagFull}). 
We define
\bsub\label{WeakRootDefs}
\begin{align}
	\msf{B} 
	\equiv&\, 
	\frac{\pi}{2}
	\left(\frac{\beta}{\bcp}\right),
	\\
	\frac{\delta}{\sqrt{4 \pi n} \Di}
	\equiv&\,
	\left\{
	\begin{array}{lr}
	\cos(\theta), & 0 \leq \delta < \sqrt{4 \pi n} \Di, \\
	\cosh(\theta), & \delta \geq \sqrt{4 \pi n} \Di.
	\end{array}
	\right.
\end{align}
\esub
We note that since $(\Dqcp)^2$ is of order $4 \pi n$ [Eq.~(\ref{DDefs})], we have
$\sqrt{4 \pi n} \Di \ll  \uc{1}$ for $\Di \ll \Dqcp$; 
the imaginary part of the root is always much smaller than the real part in this regime.
Using the above definitions, we find that $\delta$ is encoded in the transcendental equations
\begin{align}\label{WPIsoRoot1}
\begin{aligned}
	\theta
	\tan\left(\frac{\theta}{2}\right)
	=&\,
	\msf{B},
	&& 0 \leq \theta \leq \frac{\pi}{2}, \; 0 \leq \msf{B} \leq \frac{\pi}{2},
	\\
	\theta
	\tanh\left(\frac{\theta}{2}\right)
	=&\,
	|\msf{B}|,
	&& 0 < \theta,\; \msf{B} < 0.
\end{aligned}
\end{align}
The first equation applies to strong-to-weak quenches, up to the boundary of 
phase {\bf II} with phase {\bf I}, $\msf{B} = \theta = \pi/2$.
The second equation holds for weak-to-strong quenches in phases {\bf II} and {\bf III}.
Both equations have exactly one solution in their regions of validity.

The second branch of Eq.~(\ref{IsoRootsQuench})
with the plus sign [$\Etwo(u)$]		
has zero (one) isolated pair of complex conjugate roots for $\beta > \bcm$ 
($\beta < \bcm$). 
Employing Eq.~(\ref{WeakRootDefs}) to specify the imaginary part the root $\delta$
in terms of $\theta$, we find the equations
\begin{align}\label{WPIsoRoot2}
\begin{aligned}
	\theta
	\cot\left(\frac{\theta}{2}\right)
	=&\,
	|\msf{B}|,
	&& 0 \leq \theta \leq \frac{\pi}{2}, \; -2 < \msf{B} \leq -\frac{\pi}{2},
	\\
	\theta
	\coth\left(\frac{\theta}{2}\right)
	=&\,
	|\msf{B}|,
	&& 0 \leq \theta, \; \msf{B} < -2.
\end{aligned}
\end{align}
The two pairs of isolated roots that distinguish phase {\bf III} solve Eqs.~(\ref{WPIsoRoot1})
and (\ref{WPIsoRoot2}) with $\msf{B} < - \pi/2$.
Eqs.~(\ref{WPIsoRoot1}) and (\ref{WPIsoRoot2}) turn out to be identical to the corresponding
equations in the s-wave case.\cite{DzeroYuzbashyan06,BarankovLevitov06}


\subsubsection{Phase {\bf III} termination at strong final pairing, $\Df \sim \Dmr$}

For $\Di \ll \Dqcp$, the threshold Eq.~(\ref{IsoRootsThresh}) admits an additional 
solution for $\sgn(\beta) < 0$ (weak-to-strong quenches), given by
\begin{align}\label{uc2Def}
	\uc{2} = \left\{\frac{2 \Di}{\pi} \log\left[\frac{2 \Lambda}{(\Di)^2}\right]\right\}^2.
\end{align}
This threshold root in fact locates the phase {\bf III}-{\bf II} boundary near the bottom
of the phase diagram. 
Unlike $\uc{1}$ [Eq.~(\ref{uc1Def})], this additional isolated root \emph{vanishes} as $\Di \rightarrow 0$.
The reason for this is as follows. In the next section we discuss the non-equilibrium extension of the 
topological quantum critical point, indicated by the dashed curve labeled $\bqcp$ in Fig.~\ref{Fig--PhaseDiagFull}.
The $\bqcp$ line is uniquely defined in phase {\bf II} by the feature that it possesses a vanishing isolated root. 
Since the phase {\bf III}-{\bf II} boundary merges with the $\bqcp$ line when $\Di \rightarrow 0$, 
the threshold root $\uc{2}$ also vanishes in that limit.	

Via Eq.~(\ref{betaDiDf}), the quench parameter associated with the phase {\bf III} boundary is 
\begin{align}\label{bcmTerminus}
	\bcm(\Di)
	 = \bqcp(\Di) - \uc{2} \log\left(\frac{8 \pi^2 n^2}{\Lambda \uc{2}} \right),
\end{align}
valid in the same limit.
In this equation, $\bqcp(\Di)$ parameterizes the non-equilibrium topological transition line,
Eq.~(\ref{bqcpDef}) below.
At $\Di = 0$, the threshold root $\uc{2} = 0$ and $\bcm(0)$ coincides with $\bqcp(0)$.
The corresponding quench phase diagram coordinates are $\{\Di,\Df\} \simeq \{0, \Dmr\}$ 
as shown in Fig.~\ref{Fig--PhaseDiagFull}, see Eq.~(\ref{DfQCPTerminus}) in the next section.


\subsection{Non-equilibrium topological and spectral transitions \label{Sec: NEQMTrans}}

In the ground state, the topological quantum phase transition 
at	
$\{\Do,\mu\}  = \{\Dqcp,0\}$ corresponds to the isolated root configuration 
$\uoqpm$ in Eq.~(\ref{IsoRootsQCP}). This is the only 
pairing
amplitude
associated to a vanishing isolated root; for any non-zero $\Do \neq \Dqcp$,
both $\uoqpm$ have finite separation from the positive real axis.
For quenches in phase {\bf II} of Figs.~\ref{Fig--PhaseDiagBasic} and \ref{Fig--PhaseDiagFull}, 
the chemical potential $\mu(t)$ and 
the order parameter
$\Delta(t)$ respectively asymptote to constants
$\masy$ and $\Dasy$, with the latter non-zero. 
Phase {\bf II} is characterized by a single pair of isolated roots.
As discussed below in Sec.~\ref{Sec: PhaseIIGap}, $\{\Dasy,\masy\}$ have the same relation to the isolated roots
$\uqpm$ of the quench spectral polynomial $\qp(u;\beta)$ as $\{\Do,\mu\}$ have to $\uoqpm$ in the ground state, c.f.\ Eqs.~(\ref{IsoRootsGND})
and (\ref{IsoRootsPhaseII}).

We can define a non-equilibrium extension of the topological phase transition through 
the condition $\masy = 0$, which corresponds to the vanishing of an isolated root
of $\qp(u;\beta)$. For a given $\Di$, let us denote the quench parameter $\beta \equiv \bqcp$
that yields a vanishing isolated root in phase {\bf II}. 
At $u = 0$, Eqs.~(\ref{IsoRootsQuench}) and (\ref{ffiEval}) yield
\begin{align}\label{bqcpDef}
	\bqcp(\Di) 
	=
	-2 \mui \log\left[\frac{2 \Lambda}{(\Di)^2 + 2 |\mui| \, \theta(-\mui)}\right].	
\end{align}
Note that the ground state critical point satisfies this 
equation, since $\bqcp = 0$ for zero quench and $\mui = 0$ locates the ground state transition.	
As discussed in Sec.~\ref{Sec: QuenchEdge}, the Green's function winding number $W$ defined via Eq.~(\ref{W}) 
changes across this line (Fig.~\ref{Fig--PhaseDiagW}), indicating the presence or absence of edge states in the 
spectrum of the asymptotic Bogoliubov-de Gennes Hamiltonian. 
The topological transition (dashed) line drawn in Figs.~\ref{Fig--PhaseDiagBasic} and \ref{Fig--PhaseDiagFull} was
obtained by solving Eq.~(\ref{betaDiDf}) numerically to calculate $\Df$ from $\Di$, using Eq.~(\ref{bqcpDef}).
 
In the limit $\Di \rightarrow 0$ (very weak initial pairing), the topological transition line 
$\bqcp$
terminates 
at a particular value of $\Df$ in Fig.~\ref{Fig--PhaseDiagFull}.
Eqs.~(\ref{bqcpDef}) and (\ref{betaDiDf}) imply that for $\beta(\Di,\Df) = \bqcp(\Di)$,
\begin{align}\label{DfQCPTerminus}
	\lim_{\Di \rightarrow 0} 
	\Df \simeq \Dmr,
\end{align}
up to terms of size $\sqrt{n}[\log(\Lambda/2 \pi n)]^{-5/2}$.
By contrast, the ground state
pairing amplitudes
$\Dcoh$ and $\Dqcp$ differ from 
$\Dmr$ by terms of order $\sqrt{n}[\log(\Lambda/2 \pi n)]^{-3/2}$.
[Eq.~(\ref{DDefs}) gives explicit formulae for $\{\Dcoh,\Dmr,\Dqcp\}$].

We conclude that the topological transition in the non-equilibrium phase diagram deviates
from the equilibrium line $\Df = \Dqcp$ for $\Di \ll \Dqcp$.
At $\Di = 0$, the transition $\bqcp$ is such that $\Df \simeq \Dmr$, which is of the same
order as, but smaller than $\Dqcp$. The boundary $\bcm$ separating dynamical phases {\bf II}
and {\bf III} also terminates at this point. 

In the ground state, the quasiparticle energy gap $\Emin$ 
occurs at non-zero (zero) momentum for 
$\Do < \Dcoh$
($\Do \geq \Dcoh$) [Eqs.~(\ref{EminDef--WP}) and (\ref{EminDef--SP})]. 
The isolated ground state roots $\uoqpm$
in Eq.~(\ref{IsoRootsGND}) are purely imaginary at the transition $\Do = \Dcoh$,
marked (2) in Fig.~\ref{Fig--Roots and TDOS}(a).
The dotted curve labeled $\bcoh$ in Fig.~\ref{Fig--PhaseDiagFull} is the non-equilibrium extension,
obtained via the numerical solution of Eqs.~(\ref{IsoRootsQuench}) and (\ref{ffiEval}) ($\zeta = - 1$) 
locating one pair of purely imaginary isolated roots for a given $\Di$. 
Along this curve, the asymptotic 
values
$\{\Dasy,\masy\}$ satisfy 
\begin{align}\label{bcohDef}
	\masy = (\Dasy)^2
\end{align}
[c.f.\ Eq.~(\ref{IsoRootsPhaseII})].

Throughout phase {\bf II}, the order parameter
approaches its asymptotic value via a power-law-damped oscillation. 
The precession of the pseudospins in the asymptotically constant field $\vec{B}_i$ 
[Eq.~(\ref{GndField})] implies that the self-consistent time-evolution of $\Delta$ can be 
expressed as 
\begin{align}\label{DeltaAsymEarly}
	\Delta(t) - \Dasy
	\propto
	\int
	d \e
	\,
	\alpha(\e)
	\exp\left[ - i 2 E_\infty(\e) t \right],
\end{align}
where $E_\infty(\e) = E(\e;\Dasy,\masy)$ is the asymptotic dispersion relation
[Eq.~(\ref{QPNRG})]. In Sec.~\ref{Sec: SSApproachII}, we show that Eq.~(\ref{DeltaAsymEarly}) 
evaluates to
\[
	\Delta(t) = \Dasy + c t^{-\alpha} \cos(\Omega t + \phi),
\]
where $c$ is a constant and we ignore ``non-universal'' corrections of order $1/\Lambda$.
We will find that $\alpha = 1/2$ and $\Omega = 2 \Emin$ to the left of the 
$\bcoh$ curve. For quenches to the right of the $\bcoh$ line in Fig.~\ref{Fig--PhaseDiagFull},
$\alpha = 2$ and $\Omega = 0$ (excluding certain special cases).
The changes in $\alpha$ and $\Omega$ are associated to the transition in the asymptotic 
dispersion relation,	
as determined by $\{\Dasy,\masy\}$. For $\Dasy < (\masy)^2$
(left of $\bcoh$), the minimum in $E(\e;\Dasy,\masy) \equiv \Emin$ at non-zero $\e$ results in a 
non-trivial saddle-point for the dynamics. This disappears when $\Emin$
moves to $\e = 0$ [$\Dasy \geq (\masy)^2$, quenches to the right of the $\bcoh$ curve].

As discussed in Sec.~\ref{Sec: GNDRoots}, the 
amplitude
$\Dmr$ corresponds to 
a doubly-degenerate, negative real root pair in the ground state polynomial $\qp(u)$.
Such a pair is marked (3) in Fig.~\ref{Fig--Roots and TDOS}(a). 
We look for a non-equilibrium extension in the form of a doubly-degenerate, negative real 
isolated root $u = -v < 0$ satisfying 
\[
	\qp(-v;\beta) = \frac{d}{d v} \qp(-v;\beta) = 0.
\]
From Eqs.~(\ref{qpQuench}) and (\ref{IsoRootsQuench}), these conditions become
\bsub\label{OrangeLine}
\begin{align}
	\beta
	=&\,
	\left[
	\left(\frac{v}{2} + \mui\right)
	-
	\Di \sqrt{v} 
	\right]
	\ffi(-v;-1),
	\\
	=&\,
	\frac{	
	\frac{d}{d v}
	\left[\Ei(-v) \ffi(-v;-1)\right]^2
	}{
	\ffi(-v;-1)
	+
	\left(v + 2 \mui \right)
	\frac{d}{d v}
	\ffi(-v;-1)
	}.
\end{align}
\esub
On the first line, we have selected the branch of Eq.~(\ref{IsoRootsQuench}) 
that includes the ground state solution $\{\Do,\mu,\beta\} = \{\Dmr,\Dmr^2/2,0\}$;
the other branch gives $v \sim \ord{1/\Lambda}$, which is beyond
the logarithmic accuracy employed here. In the limit of weak initial pairing
$\Di \ll \Dqcp$, Eq.~(\ref{OrangeLine}) reduces to
\[
	\sqrt{v} \log\left(\frac{e \Lambda v}{8 \pi^2 n^2 }\right) \simeq \Di \log\left[\frac{2 \Lambda}{(\Di)^2}\right].
\]
This has the solution
\begin{align}
	\sqrt{v_c} 
	= 
	\frac{
	\Di\log\left[\frac{2 \Lambda}{(\Di)^2}\right]}{
	2 \W_{0}\left\{
	\log\left[\frac{2 \Lambda}{(\Di)^2}\right]
	\frac{\Di}{4 \pi n}
	\sqrt{\frac{e \Lambda}{2}}
	\right\}
	},
\end{align}
valid for $v \geq 8 \pi^2 n^2/e \Lambda$.
In this equation, $\W_0(z)$ denotes the $k = 0$ branch of Lambert's W function.
The quench parameter is
\begin{align}\label{bmr2bqcp}
	\bmr(\Di) 
	=
	\bqcp(\Di)
	+
	v_c + \sqrt{v_c} \Di \log\left[\frac{2 \Lambda}{(\Di)^2}\right].
\end{align}
Like the spectral transition line $\bcoh$, $\bmr$ converges
to $\bqcp$ as $\Di \rightarrow 0$; all three curves coalesce
at $\Df \simeq \Dmr$ [Eq.~(\ref{DfQCPTerminus})] for vanishing initial 
interaction strength, as
shown in Fig.~\ref{Fig--PhaseDiagFull}.
The dot-dashed curve in this figure marked $\bmr$ was obtained through the numerical solution
to Eq.~(\ref{OrangeLine}), using Eq.~(\ref{betaDiDf}). 

Summarizing, in phase {\bf II} of the quench phase diagram in Fig.~\ref{Fig--PhaseDiagFull},
the same patterns of isolated roots that appear in the ground state (diagonal line) 
also appear for quenches ($\beta \neq 0$). Implications of the lines $\bcoh$ and
$\bqcp$ for asymptotic quench dynamics are discussed in detail in Sec.~\ref{Sec: WindingObs}, below.

\section{Steady-state order parameter dynamics \label{Sec: GapDyn}}

In this section we determine the generalized steady-state behavior of 
$\Delta(t)$ in
the limit $t \rightarrow \infty$. We include the case exhibiting persistent oscillations,
phase {\bf III} in Fig.~\ref{Fig--PhaseDiagBasic}. 
Calculations of the \emph{approach} to the steady-state are deferred until Sec.~\ref{Sec: WindingObs}.


\subsection{Phase {\bf II}: Constant $\Delta(t) \rightarrow \Dasy$  \label{Sec: PhaseIIGap}}

In phase {\bf II} of Fig.~\ref{Fig--PhaseDiagBasic}, there is a single pair of isolated roots.
The reduced problem 
[Lax reduction, Eqs.~(\ref{LaxRedDef}) and (\ref{LaxRed})]
has one collective spin $\vec{\sigma}$ and zero separation variables. 
The 
order parameter
solves a version of Eq.~(\ref{GapEOM}) with $N = 1$. The solution is
\begin{align}\label{DasySol}
	\Delta(t) =&\, \Dasy \exp(-2 i \masy t + i \phi_0),
	\\
	\masy - \frac{\chi}{2} =&\, G \chi \sigma^z,
\end{align}
where $\chi$ is the mode energy. We relate $\{\Dasy,\masy\}$ to the roots of the reduced
spectral polynomial.
Eq.~(\ref{LaxDef}) implies that
\[
	L_{\sigma}^z(u) = \frac{1}{G}\left[\frac{\left(\frac{\chi}{2} - \masy \right)}{u - \chi} + \frac{1}{2} \right].
\]
Using Eqs.~(\ref{L2Def}), (\ref{qpDef}), and (\ref{LaxRedDef}),	
the spectral polynomial is 
\[
	\q_{2}(u) = \frac{1}{4}(u - \uqa{+})(u - \uqa{-}),
\]	
with roots
\begin{align}\label{IsoRootsPhaseII}
	\uqpm
	=
	2 \left[\masy - (\Dasy)^2 \pm \Dasy \sqrt{(\Dasy)^2 - 2 \masy}\right].
\end{align}

We interpret $\masy$ as the out-of-equilibrium
chemical potential because the $N$-spin ground state has 
$\Delta(t) = \Do \exp(-2 i \mu t)$.
This follows from using Eq.~(\ref{spinGND}) as the initial condition to Eq.~(\ref{spinEOM}):
Due to the mismatch between $\vec{B}_i$ in the EOM and the ground state field $\vec{B}_i + 2 \mu\hat{z}$,
the pseudospins uniformly precess $s_i^{-}(t) = s_i^{-}(0) \exp(-2 i \mu t)$.
This can be eliminated by moving to a rotating frame.\cite{footnote--NotAlignGND} 
$\{\Dasy,\masy\}$ have the same relation to the isolated roots $\uqpm$ of the quench spectral polynomial $\qp(u;\beta)$ 
as $\{\Do,\mu\}$ have to $\uoqpm$ in the ground state, Eq.~(\ref{IsoRootsGND}).

For a quench, the 
pairing
amplitude and chemical potential evolve from the initial pre-quench state.
The isolated root pair for $\qp(u;\beta)$ determine 
${\displaystyle{\lim_{t \rightarrow \infty}}} \{\Delta(t),\mu(t)\} = \{\Dasy,\masy\}$ via Eq.~(\ref{IsoRootsPhaseII}).
By contrast, the approach to the asymptotic steady-state (typically a power-law-damped oscillation)
is governed by the full $N$-spin distribution function. This is computed exactly in the thermodynamic limit
in Sec.~\ref{Sec: SpinDist}, below.


\subsection{Phase {\bf III}: Oscillating order parameter  \label{Sec: PhaseIIIGap}}

In phase {\bf III}, $\qp(u;\beta)$ exhibits two isolated pairs of roots.
The pair confined to this region nucleates along the boundary marked $\bcm$ in 
Fig.~\ref{Fig--PhaseDiagFull}. The second pair persists into phase {\bf II}. 
At weak initial and final coupling $\{\Di,\Df\} \ll \Dqcp$, these
solve Eqs.~(\ref{WPIsoRoot2}) and (\ref{WPIsoRoot1}), respectively. 

The isolated roots entirely confined to {\bf III} always appear as a complex conjugate
pair, with a positive real part. In what follows, we denote this pair as 
\begin{align}\label{u1pmDef}
	u_{1,\pm} \equiv u_{1,\rr} \pm i u_{1,\ii}, \;\; u_{1,\{\rr,\ii\}} \geq 0.
\end{align}
The isolated pair that persists into phase {\bf II}
also occurs as a complex conjugate pair throughout the bulk of phase {\bf III};
we denote this pair as
\begin{align}\label{u2pmDef}
	u_{2,\pm} \equiv u_{2,\rr} \pm i u_{2,\ii}, \;\; u_{2,\ii} \geq 0.
\end{align}

In Fig.~\ref{Fig--PhaseDiagFull}, there is a very narrow sliver in phase {\bf III} 
bounded on the left (right) by the $\bmr$ ($\bcm$) curve,
of width 
$\sim \sqrt{n}[\log(\Lambda/2 \pi n)]^{-5/2} \ll \{\Dcoh,\Dmr,\Dqcp\}$.
[C.f.\ the text surrounding Eq.~(\ref{DfQCPTerminus}).]
Within this sliver, the second pair of roots
is negative real (to the right of the $\bmr$ line). 
In this subsection, we consider quenches in the bulk of phase {\bf III}, wherein
the isolated roots always occur in two complex conjugate pairs. 
The sliver with negative real roots is considered in Appendix~\ref{Sec: APP--IIIDyn}.


\subsubsection{Pairing energy EOM}

As argued in Sec.~\ref{Sec: LaxRed}, the asymptotic dynamics of $\Delta(t)$ for the quench 
will be the same as in the two-spin solution, which we now derive. 
We first decompose the complex pairing amplitude into modulus and phase components:
\begin{align}\label{DeltaPolarRT}
	\Delta \equiv \sqrt{\Rhot} \, \exp(-i \phi),
\end{align}	
where $\Rhot$ is the pairing energy.
For the BCS problem with two spins $\vec{\sigma}_{1,2}$,
conservation of the total energy $E_\sigma$ and of the 
$z$-angular momentum $J_\sigma$ imply that 
\bsub
\begin{align}
	\frac{\Rhot}{G} + E_\sigma =&\, \chi_1 \sigma_1^z + \chi_2 \sigma_2^z,
	\\
	J_\sigma =&\, \sigma_1^z + \sigma_2^z,
\end{align}
\esub
where $\chi_{1,2}$ are the mode energies [c.f.\ Eq.~(\ref{LaxRedDef})] and 
$\Rhot$ is the pairing energy defined by Eq.~(\ref{DeltaPolarRT}).	
Expressing the spins in terms of the latter, we have
\begin{align}
\begin{aligned}
	\sigma_p^z \equiv&\, 
	\mathfrak{a}_p \frac{\Rhot}{G} + \mathfrak{b}_p,
	\\
	\mathfrak{a}_{1,2} =&\, 
	\pm 
	\frac{1}{(\chi_1 - \chi_2)},
	\\
	\mathfrak{b}_{1,2} = &\, 
	\pm 
	\left(
	\frac{E_\sigma - \chi_{2,1} J_\sigma}{\chi_1 - \chi_2}
	\right).
\end{aligned}
\end{align}

According to the reduction formula Eq.~(\ref{Redspin}),
the $z$-component of the $i^{\textrm{th}}$ spin 
in the $N$-spin problem is expressed in terms of $\vec{\sigma}_{1,2}$
via	
\begin{align}\label{ResTor: s-sigma}
	s_i^z 
	=
	\frac{d_i}{\e_i}
	L_\sigma^{z}(\e_i)
	\equiv
	a_i 
	\frac{\Rhot}{G}
	+
	b_i.
\end{align}
Eq.~(\ref{diDef}) implies that the constant $a_i$ is given by
\begin{align}\label{aiDef0}
	a_i
	=
	\frac{G \e_i \zeta_i
	}{
	2 \sqrt{
	\mathcal{Q}_4(\e_i)
	}
	},
\end{align}
where $\zeta_i = \pm 1$ and we have introduced the fourth order spectral polynomial 
for the two-spin problem,
\begin{align}\label{q4Def}
	\q_4(u) \equiv G^2 (u - \chi_1)^2(u - \chi_2)^2 L_2^{(\sigma)}(u). 
\end{align}
In this equation, $L_2^{(\sigma)}(\e_i)$ denotes the reduced Lax norm 
defined below Eq.~(\ref{diDef}).

The constants $\{a_i,b_i\}$ must satisfy
\begin{align}
\begin{gathered}
	\sum_{i=1}^N a_i = 0,\;\; \sum_{i=1}^N b_i = J,\\
	\sum_{i=1}^N \e_i a_i = 1,\;\; \sum_{i=1}^N \e_i b_i = H.
\end{gathered}
\end{align}
The two equations on the first line follow from particle conservation
($J$ is the conserved total z-spin). 
The remaining equations encode energy conservation; $H$ is the Hamiltonian in Eq.~(\ref{H}).

We differentiate Eq.~(\ref{ResTor: s-sigma}) to obtain
\begin{align}
	-i 
	\frac{a_i}{\sqrt{\e_i \Rhot}} 
	\frac{\dot{\Rhot}}{G}
	=
	\exp(i \phi) s_i^- - \exp(-i \phi) s_i^+, 
\end{align}
where $\dot{x} \equiv d x / d t$ and we have used Eq.~(\ref{spinEOM}).
We also have
\begin{multline}
	\frac{d}{d t}
	\left[
	\exp(i \phi)
	s_i^-
	+
	\exp(-i \phi)
	s_i^+
	\right]
	\nonumber\\
	=
	\frac{a_i}{\sqrt{\e_i \Rhot}} 
	\frac{\dot{\phi} \dot{\Rhot}}{G}
	-
	\sqrt{\e_i}
	\frac{a_i}{\sqrt{\Rhot}} 
	\frac{\dot{\Rhot}}{G}.
\end{multline}
We define
\begin{align}\label{ADef}
	\dot{A} \equiv \frac{1}{2\sqrt{\Rhot}} \dot{\phi} \dot{\Rhot},
\end{align}
so that
\begin{align}\label{AEq}
	\exp(i \phi) s_i^- 
	=&\,
	\frac{a_i}{G\sqrt{\e_i}}
	\left(
	- i 
	\frac{1}{2 \sqrt{\Rhot}}
	\dot{\Rhot}
	+
	A
	-	
	\e_i
	\sqrt{\Rhot}
	\right)
	+
	c_i.
\end{align}
Eq.~(\ref{DeltaDef}) implies that
\begin{align}
	\sum_{i = 1}^{N} \sqrt{\e_i} c_i = 0.
\end{align}
We compute the modulus-squared of Eq.~(\ref{AEq}):
\begin{align}\label{RhotdotMod}
	&\,
	-
	\frac{1}{4 \Rhot}
	\dot{\Rhot}^2
	=
	\e_i
	\Rhot^2
	+
	\e_i
	\left(
	\e_i
	+
	2 G 
	\frac{b_i}{a_i}
	\right)
	\Rhot
\nonumber\\
&\qquad
	-
	2 \e_i
	\left(
	\frac{G \sqrt{\e_i} c_i}{a_i}
	+
	A
	\right)
	\sqrt{\Rhot}
	+
	2
	\frac{G \sqrt{\e_i} c_i}{a_i}
	A
	+
	A^2
\nonumber\\
&\qquad
	+
	\frac{G^2 \e_i}{a_i^2}
	\left(
	b_i^2
	+
	c_i^2
	-
	\frac{1}{4}
	\right).
\end{align}
Multiplying both sides by $a_i$ and summing over $i$, we solve
for $A$ to find
\begin{align}\label{ASol}
	A
	=
	\frac{1}{2}
	\Rhot^{3/2}
	+
	2 \muco
	\sqrt{\Rhot}
	+	
	\frac{\psi}{\sqrt{\Rhot}}
	+
	\gamma,
\end{align}
where
\begin{align}\label{ASolCoeffs}
\begin{aligned}
	2 \muco 
	\equiv&\,
	\sum_{i} 
	\frac{a_i \e_i^2}{2}
	+
	G H,
	\\
	\psi
	\equiv&\,
	\sum_{i}
	\frac{G^2 \e_i}{2 a_i}
	\left(
	b_i^2
	+
	c_i^2
	-
	\frac{1}{4}
	\right),
	\\
	\gamma
	\equiv&\,
	-
	\sum_{i}
	G \e_i^{3/2} c_i.
\end{aligned}
\end{align}
Eq.~(\ref{RhotdotMod}) becomes
\begin{align}
	-
	\frac{1}{4 \Rhot}
	\dot{\Rhot}^2
	=&\,
	\frac{\Rhot^3}{4}
	+
	2 \muco \Rhot^2
	+
	\left(
	\frac{c_i \sqrt{\e_i} G}{a_i} + \gamma
	\right)
	\Rhot^{3/2}
	\nonumber\\
	&\,
	+
	\ord{\Rhot}.
\end{align}
We therefore require that 
$\frac{c_i \sqrt{\e_i} G}{a_i} \equiv \kappa$, 
independent of $i$.
Eq.~(\ref{ASolCoeffs}) implies that
\begin{align}
	\gamma
	=
	-
	\kappa.
\end{align}
Eq.~(\ref{RhotdotMod}) reduces to
\begin{align}\label{RhotEOM}
	-
	\dot{\Rhot}^2
	=
	\Rhot^4
	+
	8 \muco  \Rhot^3
	+
	8 \rho \Rhot^2
	+
	4 \sigma \Rhot
	+
	4 \psi^2,
\end{align}
where
\begin{align}
\begin{aligned}
	\rho
	\equiv&\,
	\frac{
	G \e_i b_i  
	}{
	a_i
	}
	+
	\frac{1}{2}
	\left( \psi + \et_i^2 \right)
	\\
	\sigma
	\equiv&\,
	\frac{G^2 \e_i}{a_i^2}
	\left(
	b_i^2
	-
	\frac{1}{4}
	\right)
	-
	2 \psi \et_i,
	\\
	\et_i
	\equiv&\,
	\e_i - 2 \muco.
\end{aligned}
\end{align}
Solving for $a_i$ and $b_i$, we obtain
\bsub
\begin{align}
	a_i
	=&\,
	\frac{G \e_i \zeta_i}{	
	2
	\sqrt{
	\left[
	\frac{1}{2} 
	\left(\ek_i^2 + \psi\right)
	-
	\rho
	\right]^2
	-
	\e_i
	\left(
	2 \psi \ek_i
	+
	\sigma
	\right)
	}
	},
	\label{aiDef}
	\\
	b_i
	=&\,
	\frac{a_i}{G \e_i}
	\left[
	\rho
	-
	\frac{1}{2} 
	\left(\ek_i^2 + \psi\right)
	\right],
\end{align}
\esub
where $\zeta_i = \pm 1$.
Eq.~(\ref{RhotEOM}) is an elliptic equation of motion for the pairing energy $\Rhot$.

Comparing Eqs.~(\ref{aiDef0}) and (\ref{aiDef}), we determine that the two-spin spectral polynomial
can be expressed as
\begin{align}\label{q4Final}
	\q_4(u) 
	=
	4
	\left\{
	\begin{aligned}
	&
	\left[
	\left(\frac{u}{2} - \muco\right)^2 
	+ 
	\frac{\psi}{4} 
	-
	\frac{\rho}{2}
	\right]^2
	\\
	&\qquad\qquad
	-
	u
	\left[
	\psi 
	\left(\frac{u}{2} - \muco\right)
	+
	\frac{\sigma}{4}
	\right]
	\end{aligned}
	\right\}.
\end{align}
Eqs.~(\ref{RhotEOM}) and (\ref{q4Final}) express the pairing energy
dynamics and the reduced spectral polynomial in terms of a common set of 
parameters $\{\muco,\rho,\sigma,\psi\}$. 

For most quenches in phase ${\bf III}$ (to the left of the $\bmr$ line 
in Fig.~\ref{Fig--PhaseDiagFull}), the isolated roots $u_{1,\pm}$
and $u_{2,\pm}$ take the form of two complex conjugate pairs
[Eqs.~(\ref{u1pmDef}) and (\ref{u2pmDef})]. 
Expanding Eq.~(\ref{q4Final}) and matching powers of $u$ to the anticipated
form, we find that 
\begin{align}\label{4Coeffs-1}
\begin{aligned}
	\muco 
	=&\,
	\frac{1}{4}
	\left(u_{1,\rr} + u_{2,\rr}\right),
	\\
	\rho
	=&\, 
	-
	\frac{1}{4}(u_{1,\ii}^2 + u_{2,\ii}^2)
	+
	\frac{1}{2}
	U_{\rr}^2
	-
	\frac{3\psi}{2},
	\\
	\sigma
	=&\,
	\frac{U_{\rr}}{2}
	\left(
	u_{2,\ii}^2
	-
	u_{1,\ii}^2
	\right)
	-
	4 \psi \muco,
\end{aligned}
\end{align}
where 
\begin{align}\label{4Coeffs-2}
	U_{\rr} =&\,
	\frac{1}{2}
	\left(u_{1,\rr} - u_{2,\rr}\right).
\end{align}
The parameter $\psi$ has two solutions in terms
of the roots, 
\begin{align}\label{4Coeffs-3}
	\psi_{\pm}
	=&\,
	\frac{1}{8}
\left[
	-
	(u_{1,\ii}^2 + u_{2,\ii}^2) 
	-
	2 u_{1,\rr} u_{2,\rr}
	\pm
	2
	|u_{1}||u_{2}|
\right],
\end{align}
where $|u| = \sqrt{u_{\rr}^2 + u_{\ii}^2}$ is the modulus the complex root $u$.
The physical solution is $\psi = \psi_+$, since this gives positive turning points for the 
positive-definite pairing energy $\Rhot$, as shown below.


\subsubsection{Pairing energy dynamics}

We first consider a quench confined to the weak-pairing BCS region with 
$\{\Di,\Df\} \ll \Dqcp$.
In phase {\bf III}, the corresponding roots take the form
$u_{\{1,2\},\pm} \simeq 2 \mui \pm 2 i \delta_{1,2}$, where
$\delta_{1,2}$ is of order $\sqrt{\mui}\Di \ll \mui$.
Here $\mui \simeq 2 \pi n$ denotes the chemical potential
in the initial state; see Sec.~\ref{Sec: ThreshWeak} for details. 
To leading order, Eqs.~(\ref{4Coeffs-1})--(\ref{4Coeffs-3}) simplify as follows:
\begin{gather}
	\muco 
	\simeq 
	\mui,
	\;\;
	\rho
	\simeq
	-
	\frac{1}{4}(u_{1,\ii}^2 + u_{2,\ii}^2),
	\;\;
	\sigma
	\simeq
	\frac{(u_{1,\ii}^2 - u_{2,\ii}^2)^2}{32 \mui},
	\nonumber\\
	U_{\rr} 
	\sim
	\ord{\Di}^2,
	\;\;
	\psi_+
	\sim
	\ord{\Di}^4.
\end{gather}
Given that $\Rhot \sim \ord{\Di}^2$ and retaining only the leading terms,
Eq.~(\ref{RhotEOM}) reduces to 
\begin{align}\label{RhotWPEOM}
\begin{aligned}[t]
	\dot{\Rhot}^2
	\simeq&\,
	8 \mui \,
	\Rhot 
	\left(
	\Rhot_+
	-
	\Rhot
	\right)
	\left(
	\Rhot
	-
	\Rhot_-
	\right),
	\\
	\Rhot_{\pm}
	=&\,
	\frac{1}{8 \mui}
	\left(u_{1,\ii} \pm u_{2,\ii}\right)^2.
\end{aligned}
\end{align}
This has the same structure as the previously-studied 
s-wave case.\cite{Barankov04,YuzbashyanAltshuler05,YuzbashyanAltshuler06,BarankovLevitov06} 
The turning points of the 
modulus
$|\Delta_{\pm}| \equiv \sqrt{\Rhot_{\pm}}$ are proportional to 
the sum and difference of the isolated root pairs' imaginary parts.
At the boundary of phase {\bf III} marked $\bcm$ in Fig.~\ref{Fig--PhaseDiagFull},
the imaginary part of pair one vanishes $|u_{1,\ii}| \rightarrow 0$, 
leading to the collapse of the oscillatory amplitude.

Eq.~(\ref{RhotWPEOM}) has the solution
\begin{align}
\begin{aligned}[t]
	|\Delta|(t)
	=&\,
	\frac{u_{\ii}}{\sqrt{2 \mui}}
	\dn
	\bigg(
	u_{\ii} t
	\bigg|
	{{\frac{u_{1,\ii} u_{2,\ii}}{u_{\ii}^2}}}	
	\bigg),
	\\
	u_{\ii}
	\equiv&\,
	{\ts{\frac{1}{2}}}
	\left(u_{1,\ii} + u_{2,\ii}\right),
\end{aligned}
\end{align}
where $|\Delta| = \sqrt{\Rhot}$ and
$\dn\left( z | M \right)$ denotes the Jacobi elliptic function 
($M = k^2$ is the modulo parameter). 
Just inside of phase {\bf III} near the boundary with {\bf II},
the period of $|\Delta|(t)$ is $T \simeq 2 \pi/u_{2,\ii} \sim \ordb{\sqrt{\mui}\Di}^{-1}$,
valid in the weak pairing limit $\{\Di,\Df\} \ll \Dqcp$.

Next we consider general phase {\bf III} quenches. 
Using Eqs.~(\ref{4Coeffs-1})--(\ref{4Coeffs-3}) and taking $\psi = \psi_+$, the 
fourth-order polynomial in Eq.~(\ref{RhotEOM}) can be factored. 
The result is 
\bsub
\begin{align}\label{RhotEOMF}
	\dot{\Rhot}^2
	=
	(\Rhot_{+} - \Rhot)(\Rhot - \Rhot_{-})
	(\Rhot + \wRhot_{+})
	(\Rhot + \wRhot_{-}),
\end{align}
where
\begin{align}\label{RhotParams}
\begin{aligned}
	\Rhot_{\pm}
	\equiv&\,
	\frac{1}{2} 
	\left[
	\sqrt{\left(|u_1| - u_{1,\rr}\right)} \,
	\pm 
	\sqrt{\left(|u_2| - u_{2,\rr}\right)} \,
	\right]^2,
	\\
	\wRhot_{\pm} 
	\equiv&\,
	\frac{1}{2} 
	\left[
	\sqrt{\left(|u_1| + u_{1,\rr}\right)} \,
	\pm 
	\sqrt{\left(|u_2| + u_{2,\rr}\right)} \,
	\right]^2.
\end{aligned}
\end{align}
\esub
The above is an elliptic EOM for the pairing energy $\Rhot$, which executes undamped 
periodic motion between the turning points $\Rhot_{-} \leq \Rhot \leq \Rhot_{+}$.

\begin{figure}
   \includegraphics[width=0.35\textwidth]{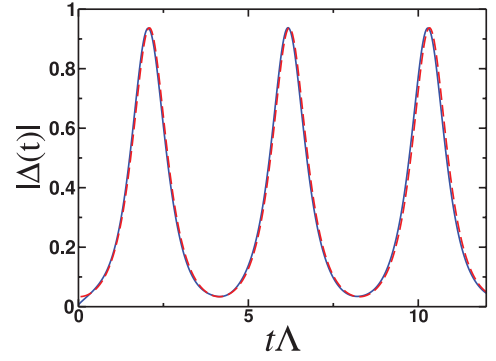}
   \caption{
	Persistent 
	order parameter
	oscillations following a quench.
	The same as Fig.~\ref{Fig--RegionIIIOsc1}, but for	
	quench coordinates
	$\{\Di,\Df\} = \{0.00503, 0.961\}$.
	}
   \label{Fig--RegionIIIOsc2}
\end{figure}

In Figs.~\ref{Fig--RegionIIIOsc1} and \ref{Fig--RegionIIIOsc2}, representative
order parameter
oscillations for phase {\bf III} quenches are shown. The
blue solid curves are the results of numerical simulations of the BCS Hamiltonian
in Eq.~(\ref{H}) for 5024 classical Anderson pseudospins. The red dashed curves 
in these figures are solutions to Eq.~(\ref{RhotEOMF}), with parameters in Eq.~(\ref{RhotParams})
extracted from the roots.

We define
\begin{align}
	\Rhot_{0}
	\equiv
	\frac{\Rhot_{+} + \Rhot_{-}}{2},\;\;
	\loa
	\equiv
	\frac{\Rhot_{+} - \Rhot_{-}}{2},
\end{align}
and introduce dimensionless amplitude $y$ via
\begin{gather}\label{yDefs}
\begin{gathered}[t]
	\Rhot(t) 
	\equiv
	\Rhot_{0} + \loa \, y(t),
	\\
	y_1
	\equiv
	\frac{\Rhot_{0} + \wRhot_{-}}{\loa},\;\;
	y_2
	\equiv
	\frac{\Rhot_{0} + \wRhot_{+}}{\loa}.	
\end{gathered}
\end{gather}
The relative amplitude $y$ is constrained to $-1 \leq y \leq 1$,
while $1 \leq y_1 \leq y_2$.
Eq.~(\ref{RhotEOMF}) becomes
\begin{align}\label{yEOM}
	\dot{y}^2
	=&\,
	\loa^2
	(1 - y^2)
	(y + y_1)
	(y + y_2).
\end{align}
The solution may be written as 
\begin{align}\label{ySol}
	y(t)
	=&\,
	\frac{
	2 y_2 \cn^2\left( \alpha t | M \right) - (y_2 + 1)
	}{
	(y_2 + 1) - 2 \cn^2\left( \alpha t | M \right)
	},
\end{align}
where $\cn\left( z | M \right)$ denotes the Jacobi elliptic function 
($M = k^2$ is the modulo parameter).
In terms of the roots, 
\begin{align}\label{ySolRootParams}
\begin{aligned}
	\Rhot_{0}
	=&\,
	\frac{1}{2}
	\left(
	|u_1| + |u_2| - u_{1,r} - u_{2,r}
	\right),
	\\
	\loa
	=&\,
	\sqrt{|u_1| - u_{1,\rr}}\sqrt{|u_2| - u_{2,\rr}},
	\\
	y_2
	=&\,
	\frac{|u_1| + |u_2| + \sqrt{|u_1|+u_{1,\rr}}\sqrt{|u_2|+u_{2,\rr}}}{\loa},
	\\
	M
	=&\,
	\frac{u_{1,\ii} u_{2,\ii}}{\alpha^2},
	\\
	\alpha
	=&\,
	\frac{1}{2}
	\sqrt{(u_{1,\rr} - u_{2,\rr})^2 + (u_{1,\ii} + u_{2,\ii})^2}.
\end{aligned}
\end{align}
The physical period $T$ of $\Rhot(t)$ is
\begin{align}\label{period}
	T 
	=
	\frac{4 K(M)}{2 \alpha},
\end{align}
where $K(M)$ is the complete elliptic integral of the first kind (and $M = k^2$).

Consider quenches near the phase boundary within {\bf III} such that $u_{1,\ii} \rightarrow 0$,
i.e.\ quench coordinates $\{\Di,\Df\}$ lying just below the curve $\bcm$ in Fig.~\ref{Fig--PhaseDiagFull}.
Here Eqs.~(\ref{yDefs}) and (\ref{ySol}) simplify to 
\begin{align}\label{R(t) PhaseBND}
\begin{aligned}[t]
	R(t) \simeq&\, R_0 + R_d \cos(\Omega_{\msf{c}} t),
	\\
	R_0 \simeq&\, \frac{|u_2| - u_{2,\rr}}{2} + \ord{u_{1,\ii}}^2,
	\\
	R_d \simeq&\, \sqrt{\frac{|u_2| - u_{2,\rr}}{2 u_{1,\rr}}} \, u_{1,\ii} + \ord{u_{1,\ii}}^3,
	\\
	\Omega_{\msf{c}} \simeq&\, \sqrt{(u_{1,\rr} - u_{2,\rr})^2 + u_{2,\ii}^2} + \ord{u_{1,\ii}}.
\end{aligned}
\end{align}
The orbit collapses for $u_{1,\ii} \rightarrow 0$, wherein isolated pair one merges with the continuum along the 
real axis. This is the phase boundary marked $\bcm$ in Fig.~\ref{Fig--PhaseDiagFull}. 

Finally, we consider quenches along the small segment of the $\bmr$ line in Fig.~\ref{Fig--PhaseDiagFull}
which intrudes into phase {\bf III} near $\{\Di,\Df\} = \{0,\Dmr\}$. 
For a discussion of the $\bmr$ line and its termination
at $\Di = 0$, see Sec.~\ref{Sec: NEQMTrans}, Eqs.~(\ref{DfQCPTerminus}) and (\ref{bmr2bqcp}).
Along this line, the second root pair becomes negative real and degenerate: $u_{2,\pm} = - v_2 < 0$. 
The solution in Eq.~(\ref{ySol}) reduces to
\begin{align}\label{ySolbmr}
\begin{aligned}[t]
	y(t)
	=&\,
	\frac{
	y_2  \cos(\Omega_{\msf{MR}} t) - 1
	}{
	y_2 - \cos(\Omega_{\msf{MR}} t)
	},
	\\
	\Omega_{\msf{MR}}
	=&\,
	\sqrt{(u_{1,\rr} + v_2)^2 + u_{1,\ii}^2}.
\end{aligned}
\end{align}


\subsubsection{Order parameter phase dynamics}

Eqs.~(\ref{ADef}) and (\ref{ASol}) imply that the 
pairing amplitude
phase $\phi$ in Eq.~(\ref{DeltaPolarRT}) 
satisfies 
\begin{align}\label{PhaseEOM}
	\dot{\phi} 
	=
	\frac{3}{2}
	\Rhot
	+
	2 \muco
	-	
	\frac{\psi_+}{\Rhot}.
\end{align}
From Eqs.~(\ref{4Coeffs-1}), (\ref{4Coeffs-3}), and (\ref{R(t) PhaseBND}), one can show that at the boundary separating 
phases {\bf II} and {\bf III} where $u_{1,\ii} = 0$, 
\[
	\dot{\phi}
	=
	|u_2|
	=
	2 \masy,
\]
which is the expected result. Here we have used Eq.~(\ref{IsoRootsPhaseII}) to relate the remaining pair of isolated roots $u_{2,\pm}$ to
$\masy$.

\section{Non-equilibrium winding numbers and observables \label{Sec: WindingObs}}

In this section we calculate the long-time asymptotic spin distribution function from the conservation
of the Lax norm. Using this result, we compute the winding numbers $Q$ and $W$ and the \emph{approach} of 
$\Delta(t)$ to its asymptotic constant value in phases {\bf I} and {\bf II} of the phase diagram (Fig.~\ref{Fig--PhaseDiagBasic}).
We relate the parity of zeroes in the Cooper pair distribution (introduced in Sec.~\ref{Sec: CPRF}) to $Q$ and $W$, 
and to the RF spectroscopy amplitude in Eq.~(\ref{A}). Additional results, including the Bogoliubov amplitudes 
$u_{\vex{k}}(t)$ and $v_{\vex{k}}(t)$ as well as single-particle Green's functions are relegated to 
Appendix~\ref{Sec: APP--GFs}.

As discussed in Sec.~\ref{Sec: PhaseIIGap}, in phase {\bf II} the asymptotic behavior of the 
order parameter
is $\Delta(t) = \Dasy \exp(-2 i \masy t)$.
It is the modulus of 
$\Delta$
that goes to a constant, but the phase winds at the frequency $2 \masy$. This includes the ground state (zero quench)
with $\masy = \mui$, due to the chemical potential shift of the field relative to $\vec{B}_i$ in Eq.~(\ref{spinEOM}).  
Unless otherwise noted, in this section we will work in the rotating 
frame $s^-_i(t) \rightarrow s^-_i(t) \, \exp(2 i \masy t)$ when discussing phase {\bf II}, c.f.\ Ref.~\onlinecite{footnote--NotAlignGND}. 
In this frame, $\Delta(t) \rightarrow \Dasy$ (constant).


\subsection{Pseudospin distribution function \label{Sec: SpinDist}}

In phases {\bf I} and {\bf II} of the quench phase diagram in Fig.~\ref{Fig--PhaseDiagBasic}, 
$\Delta(t)$
asymptotes to a constant $\Dasy$ (equal to zero in {\bf I}). 
In the long-time limit, the effective magnetic field $\vec{B}_i$ seen by Anderson pseudospin $\vec{s}_i$ 
is given by Eq.~(\ref{GndField}). This is identical to the field in an ``effective'' ground state with 
pairing amplitude
$\Dasy$ and chemical potential $\masy$. In the actual BCS or BEC ground state, each spin is parallel to its associated
field.\cite{footnote--NotAlignGND} For a quench, the situation is different. As $t \rightarrow \infty$,
each spin \emph{precesses} about its field with an energy-dependent frequency, as in Eq.~(\ref{GappedPrec}). 
The pseudospin
distribution function $\gamma_i$ determines the projection of the spin onto the field in this
equation. [In phase {\bf I}, we should take $\hat{B}_i = - \hat{z}$ and $\masy = 0$; this reconciles Eqs.~(\ref{GaplessPrec}) and (\ref{GappedPrec}).
In phase {\bf II}, $\gamma$ is referred to as the Cooper pair distribution in Sec.~\ref{Sec: CPRF}.]

The conservation of the Lax norm (spectral polynomial) allows the determination of $\gamma_i$ in Eq.~(\ref{GappedPrec}).
As $t \rightarrow \infty$, the Lax components in Eq.~(\ref{LaxDef}) become  
\begin{align}\label{LaxInf}
\begin{aligned}
	L^{\pm}(u;t) 
	=&\,
	\dos
	\intem
	\frac{
	\sqrt{\e}
	\left[s^{x}(\e;t) \pm i s^{y}(\e;t)\right]}{\e - u},
	\\	
	L^{z}(u;t) 
	=&\,
	\dos
	\intem
	\left[
	\frac{
	\e
	s^z(\e;t)}{\e - u}
	\right]
	+ 
	\frac{\pi \dos}{\gf}.
\end{aligned}
\end{align}
In these equations, we have converted to the continuum via Eq.~(\ref{ContDisc});
$\gf$ is the post-quench coupling strength [Eq.~(\ref{gtoG})],
and $s^{a}(\e_i;t) \equiv \hat{a}\cdot\vec{s}_i(t)$  [$a \in \{x,y,z\}$] 
is the continuum version of the precessing spin in Eq.~(\ref{GappedPrec}).
The energy cutoff in Eq.~(\ref{LaxInf}) is 
\begin{align}\label{emaxDef}
	\emax \equiv 2\left[\Lambda + \mui\right]
\end{align}
[see Eqs.~(\ref{ContDisc}) and (\ref{ffiint})].

For $\Dasy \neq 0$, all three Lax components in Eq.~(\ref{LaxInf}) contain both oscillating
and constant terms. Our procedure to determine $\gamma(\e)$ is as follows:
We sit at some fixed complex value of $u$ away from the positive real axis.
Next, we take $t \rightarrow \infty$. The oscillatory terms in $L^{\{\pm,z\}}(u;t)$
vanish in this limit, as can be seen through repeated integration-by-parts.\cite{footnote--LaxIBP}
The result is
\begin{align}\label{LaxInf-1}
\begin{aligned}
	L^{x}
	(u;\infty) 
	=&\,
	\frac{\dos}{2}
	\intem
	\gamma(\e)
	\frac{\e \Dasy}{(\e - u) \Easy(\e)},
	\\
	L^y
	(u;\infty) 
	=&\,
	0,
	\\
	L^{z}
	(u;\infty) 
	=&\,
	\frac{\dos}{2}
	\intem
	\gamma(\e)
	\frac{\e \left(\frac{\e}{2} - \masy\right)}{(\e - u) \Easy(\e)}
	+	
	\frac{\pi \dos}{\gf}.
\end{aligned}
\end{align}
Finally, we let $u$ approach the positive real axis and evaluate the Lax components at
$u \rightarrow u \pm i \eta$, with $u$ and $\eta$ positive and real on the right-hand side. 
We obtain
\bsub\label{LaxInf-2}
\begin{align}
\begin{aligned}
	L^{x}
	(u \pm i \eta;\infty) 
	=&\,
	\frac{\dos}{2}
	\left[
	\pp(u)
	\pm
	i \pi 
	\frac{u \Dasy}{\Easy(u)}
	\gamma(u)
	\right],
	\\
	L^{z}
	(u \pm i \eta;\infty) 
	=&\,
	\frac{\dos}{2}
	\left\{
	\qq(u)
	\pm
	i \pi
	\left[
	\frac{u\left({\textstyle{\frac{u}{2}}} - \masy\right)}{\Easy(u)}
	\right]
	\gamma(u)
	\right\},
\end{aligned}
\end{align}
where
\begin{align}
\begin{aligned}
	\pp(u)
	\equiv&\,
	P
	\intem
	\left(\frac{1}{\e - u}\right)
	\frac{\e \gamma(\e) \Dasy}{\Easy(\e)},
	\\
	\qq(u)
	\equiv&\,
	P
	\intem
	\left(\frac{1}{\e - u}\right)
	\frac{\e \gamma(\e) \left({\textstyle{\frac{\e}{2}}} - \masy\right)}{\Easy(\e)}
	+ 
	\frac{2 \pi}{\gf}.
\end{aligned}
\end{align}
\esub
In these equations, 
$P$ 
denotes the principal value.
Combining Eqs.~(\ref{LaxInf-2}) and (\ref{L2Def}) determines the Lax norm at infinite time,
$L_2(u \pm i \eta;t = \infty)$.
We equate this to Eq.~(\ref{L2quench}), which gives $L_2(u)$ in terms of the pre-quench 
state, leading to
\begin{align}\label{L2Cons}
	&
	u 
\left\{
	\pp(u)
	\pm
	i \pi 
	\frac{u \Dasy}{\Easy(u)}
	\gamma(u)
\right\}^2
	\nonumber\\
	&\quad
	+
\left\{
	\qq(u)
	\pm
	i \pi
	\left[
	\frac{u\left({\textstyle{\frac{u}{2}}} - \masy\right)}{\Easy(u)}
	\right]
	\gamma(u)
\right\}^2
	=
	I_{\mp}(u),
\end{align}
where the initial state is encoded in
\begin{align}
	I_{\mp}(u)
	\equiv
	&\,
	\beta^2 + \left[ \Ei(u) \ffi(u;1) \mp i \pi u \right]^2
	\nonumber\\
	&
	+
	\beta
	\frac{\left(u - 2 \mui \right)}{\Ei(u)}
	\left[ \Ei(u) \ffi(u;1) \mp i \pi u \right]\!.
\end{align}
The form of $\ffi(u;1)$ is given by Eq.~(\ref{ffiEval}). 
Eq.~(\ref{L2Cons}) implies that 
\begin{multline}\label{L2Cons-2}
	s_{\pm}
	\sqrt{
	I_{\mp}(u)
	-
	\frac{
	u \left[ 
	\pp(u) \left(\frac{u}{2} - \masy\right) - \qq(u) \Dasy
	\right]^2
	}{\Easy^2(u)}
	}	
	\\
	=
	\frac{\left({\textstyle{\frac{u}{2}}} - \masy\right) \qq(u) + u \Dasy \pp(u)}{\Easy(u)}
	\pm
	i \pi u \gamma(u),
\end{multline}
with $\{s_{+},s_{-}\} \in \pm 1$. 

We take the difference of the $\pm i \eta$ prescriptions in Eq.~(\ref{L2Cons-2}) to obtain
\begin{align}\label{gamma-PL}
	\gamma(u)
	=
	\frac{1}{2 i \pi u}
	\left\{
\begin{aligned}
	&\,
	s_{+}
	\sqrt{
	I_{-}(u)
	-
	u
	\left[
	\frac{\Xi(u)}{\Easy(u)}
	\right]^2
	}
	\\
	&\,
	-
	s_{-}
	\sqrt{
	I_{+}(u)
	-
	u
	\left[
	\frac{\Xi(u)}{\Easy(u)}
	\right]^2
	}
\end{aligned}
	\right\},
\end{align}
where
\begin{align}\label{XiDef}
	\Xi(u)
	\equiv&\,
	\pp(u) \left(\frac{u}{2} - \masy\right) - \qq(u) \Dasy
	\nonumber\\
	=&\,
-
\Dasy
\left\{
	\frac{2\pi}{\gf}
	-
	\intem
	\,
	\left[-\gamma(\e)\right]
	\frac{\e}{2 \Easy(\e)} 
\right\}
	\nonumber\\
	=&\,
	0.
\end{align}
That $\Xi = 0$ is explained as follows.
Clearly this holds in phase {\bf I}, wherein $\Dasy = 0$.
To see why $\Xi$ vanishes for $\Dasy > 0$ (phase {\bf II}),  we note that 
the term in brackets on the second line of Eq.~(\ref{XiDef}) is the
continuum version of the 
BCS
equation, Eqs.~(\ref{BCSEqGap}) and (\ref{BCSEqg}),
for effective spins of ``length'' $-\gamma(\e)/2$.
Indeed, the Lax components at infinite time in Eq.~(\ref{LaxInf-1}) 
appear as though evaluated for a ground state with $\{\Do,\mu\} = \{\Dasy,\masy\}$,
for effective spins aligned \emph{along} the field as in Eq.~(\ref{spinGND}), but
with a renormalized spin length set by $\gamma(\e)$ (which is the projection onto the
field of the physical, precessing pseudospins).

We therefore conclude that
\begin{align}\label{gammaFinalv1}
	\gamma(\e)
	=
	\frac{s}{2 i \pi \e}
	\left[
	\sqrt{I_{-}(\e)}
	-
	\sqrt{I_{+}(\e)}
	\right],
\end{align}
which is \emph{independent} of $\{\Dasy,\masy\}$.
Relative to Eq.~(\ref{gamma-PL}), we set $s_+ = s_- \equiv s \in \pm 1$
to obtain a real amplitude.
Eq.~(\ref{gammaFinalv1}) holds throughout phases {\bf I} and {\bf II}.
Note that $\gamma(\e) \rightarrow -1$ 
as $\e \rightarrow \infty$ for any quench, since the particle density is finite and all spins 
are aligned along $-\hat{z}$ for sufficiently large energies.
Subject to this boundary condition, the physical branch (sign $s$) of Eq.~(\ref{gammaFinalv1}) 
changes at an energy $\e$ whenever $\gamma(\e) \rightarrow 0$ with a non-zero slope, 
so as to produce a continuous distribution function.

A more useful but equivalent expression is  
\bsub\label{gammaFinalv2Group}
\begin{align}\label{gammaFinalv2}
	\gamma(\e)
	=&\,
	s
	\sqrt{
	1
	-
	\frac{1}{2 (\pi \e)^2}
	\left[
	\nn(\e)
	-
	\sqrt{
	\left\{
	\begin{aligned}
	&\nn^2(\e)
	\\&
	-
	\e
	\left[
	\frac{2 \pi \e \Di \beta}{\Easy(\e)}
	\right]^2
	\end{aligned}
	\right\}
	}
	\right],
	}
\end{align}
where
\begin{align}
	\nn(\e)
	\equiv&\,
	\left[
	\left(\frac{\e}{2} - \mui\right) \ffi(\e;1) + \beta
	\right]^2
	\nonumber\\
	&\,
	+
	\e \left[ \Di \ffi(\e;1) \right]^2
	+
	(\pi \e)^2.
\end{align}
\esub
Eq.~(\ref{gammaFinalv2}) gives a manifestly real formula for $\gamma(\e)$; 
one must still choose the branch $s \in \pm 1$ as a function of energy
so as to produce a continuous distribution function. 

Finally, we note that the expression for $\gamma(\e)$ in Eqs.~(\ref{gammaFinalv1}) or
(\ref{gammaFinalv2}) also applies in phase {\bf III}, if suitably interpreted. 
In this case, $-\gamma(\e)/2$ denotes the projection of spin $\vec{s}(\e)$ in the post-quench
asymptotic state onto the \emph{reduced} spin solution $\vec{s}_{\sss{\msf{red}}}(\e)$. 
The reduced spin solution is defined such that $\vec{s}_{\sss{\msf{red}}}(\e)$ satisfies Eq.~(\ref{Redspin}) 
in terms of the collective variables $\vec{\sigma}_{1,2}$, as discussed in Sec.~\ref{Sec: PhaseIIIGap}.


\subsection{Winding numbers \label{Sec: WN}}

\subsubsection{Green's function winding $W$}

The winding number $W$ in Eq.~(\ref{W}) depends upon the asymptotic form of the 
retarded Green's function $\G_{\vex{k}}(t,t')$. 
As discussed in Sec.~\ref{Sec: QuenchEdge}, this function satisfies the Bogoliubov-de Gennes equation
(\ref{Gasy}), subject to the initial condition in (\ref{GasyIC}). 
It is therefore independent of the distribution function $\gamma(\e)$, which
does not appear in these equations.
This is confirmed by a calculation in Appendix~\ref{Sec: APP--GFs}, which yields 
the explicit form for $\G_{\vex{k}}(t,t')$ in Eq.~(\ref{GEval}).
This result is identical to that for a system in its ground state, except
that here the 
order parameter
$\Dasy$ and the chemical potential $\masy$ are determined by the quench
through the single isolated pair of roots [Eq.~(\ref{IsoRootsPhaseII})].

The winding number $W$ therefore depends only upon $\sgn(\masy)$ in phase {\bf II},
and takes the values shown in Fig.~\ref{Fig--PhaseDiagW}, as discussed in 
Sec.~\ref{Sec: NEQMTrans}. By the argument in Sec.~\ref{Sec: QuenchEdge},
$W = 1$ ($W = 0$) signals the presence (absence) of edge states in the Bogoliubov-de Gennes 
quasiparticle spectrum following a quench in phase {\bf II}. 
By contrast, $W$ is ill-defined in the gapless phase {\bf I}.

\subsubsection{Pseudospin winding $Q$}

As explained in Sec.~\ref{Sec: Topogapless}, starting from an initial $p+ip$ state in either
the BCS or BEC phases, the evolving spin distribution can be parameterized at any time $t$
as in Eq.~(\ref{PseudoTex}), where $\varrho(k)$ and $\Theta(k)$ are time-dependent. 
Eq.~(\ref{Q}) then implies that the pseudospin winding number $Q$ is given by
\begin{align}\label{Q-Eval}
	Q
	=
	{\textstyle{\frac{1}{2}}}
	\left\{
	\sgn[\varrho(k = 0)]
	+
	1
	\right\}.
\end{align}
As $t \rightarrow \infty$, the momentum space pseudospin texture is reconstructed from 
Eqs.~(\ref{GaplessPrec}) and (\ref{GappedPrec}) in phases {\bf I} and {\bf II}, respectively. 
The latter is transcribed in Eq.~(\ref{SpinDist}) of Appendix~\ref{Sec: APP--GFs}. 

We consider first the gapless phase {\bf I}.
The winding $Q = {\textstyle{\frac{1}{2}}}\left\{\gamma(0) + 1\right\}$.
From Eq.~(\ref{gammaFinalv2Group}), one can check that 
\begin{align}\label{|gamma0|}
	\lim_{\e \rightarrow 0} 
	|\gamma(\e)|
	=
	\left\{
	\begin{array}{ll}
	1, & \mui \neq 0 \\
	0, & \mui = 0
	\end{array}
	\right.,
\end{align}
where $\mui$ is the pre-quench chemical potential, with $\mui > 0$ ($\mui < 0$) 
indicating a BCS (BEC) initial state.
We conclude that $Q$ is well-defined throughout the gapless phase {\bf I},
except for a quench starting from the quantum critical point $\Di = \Dqcp$. 

To compute $Q$, we must determine the branch of Eq.~(\ref{gammaFinalv2}) (i.e., $s = \pm 1$)
relevant for $\e \rightarrow 0$. We know that $\lim_{\e \rightarrow \infty} \gamma(\e) = -1$, so
that the branch is $s = -1$ at large $\e$. The branch switches every time $\gamma$ goes to zero
with a non-zero slope, so as to preserve the continuity. We find that in the gapless phase 
{\bf I}, 
\begin{align}\label{QFinal}
	\lim_{\e \rightarrow 0}
	\gamma(\e)
	=
	\left\{
	\begin{array}{lll}
	-1, & \mui < 0 & \Rightarrow Q = 0 \\
	+1, & \mui > 0 & \Rightarrow Q = 1 
	\end{array}
	\right..
\end{align}
In other words, $Q$ is conserved in the gapless phase. 
This can be understood in various ways.
The pseudospin winding number cannot change unless
{\bf (a)} the spin distribution develops a discontinuity or a diabolical point,\cite{Volovik} or
{\bf (b)} a skyrmion-number changing process (hedgehog) occurs in \emph{momentum-time}.
Scenario {\bf (a)} cannot occur within a finite time interval, because the time evolution is 
a smooth deformation.
Scenario {\bf (b)} cannot happen for the reduced p-wave BCS Hamiltonian dynamics
[which are identical for Eqs.~(\ref{Hactual}) and (\ref{Hchiral}), as shown in Appendix~\ref{Sec: APP--ChiralP}].

The conservation of $Q$ in phase {\bf I} wherein $\Delta(t) \rightarrow 0$ leads to the notion of a 
``gapless topological phase.'' This occurs for quenches in the region marked {\bf B}, Fig.~\ref{Fig--PhaseDiagQ}.
Those in {\bf A} are topologically trivial. 
Corresponding topological and trivial pseudospin textures appear similar to those in Figs.~\ref{Fig--Skyrm}(a) and \ref{Fig--Skyrm}(c),
but now these textures undulate in time: 
The spins at radius $k$ precess about $\hat{z}$ with frequency $k^2$ [Eq.~(\ref{GaplessPrec})].
In Figs.~\ref{Fig--Gapless-A} and \ref{Fig--Gapless-B}, $\gamma(\e)$ is plotted against $k = \sqrt{\e}$ 
for representative quenches in {\bf A} and {\bf B}, respectively.

Next we consider phase {\bf II}.
As discussed above and in Sec.~\ref{Sec: QuenchEdge}, in the limit $t \rightarrow \infty$, the 
retarded Green's function winding number $W$ is completely determined by $\masy$. This appears
in $\vec{B}(\e)$, the continuum version of Eq.~(\ref{GndField}). The latter can be viewed as an effective ground state
field, which ``winds'' whenever $W \neq 0$
(recall that spins are aligned along the field in the actual ground state).
For $W = 1$, we have $\masy > 0$ and $\hat{B}(0) = \hat{z}$ (``winding''),
while $W = 0$ implies that $\masy < 0$ and $\hat{B}(0) = - \hat{z}$ (``non-winding'').
As shown in Fig.~\ref{Fig--PhaseDiagW}, $W$ undergoes a dynamical topological transition  
for quenches across the quantum critical point. In particular, $W$ evolves from trivial to non-trivial 
or vice-versa for quenches in the regions marked {\bf C} and {\bf H} in Fig.~\ref{Fig--PhaseDiagSec}.

The pseudospin winding $Q$ is determined by $s^z(0)$. Since $|\gamma(0)| = 1$ for
$\mui \neq 0$ [Eq.~(\ref{|gamma0|})], Eq.~(\ref{GappedPrec}) implies that $Q = 0$
for $\gamma(0) B^z(0) =1$ and $Q = 1$ for $\gamma(0) B^z(0) = -1$.
For a given phase {\bf II} quench $\{\Di,\Df\}$, let us denote the
initial value of the Green's function winding as $W_0$, while 
$W_\infty$ is the asymptotic value as $t \rightarrow \infty$. 
Imposing continuity on the function $\gamma(\e)$, we find that 
\begin{align}\label{WiWf}
	\gamma(0) 
	=
	\left\{
	\begin{array}{ll}
	+1, & W_\infty \neq W_0\\
	-1, & W_\infty = W_0
	\end{array}
	\right..
\end{align}
In other words, the Cooper pair distribution $\gamma(\e)$ ``winds'' from $-1$ at $\e \rightarrow \infty$ to
$+1$ at $\e = 0$ whenever $W$ in the asymptotic post-quench state differs from its value in the initial state.
As a result, we determine that $Q$ is conserved for all quenches in phase {\bf II}, so that
Eq.~(\ref{WiWf}) can be rewritten as  
\begin{align}
	\gamma(0) 
	=
	\left\{
	\begin{array}{ll}
	+1, & W \neq Q\\
	-1, & W = Q
	\end{array}
	\right.,
\end{align}
where both winding numbers are computed in the asymptotic steady-state.

\begin{figure}
   \includegraphics[width=0.35\textwidth]{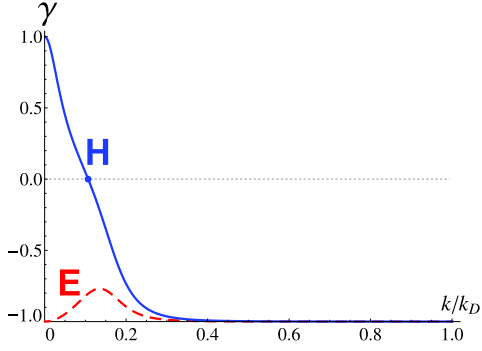}
   \caption{
	The Cooper pair distribution function $\gamma(k)$ as in Fig.~\ref{Fig--ZeroParity1}, 
	but plotted for representative quenches in 
	regions {\bf E} (red dashed line) and {\bf H} (blue solid line) of phase {\bf II}.	
	The quench coordinates for {\bf E} are 
	$\{\Di,\Df\} = \{0.972, 1.25\}$, while
	the quench coordinates for {\bf H} are
	$\{\Di,\Df\} = \{0.972, 1.55\}$.
	Regions {\bf E} and {\bf H} are specified in Fig.~\ref{Fig--PhaseDiagSec}.	
}
   \label{Fig--ZeroParity2}
\end{figure}

Plots of $\gamma(\e)$ for phase {\bf II} quenches in regions $\{\bm{\mathrm{C}},\bm{\mathrm{D}},\bm{\mathrm{E}},\bm{\mathrm{H}}\}$ marked in 
Fig.~\ref{Fig--PhaseDiagSec} appear in Figs.~\ref{Fig--ZeroParity1} and \ref{Fig--ZeroParity2}.
Because quenches in regions {\bf C} and {\bf H} have $W \neq Q$ and therefore $\gamma(0) = +1$,
we deduce that the number of zeroes in $\gamma(\e)$ is odd for quenches in these regions. 
By contrast, $\gamma(\e)$ must exhibit an even number of zeroes for quenches
wherein $W = Q$, including those in {\bf D} and {\bf E}. 
A quench therefore imprints a new $\mathbb{Z}_2$ index upon the Cooper pair distribution 
function, in the form of its parity of zeroes. By contrast, $\gamma(\e) = -1$ 
for all mode energies $\e$ 
in both the BCS and BEC ground states. 

For an ultracold atomic realization of the $p+ip$ superfluid, the Cooper pair distribution can in 
principle be measured in RF spectroscopy, as discussed in Sec.~\ref{Sec: CPRF}.


\subsection{Approach to the steady-state \label{Sec: SSApproach}}

\subsubsection{Phase {\bf I}: Decay to zero \label{Sec: SSApproachI}}

In the continuum limit, the post-quench 
order parameter
in Eq.~(\ref{DeltaDef}) is given by 
\begin{align}\label{DeltaContgapless}
	\Delta(t)
	=&\, 
	-
	\frac{\gf}{2 \pi}
	\intem
	\sqrt{\e} \,
	s^-(\e;t).
\end{align}
The 
order parameter
can be self-consistently determined by linearizing the spin equations
of motion in Eq.~(\ref{spinEOM}). 
In phase {\bf I}, $\Delta(t)$ decays to zero as $t \rightarrow \infty$, 
and $s^-(\e;t)$ can be written as a sum of the pure precession in 
Eq.~(\ref{GaplessPrec}), plus a fluctuation:
\begin{align}\label{sminusDecomp}
\begin{aligned}
	s^-(\e;t)
	\equiv&\, 
	s_\infty^-(\e;t)
	+
	\delta s^-(\e;t), 
	\\
	s_\infty^-(\e;t)
	=&	
	{\textstyle{\frac{1}{2}}}
	\sqrt{1 - \gamma^2(\e)} 
	\exp\left\{- i \left[\e t + \Theta(\e)\right]\right\}.
\end{aligned}
\end{align}

The continuum limit of Eq.~(\ref{spinEOM}) can be written as 
\begin{align}\label{spinEOMc}
\begin{aligned}
	\dot{s}^z(\e;t)
	=&\,
	i \sqrt{\e}
	\left[ \Delta^{*} s^-(\e;t) - \Delta s^+(\e;t) \right],
	\\
	\dot{s}^-(\e;t)
	=&\,
	i
	\left[ 
	2 \sqrt{\e} \Delta s^z(\e;t) 
	- 
	\e s^-(\e;t) \right],
\end{aligned}
\end{align}
where $\dot{x} = d x / d t$.
Eq.~(\ref{spinEOMc}) is invariant under the \emph{effective} time-reversal transformation
\begin{gather}
	s^z(\e;t) \rightarrow s^z(\e;-t),
	\;\;
	s^{\pm}(\e;t) \rightarrow s^{\mp}(\e;-t),
	\nonumber\\
	\Delta(t) \rightarrow \Delta^{*}(-t).
	\nonumber
\end{gather}
These relations are also satisfied by the initial condition, Eq.~(\ref{spinGND})
for the pre-quench $p+ip$ state with real $\Do$.
Therefore we can set $\Theta(\e) = 0$ in Eqs.~(\ref{GaplessPrec}), (\ref{GappedPrec}), and (\ref{sminusDecomp}).

To linear order in the smallness of $\Delta(t)$, 
\bsub
\begin{align}
	\delta \dot{s}^-(\e;t) 
	=&\, 
	i
	\left[
	\sqrt{\e} 
	\gamma(\e)
	\Delta(t)
	- 
	\e 
	\delta s^-(\e;t)
	\right],
	\label{deltasdot}
	\\
	\Delta(t)
	=&\,
	\Delta_\infty(t)
	+
	\delta\Delta(t),
	\label{DeltaDecompose}
	\\
	\Delta_\infty(t)
	\equiv&\, 
	-
	\frac{\gf}{4 \pi}
	\intem
	\Aa(\e)
	\,
	e^{-i \e t},
	\label{DInfI}
	\\
	\delta\Delta(t)
	=&\, 
	-
	\frac{\gf}{2 \pi}
	\intem
	\sqrt{\e} \,
	\delta s^-(\e;t),
	\label{deltaDeltaDef}
	\\
	\Aa(\e)
	\equiv&\,
	\sqrt{\e}
	\sqrt{1 - \gamma^2(\e)}.
	\label{AaDef}
\end{align}
\esub	
In these equations, $\Delta_\infty(t)$, $\delta\Delta(t)$, and $\delta s^-(\e;t)$ all 
vanish by assumption as $t \rightarrow 0$.

We first calculate $\Delta_\infty(t)$, which exhibits the same
power-law decay as the full $\Delta(t)$. The latter is also computed explicitly below.
For an initial state not at the quantum critical point ($\mui \neq 0$), 
Eqs.~(\ref{gammaFinalv2Group}) and (\ref{AaDef}) imply that
\begin{align}\label{AaEval}
	\Aa(\e)
	=&\,
	c \, 
	\e
	+ 
	\ord{\e^2},
\end{align}
\begin{align}
	c
	\equiv&\,
	\frac{\beta \Di
	}{
	|\mui| 
	\left(
		\beta + 2 \mui  \log\left[\frac{2 \Lambda}{(\Di)^2 + 2 |\mui| \theta(-\mui)}\right]
	\right)},
	\nonumber
\end{align}
where $\{\Di,\mui\}$ characterize the initial state and $\beta > 0$ is the quench parameter [Eq.~(\ref{betaDef})].
Because $\Aa(\e)$ is a regular (if complicated) function over the integration interval, we can evaluate
Eq.~(\ref{DInfI}) via repeated integration-by-parts.
The leading result is
\[
	\Delta_\infty(t)
	=
	\frac{\gf}{4 \pi}
	\left[
	\frac{c}{t^2}
	+
	\frac{1}{i t} \Aa(\emax) \exp(-i \emax t)
	\right].
\]
The energy cutoff $\emax$ was defined in Eq.~(\ref{emaxDef}).
Via Eqs.~(\ref{gammaFinalv2Group}) and (\ref{AaDef}), one can show that 
$\Aa(\emax) = d/\Lambda + \ord{\Lambda}^{-2}$ with $d$ a cutoff-independent constant, leading to
\begin{align}\label{DDecayGL-G}
	\Delta_\infty(t)
	=
	\frac{\gf}{4 \pi}
	\left[
	\frac{c}{t^2}
	-
	i
	\frac{d}{\Lambda \, t}
	\exp(-i \emax t)
	\right].
\end{align}
This has the form given by Eq.~(\ref{DDecayG}), which 
consists of a cutoff-independent $1/t^2$ decay plus a ``non-universal'' oscillating term proportional to
$1/\Lambda t$. The latter is technically beyond the logarithmic accuracy to 
which we have been working throughout, but can be important when comparing against numerics.

By contrast, for a quench starting from the quantum critical point (with $\Di = \Dqcp$ and $\mui = 0$),
one finds that
\begin{align}\label{AaEvalQCP}
	\Aa(\e)
	=
	\sqrt{\e}
	-
	\tilde{c} \,
	\e^{3/2}
	+
	\ord{\e^{5/2}},
\end{align}
with $\tilde{c}$ a constant. 
The square root leads to the slower $t^{-3/2}$ decay law in Eq.~(\ref{DDecayQCP});
the ultraviolet gives the same non-universal contribution.

We can also determine the precise form of $\Delta(t)$.
Ignoring the cutoff-dependent term, the $1/t^2$ decay of $\Delta_\infty(t)$ 
in Eq.~(\ref{DDecayGL-G}) enters as a source in the right-hand side of Eq.~(\ref{deltasdot}).
As a result, $\delta s^-(\e;t) \propto 1/t^2$ and we can drop the time-derivative on the left-hand
side, which decays faster. We thereby obtain	 
\[
	\delta s^-(\e;t)
	=
	\frac{1}{\sqrt{\e}}
	\gamma(\e)
	\Delta(t),
\]
and using Eq.~(\ref{deltaDeltaDef})
\begin{align}\label{deltaDelta2}
	\delta\Delta(t)
	=&\, 
	\frac{\gf \, \Delta(t)}{2 \pi}
	\left(
	\emax - 8 \pi n
	\right).
\end{align}
Here we have expressed the conserved particle density $n$ through the distribution function,
\[
	n 
	=
	\frac{1}{8 \pi}
	\intem
	\left[
	1
	+
	2 s^z(\e)
	\right]
	=
	\frac{1}{8 \pi}
	\intem
	\left[
	1
	+
	\gamma(\e)
	\right].
\]
Using Eq.~(\ref{DeltaDecompose}), Eq.~(\ref{deltaDelta2}) simplifies to
\begin{align}\label{Delta(t)RegionI}
	\Delta(t)
	=&\, 
	\frac{\Delta_\infty(t)}{
	\left[
	1
	+
	\frac{\gf}{2 \pi}
	\left(
	8 \pi n - \emax
	\right)
	\right]}
	\nonumber\\
	=&\, 
	\frac{1}{
	\left\{
	\muf \log\left[\frac{2 \Lambda e}{(\Df)^2}\right]
	-
	\mui
	\right\}}
	\frac{c}{4 t^2}.
\end{align}
On the second line, we have employed the BCS equation (\ref{BCSEqF}).
Here $\mui$ ($\muf$) denotes the chemical potential associated to $\Di$ ($\Df$) in the
BCS ground state. For a quench in phase {\bf I}, we have $\muf \simeq 2 \pi n$,
while $\mui \leq 2 \pi n$. 
We note that only logarithmic dependence upon the cutoff appears in the final expression.

\subsubsection{Phase {\bf II}: Decay to $\Dasy > 0$ \label{Sec: SSApproachII}}

We will evaluate the counterpart of Eq.~(\ref{DInfI}) for quenches wherein the
order parameter
asymptotes to a non-zero constant.
The precessing spin in Eq.~(\ref{GappedPrec}) has the minus component
\begin{align}\label{SminusG}
	s^-_{\infty}(\e;t)
	=&
	\frac{1}{\sqrt{\e}}
	\left\{
	\begin{aligned}
	&
	\alpha_1(\e)
	\cos\left[2 \Easy(\e) t \right]
	\\&
	-
	i
	\alpha_2(\e)
	\sin\left[2 \Easy(\e) t \right]
	+
	\alpha_3(\e)
	\end{aligned}
	\right\},
\end{align}
where
\begin{align}
	\begin{aligned}
	\alpha_{1}(\e)
	\equiv&\,
	\Aa(\e)
	\frac{\left(\frac{\e}{2} - \masy\right)}{2 \Easy(\e)},
	\\
	\alpha_{2}(\e)
	\equiv&\,
	{\textstyle{\frac{1}{2}}}
	\Aa(\e),
	\\
	\alpha_{3}(\e)
	\equiv&\,
	\gamma(\e)  
	\frac{\e \Dasy}{2 \Easy(\e)},
	\end{aligned}
\end{align}
and the amplitude $\Aa(\e)$ was defined in Eq.~(\ref{AaDef}).

Inserting Eq.~(\ref{SminusG}) into Eq.~(\ref{DeltaContgapless}),
the static term involving $\alpha_3(\e)$ evaluates to 
\begin{align}
	-
	\frac{\gf}{2 \pi}
	\int_0^{\emax}
	d \e
	\,
	\alpha_3(\e)
	=
	\Dasy,
\end{align}
where we have used Eq.~(\ref{XiDef}). 
Thus the spin distribution reconstructed from the conservation of the 
Lax norm is consistent with the 
pairing
amplitude $\Dasy$ computed from 
the isolated root pair. 
The time-dependent part of the 
order parameter
is given by
\begin{align}\label{dDelInf}
	\delta\Delta_\infty(t)
	=&\, 
	-
	\frac{\gf}{2 \pi}
	\int_0^{\emax}
	d \e
	\left\{
	\begin{aligned}
	&
	\alpha_1(\e)
	\cos\left[2 \Easy(\e) t \right]
	\\&
	-
	i
	\alpha_2(\e)
	\sin\left[2 \Easy(\e) t \right]
	\end{aligned}
	\right\}.
\end{align}

To compute Eq.~(\ref{dDelInf}), we must distinguish two regimes.
For quenches with $\Dasy^2 < \masy$, the dominant contribution comes
from a saddle-point at non-zero $\e$. This is the region of phase {\bf II}
to the left of the line marked $\bcoh$ in Fig.~\ref{Fig--PhaseDiagFull},
as discussed above Eq.~(\ref{bcohDef}) in Sec.~\ref{Sec: NEQMTrans}. 	
For quenches in this regime, $\Easy(\e)$ reaches its minimum value 
(the non-equilibrium 
spectral 	
gap) $\Emin = \Dasy\sqrt{2 \masy - \Dasy^2}$ 
at $\esp = 2(\masy - \Dasy^2) > 0 $:  
\[
	\Easy(\e) = \Emin + \frac{(\e - \esp)^2}{8\Emin} + \ldots
\]
The saddle-point gives 
\begin{align}\label{DAppW}
\begin{aligned}
	\delta\Delta_\infty(t)
	\simeq&\,	
	-
	\gf
	\sqrt{\frac{\Emin}{2 \pi t}}
	\left[
	\begin{aligned}
	&
	\alpha_{-}
	\cos\left(2  \Emin t \right)
	\\&
	-
	\alpha_+
	\sin\left(2  \Emin t \right)
	\end{aligned}
	\right],
	\\
	\alpha_{\mp}
	\equiv&\,
	\alpha_1(\esp) \mp i \alpha_2(\esp).
\end{aligned}
\end{align}
This is qualitatively the same behavior as obtained for weak BCS-to-BCS 
quenches in the s-wave case.\cite{VolkovKogan74,YuzbashyanAltshuler06}

The saddle-point contribution in Eq.~(\ref{DAppW}) vanishes 
when $\esp \rightarrow 0$ ($\masy = \Dasy^2$).
For $\Dasy^2 > \masy$, the minimum of $\Easy(\e)$ occurs
at $\e = 0$ (see Sec.~\ref{Sec: GNDRoots} for further discussion of the spectrum).
Eq.~(\ref{dDelInf}) can then by evaluated by repeated integration-by-parts. 
Note that
\[
	\alpha_1(\e) = \alpha_{2}(\e) \left[-\sgn(\masy)  + \ord{\e}\right],
\]
leading to 
\begin{align}
	\delta\Delta_\infty(t)
	\simeq&\, 
	\frac{\gf \sgn(\masy)}{4 \pi}
	\nonumber\\
	&\,\times
	\int_0^{\emax}
	d \e
	\,
	\Aa(\e)
	\,
	e^{2 i \Easy(\e) \sgn(\masy) t}.
\end{align}
This is the phase {\bf II} generalization of Eq.~(\ref{DInfI}), valid
for $\Dasy^2 > \masy$.
Eqs.~(\ref{AaEval}) and (\ref{AaEvalQCP}) imply that the cutoff-independent
part of the decay is $1/t^2$ for $\Di \neq \Dqcp$ and 
$1/t^{3/2}$ for $\Di = \Dqcp$. At $\e = 0$, the phase factor
$e^{2 i \Easy(\e) \sgn(\masy) t} \rightarrow e^{2 i \masy t}$;
this is eliminated by moving back to the ``lab'' frame.

We conclude that Eqs.~(\ref{DDecayG}) and (\ref{DDecayQCP})
also describe the decay of $\delta \Delta(t) \equiv [\Delta(t) - \Dasy]$ for 
phase {\bf II} quenches to the right of the $\bcoh$ line in Fig.~\ref{Fig--PhaseDiagFull}.
By contrast, quenches to the left of this line within phase {\bf II}
exhibit slower, oscillatory decay according to Eq.~(\ref{DAppW}), due to the 
saddle-point contribution.

\section{Conclusion \label{Sec: End}}

\subsection{Pair-breaking processes}

In this paper, we have computed the quench dynamics of a $p+ip$ superfluid in the
collisionless regime.
This is a non-adiabatic evolution of the initial state in which pair-breaking processes are neglected. 
The preconditions necessary to observe our results in an experiment are that 
\begin{align}\label{QTimeBNDS}
	\tq \ll \frac{1}{\Emin} \ll \tpb, 
\end{align}
where $\tq$ is the duration of the quench (zero for the instantaneous quench studied here), 
$\Emin$ is the minimum quasiparticle energy (quasiparticle gap), and 
$\tpb$ is the time scale associated to inelastic pair breaking processes.
The various predictions presented in this paper describe the post-quench asymptotic steady-state. 
Provided the bounds in Eq.~(\ref{QTimeBNDS}) are met, we expect our results to hold for
times $t$ such that $\frac{1}{\Emin} \ll t \lesssim \tpb$. 

For BCS-to-BCS quenches entirely confined to the weak pairing regime, 
\[
	\Emin 
	=
	\Di \sqrt{2 \mui - (\Di)^2} \simeq \Di \sqrt{4 \pi n}.
\]
Fermi liquid theory then implies the order-of-magnitude estimate
\begin{align}
	\tpb 
	\sim&\,
	\frac{1}{\Emin} \left[ \frac{\mui}{\Emin} \right]
	\gg \frac{1}{\Emin},
\end{align}
implying the existence of a large window over which the collisionless dynamics computed
in this paper can be observed. 
The investigation of pair-breaking processes upon quenches originating or terminating 
beyond the weak-coupling regime remains an important subject for future work.

\subsection{Summary and open questions}

In summary, we have investigated quantum quenches in 2D topological p-wave superfluids.
The post-quench dynamics have been computed via classical integrability. Within the classical
approximation, our treatment is exact. Because of the 
infinite-ranged	nature of the interactions in the
reduced
BCS Hamiltonian, we expect that our results apply to the quantum model in the thermodynamic limit.

We constructed the quench phase diagram, and extracted the exact asymptotic 
order parameter $\Delta(t)$
dynamics, finding that
either (1) 
$\Delta(t)$ 
goes to zero, (2) 
$\Delta(t)$ 
goes to non-zero constant, or (3) 
$\Delta(t)$ 
exhibits persistent 
oscillations. These results are qualitatively the same as the s-wave 
case.\cite{Barankov04,amin,simons,WarnerLeggett05,YKA05,YuzbashyanAltshuler05,YuzbashyanAltshuler06,BarankovLevitov06,DzeroYuzbashyan06}

The key difference from previous work is that here we have characterized the quench-induced 
dynamics of the system topology. We found that the pseudospin winding number $Q$ is unchanged by the quench, 
leading to the prediction of a  ``gapless topological state.'' By contrast, the retarded Green's function 
winding number $W$ can undergo a dynamical transition. This happens e.g.\ for quenches across the quantum critical point separating
the topologically non-trivial BCS and trivial BEC phases.
In the asymptotic steady state wherein the 
order parameter
goes to a constant, the corresponding Bogoliubov-de Gennes
Hamiltonian is expected to possess edge states in a finite geometry whenever $W \neq 0$. 

While $W$ determines the existence of edge modes following a quench, we have not determined the \emph{occupancy} of 
these states. The difficulty is that introducing an edge breaks the integrability of our momentum-space BCS model. 
A fundamental question is whether these non-equilibrium topological steady states support the kind of quantized thermal conductance
expected in an equilibrium p-wave superconductor.\cite{KaneFisher,Capelli02} 	
A related question is the formation, preservation, or destruction of Majorana zero modes following a quench in 1D
topological superconductor; this was studied numerically for a non-interacting model in Ref.~\onlinecite{Perfetto2013}.

Another interesting open 
problem	
relates to the role of topological defects in thermalization. 
Once pair-breaking processes are included, the theory is no longer integrable. One therefore
expects thermalization at the longest times. How does this occur? One possibility is that topological defects,
which can appear either as phase vortices in real space, or hedgehog instantons in momentum-time, proliferate
and scramble the topological order.

Finally, we 
have	
determined that the parity of zeroes in the Cooper pair distribution is odd whenever $Q \neq W$,
i.e.\ whenever $W$ undergoes a dynamical transition. We have argued that the Cooper pair distribution should
be observable in RF spectroscopy in an ultracold atomic or molecular realization of the 2D $p+ip$ superfluid.  
By contrast, the same response does not distinguish the BCS from BEC phases in the ground state.

Probing the Cooper pair distribution can therefore provide a bulk signature of the topological properties of
the system when it is driven far from equilibrium by a quench. In this way, a quantum quench can be used to transfer 
topological entanglement normally hidden from experiment into a physical observable, i.e.\ a non-equilibrium distribution function.

\begin{acknowledgments}

This work was supported in part by the NSF under Grants No.~DMR-0547769 (M.S.F. and E.A.Y.),
PHY-1211914, DMR-1205303 (V.G.), 
the NSF I2CAM International Materials Institute Award, Grant No.~DMR-0844115 (M. D.),
the Ohio Board of Regents Research Incentive Program Grant No.~OBR-RIP-220573 (M. D.),
by the David and Lucile Packard Foundation (M.S.F. and E.A.Y.), and by the Welch Foundation 
under Grant No.~C-1809 (M.S.F.).

\end{acknowledgments}

\appendix

\section{Ground state \label{Sec: APP--GND}}

\subsection{BCS equations \label{Sec: BCSEq}}

In the thermodynamic limit, the BCS equations for the 
pairing
amplitude $\Do$ 
and particle density $n$ are 
\bsub\label{BCSEqs}
\begin{align}
	\frac{1}{g}
	=&\,
	\frac{1}{2\pi}
	\inte
	\frac{\e/2}{\sqrt{\left(\frac{\e}{2} - \mu\right)^2 + (\Do)^2 \e}},
	\label{BCSEqg}
	\\
	n
	=&\,
	\frac{1}{8 \pi}
	\inte
	\left[
	1
	-
	\frac{(\frac{\e}{2} - \mu)}{\sqrt{\left(\frac{\e}{2} - \mu\right)^2 + (\Do)^2 \e}}
	\right],
	\label{BCSEqn}
\end{align}
\esub
where $\e = k^2$, and we have cut these integrals off at a single particle energy
$k^2/2 = \Lambda + \mu$. (The inclusion of $\mu$ simplifies the analysis; results 
are obtained to logarithmic accuracy in $\Lambda \gg |\mu|$.)
The BCS coupling $G$ in Eq.~(\ref{Hchiral}) is related to $g$ via
Eq.~(\ref{gtoG}).
The dimensionful interaction strength $g$ is non-zero in the thermodynamic limit, and carries units of inverse density.
Eq.~(\ref{BCSEqg}) is the continuum version of Eq.~(\ref{BCSEqGap}), using Eq.~(\ref{ContDisc}).

Integrating Eq.~(\ref{BCSEqs}) and discarding terms proportional to inverse powers of $\Lambda$, 
one obtains Eq.~(\ref{muGND}) for the chemical potential and 
\begin{align}\label{BCSEqF}
	\frac{1}{g} - \frac{1}{g_{\msf{QCP}}}= \frac{\mu}{\pi} \log\left[\frac{2 \Lambda e}{\Do^2 + 2 |\mu| \theta(- \mu) }\right].
\end{align}
In this equation, 
$g_{\msf{QCP}}$
is the coupling strength at the BCS-BEC transition $\mu = 0$, 
\[
	\frac{1}{g_{\msf{QCP}}} 
	= \frac{\Lambda}{\pi} - 4 n.
\]
The linear divergence in Eq.~(\ref{BCSEqg}) has been absorbed into 
$1/g_{\msf{QCP}}$. 
Because
a quench is completely specified by the initial 
order parameter
$\Di$ and the difference of the initial and 
final coupling strengths [Eq.~(\ref{betaDef})], 
$g_{\msf{QCP}}$ plays no role in the dynamics.

\subsection{Spectral transitions and tunneling density of states \label{Sec: TDOS}}

We first note three special values of $\Do$, defined implicitly through the chemical
potential equation (\ref{muGND}):
\begin{align}\label{DeltaVals}
\begin{aligned}
	\mu(n,\Dcoh) =&\, (\Dcoh)^2, \\
	\mu(n,\Dmr) =&\, \frac{1}{2}(\Dmr)^2, \\
	\mu(n,\Dqcp)=&\, 0.
\end{aligned}
\end{align}
Since $\mu$ is a monotonically-decreasing function of $\Do$
[Eq.~(\ref{muGND}) and Fig.~\ref{Fig--MuPlot}], we have $\Dcoh < \Dmr < \Dqcp$.
Each of these 
values corresponds to a particular transition or 
anomalous point in the shape of the quasiparticle energy spectrum.
To see this, we rewrite the quasiparticle energy $E_k$ in Eq.~(\ref{Ek})
in terms of $\e = k^2$ [Eq.~(\ref{QPNRG})]:
\[
	E(\e;\Do) = \sqrt{\left(\frac{\e}{2} - \mu\right)^2 + (\Do)^2 \e}.
\]

For sufficiently weak pairing, the minimum of $E$ with respect to
$\e$ occurs slightly below $2 \ef$, where $\ef$ denotes the Fermi energy.
As $\Do$ (or equivalently, the coupling strength) is increased, this
minimum moves to smaller energies. At $\Do = \Dcoh$, it reaches
zero. For $\Do < \Dcoh$ [$\mu > (\Dcoh)^2$], the tunneling density
of states exhibits a coherence peak (van Hove singularity) above its threshold value, 
see Fig.~\ref{Fig--Roots and TDOS}(b) and Eq.~(\ref{TDOS}), below. 
The coherence peak disappears for $\Do \geq \Dcoh$. 
At the special point $\Do = \Dmr$, the curvature of $E(\e)$ vanishes
everywhere:
\[
	E(\e;\Dmr) = {\textstyle{\frac{1}{2}}}\left[\e + (\Dmr)^2 \right].
\]
The condition $\Do = \Dmr$ for variable density $n$ was termed the
``Moore-Read'' line in Ref.~\onlinecite{Sierra09}.
Finally, the BCS-BEC quantum phase transition occurs at $\mu = 0$, $\Do = \Dqcp$.
Here the spectrum exhibits a gapless Dirac node at $\e = 0$,
\[
	E(\e;\Dqcp) = \Dqcp\sqrt{\e + \e^2 / 4 (\Dqcp)^2 }.
\]
Eqs.~(\ref{muGND}) and (\ref{DeltaVals}) can be solved to obtain
\begin{gather}\label{DDefs}
\begin{gathered}
	\Dcoh
	=
	\sqrt{\Upsilon(n;1)},
	\quad
	\Dmr
	=
	\sqrt{\Upsilon(n;0)},
	\\
	\Dqcp 
	=
	\sqrt{\Upsilon(n;-1)},
\end{gathered}
\end{gather}
where 
\begin{align}
	\Upsilon(n;x)
	\equiv&\,
	- \frac{4 \pi n}{\W_{-1}\left[- \frac{2 \pi n}{\Lambda} \exp(-x) \right]}
	\nonumber\\
	\simeq&\,
	\frac{4 \pi n}{\ln\left(\frac{\Lambda}{2 \pi n} \right) + x + \ln\left[\ln\left(\frac{\Lambda}{2 \pi n}\right)+x\right]}.
	\nonumber
\end{align}
Here, $\W_{-1}(z)$ is the $k = -1$ branch of Lambert's W function.

The tunneling density of states measured at a tip potential $V$ is given by\cite{Schrieffer} 
\begin{align}\label{TDOSDef}
	\nu(V) 
	\equiv&\,
	\int \frac{d^2 \vex{k}}{(2 \pi)^2}
	|u_{\vex{k}}|^2
	\delta(E_k - V)
	\nonumber\\
	=&\,
	\frac{1}{8 \pi V}
	\int d \e
	\left(\frac{\e}{2} - \mu + V\right)
	\delta[E(\e;\Do) - V],
\end{align}
where $u_{\vex{k}} = \frac{1}{\sqrt{2}}\sqrt{1+(k^2/2 - \mu)/E_k}$ is a coherence factor.
Performing the integration, one obtains
\begin{align}\label{TDOS}
	\nu(V)
	=&\,
	\frac{1}{2 \pi}
	\left[
	\frac{V - (\Do)^2}{\sqrt{V^2 - \Vmin^2}}
	\right]
	\theta(V - \Vmin)\,
	\theta(\mu - V)\,
	\nonumber\\
	&\, 
	\times
	\theta\left(\Dcoh - \Do\right)
	\nonumber\\
	&\,+
	\frac{1}{4 \pi}
	\theta(V  - |\mu|)
	\left[
	1
	+
	\frac{V - (\Do)^2}{\sqrt{V^2 - \Vmin^2}}
	\right].
\end{align}
In this equation, $\theta(\e)$ denotes the unit step function.
The first term in Eq.~(\ref{TDOS}) is non-zero only for weak pairing
strengths such that $\Do \leq \Dcoh$. In this range, the single particle
excitation gap is 
[Eq.~(\ref{EminDef--WP})] 
\begin{align}
	\Vmin 
	=&\,
	\Do\sqrt{2\mu - (\Do)^2},
\end{align}
and $\nu(V)$ exhibits a coherence peak above this energy, as shown
in Fig.~\ref{Fig--Roots and TDOS}(b). 
For $\Do > \Dcoh$, only the second term in Eq.~(\ref{TDOS}) contributes.
The minimum of $E(\e)$ occurs at $\e = 0$, where the single particle
excitation gap is $|\mu|$
[Eq.~(\ref{EminDef--SP})].
This is non-zero on both sides of the BCS-BEC
transition. On the BCS side ($\Dcoh < \Do < \Dqcp$), $\nu(V)$ in 
Eq.~(\ref{TDOS}) vanishes continuously at $V = \mu$; on the BEC side ($\Do > \Dqcp$),
there is a discontinuous jump, see Fig.~\ref{Fig--Roots and TDOS}(b). 
The difference is a coherence factor effect due to $|u_{\vex{k}}|^2$ in Eq.~(\ref{TDOSDef}).

\section{Classical dynamics in the chiral p-wave model \label{Sec: APP--ChiralP}}

In this Appendix, we establish the equivalence of dynamics generated from a $p + i p$ initial state
using the ``real'' p-wave Hamiltonian in Eq.~(\ref{Hactual}) and the chiral one in Eq.~(\ref{Hchiral}).
For the Hamiltonian in Eq.~(\ref{Hactual}), the equations of motion are
\begin{align}\label{EOM-real}
\begin{aligned}
	\dot{s}_{\vex{k}}^z =&\, \frac{i}{2} \left[B_{\vex{k}} s^+_{\vex{k}} - B^*_{\vex{k}} s^-_{\vex{k}} \right],
	\\
	\dot{s}_{\vex{k}}^- =&\, - i \left[k^2 s^-_{\vex{k}} + B_{\vex{k}} s_{\vex{k}}^z \right],
\end{aligned}
\end{align}
where
\begin{align}
	B_{\vex{k}} 
	\equiv 
	4 G	
	\sum'_{\vex{q}} \vex{k} \cdot \vex{q} \, s_{\vex{q}}^-.
\end{align}
For a time-dependent $p + i p$ state, we can write
\begin{align}
	s_{\vex{k}}^- \equiv e^{-i \phi_k} s_{k}^-, 
	\;\;
	s_{\vex{k}}^z \equiv s_{k}^z.
\end{align}
Eq.~(\ref{EOM-real}) becomes
\begin{align}\label{EOM-realp+ip}
\begin{aligned}
	\dot{s}_{k}^z 
	=&\, 
	i G 	
	\sum'_{\vex{q}} k  q 
	\left[s_{q}^- s^+_{k} - s^-_{k} s_{q}^+ \right],
	\\
	\dot{s}_{k}^- 
	=&\, 
	-
	i \left( 
	k^2 s^-_{k} 
	+ 
	2 G 	
	\sum'_{\vex{q}} k q \,
	s_{q}^-
	s_{k}^z 
	\right),
\end{aligned}
\end{align}
where we have used the fact that
\[
	\sum_{\vex{q}}' \exp(-2 i \phi_q) 
	\rightarrow
	\dos \int d \e 
	\int_0^{\pi} \frac{d \phi_q}{\pi}
	\exp(-2 i \phi_q)
	=
	0.
\]
In this last equation, we convert to the continuum via Eq.~(\ref{ContDisc}).
Defining $\vec{s}_i \equiv \vec{s}_{k_i}$, 
Eq.~(\ref{EOM-realp+ip}) takes the form
\begin{align}
\begin{aligned}
	\dot{s}^z_i 
	=&\, 
	i 
	\sqrt{\e_i} 
	\left[
	\Delta^{*}
	s^-_i
	-	
	\Delta
	s^+_i
	\right],
	\\
	\dot{s}^-_i
	=&\, 
	i 
	\left\{
	2
	\sqrt{\e_i}
	s^z_i 
	\Delta
	- 
	\e_i s^-_i 
	\right\},
	\\
	\Delta
	\equiv&\,
	- 
	G	
	\sum_j 
	\sqrt{\e_j}
	s_j^-.
\end{aligned}
\end{align}
These are identical to Eq.~(\ref{spinEOM}).

\section{Phase III dynamics for negative real roots \label{Sec: APP--IIIDyn}}

In Sec.~\ref{Sec: PhaseIIIGap}, we computed the asymptotic dynamics for 
$\Delta(t)$ through the bulk of phase {\bf III}.
Eqs.~(\ref{yDefs})--(\ref{period}) give the evolution of the 
squared
modulus $\Rhot(t) \equiv |\Delta(t)|^2$
everywhere in {\bf III} to the left of the line marked $\bmr$ in Fig.~\ref{Fig--PhaseDiagFull}.
All coefficients are determined by the two pairs of isolated roots $u_{1,\pm}$ and $u_{2,\pm}$,
which come in complex conjugate pairs [Eqs.~(\ref{u1pmDef}) and (\ref{u2pmDef})]. 
For a given quench $\{\Di,\beta\}$, these solve Eq.~(\ref{IsoRootsQuench}).

These results do not apply to a very narrow phase {\bf III} sliver of width 
$\sim \sqrt{n}[\log(\Lambda/2 \pi n)]^{-5/2} \ll \{\Dcoh,\Dmr,\Dqcp\}$
in Fig.~\ref{Fig--PhaseDiagFull}.
This is the region bounded on the left (right) by the $\bmr$ ($\bcm$) curve.
Within this sliver, the roots $u_{2,\pm}$ are non-degenerate, negative, and real.
Quenches in phases {\bf II} and {\bf III} 
between the lines marked $\bmr$ and $\bqcp$ in Fig.~\ref{Fig--PhaseDiagFull}
are non-equilibrium versions of the BCS ground state with $\Dmr < \Do < \Dqcp$.
The corresponding root configurations lie between those marked (3) and (4) in Fig.~\ref{Fig--Roots and TDOS}(a).

In this Appendix, we transcribe the 
order parameter
dynamics for quenches in this sliver.
Instead of Eq.~(\ref{u2pmDef}), the second isolated pair is
\[
	u_{2,\{a,b\}} \equiv - v_{2,\{a,b\}} < 0.
\]
Eqs.~(\ref{RhotEOMF}) and (\ref{RhotParams}) are replaced by
\bsub
\begin{align}\label{RhotEOMF-sliv}
	\dot{\Rhot}^2
	=
	(\Rhot_{+} - \Rhot)(\Rhot - \Rhot_{-})
	\left[(\Rhot - \Rhot_{\rr})^2 + \Rhot_{\ii}^2\right],
\end{align}
\begin{align}
	\label{RhotParams-sliv}
\begin{aligned}
	\Rhot_{\pm}
	\equiv&\,
	\frac{1}{4} 
	\left[
	\sqrt{v_{2,b}} + \sqrt{v_{2,a}}
	\pm 
	\sqrt{2\left(|u_1| - u_{1,\rr}\right)} \,
	\right]^2,
	\\
	\Rhot_{\rr} 
	+ 	
	i
	\Rhot_{\ii} 	
	\equiv&\,
	\frac{1}{4} 
	\left[
	\sqrt{v_{2,b}} - \sqrt{v_{2,a}}
	+
	i
	\sqrt{2\left(|u_1| + u_{1,\rr}\right)} \,
	\right]^2.
\end{aligned}
\end{align}
\esub
The solution is
\begin{align}\label{zDefs-sliv}
	\Rhot(t) 
	\equiv
	\Rhot_{0} + \loa \, z(t),
\end{align}
\begin{align}
	z(t)
	=&\,
	\frac{
	\left(1+z_0^2\right) 
	\left\{
\begin{aligned}
	&
	z_{c,\rr} 
	\left[
	1 + \cn^2(\theta t | N) 
	\right]
	\\&\,
	+
	\left( 1 + |z_{c}|^2 \right)
	\cn(\theta t | N)
\end{aligned}
	\right\}
	}{
	\left\{
\begin{aligned}
	&\,
	(1 + z_0 z_{c,\ii})^2
	+
	2 z_{c,\rr} \left(1 + z_0^2\right)  
	\cn(\theta t | N) 
	\\&\,
	+
	z_0^2 z_{c,\rr}^2 
	+
	\left[ (z_0 - z_{c,\ii})^2 + z_{c,\rr}^2 \right]
	\cn^2(\theta t | N)
\end{aligned}
	\right\}
	},
\end{align}
where $z_{c} \equiv z_{c,\rr} + i z_{c,\ii}$.
In terms of the roots,
\begin{align}
\begin{aligned}
	\Rhot_0
	=&\,
	\frac{1}{4} 
	\left[
	2 
	\left(
	|u_1|
	-
	u_{1,\rr} 
	\right)
	+
	\left(\sqrt{v_{2,a}}+\sqrt{v_{2,b}}\right)^2
	\right],
	\\
	\loa
	=&\,
	\frac{1}{\sqrt{2}}
	\sqrt{|u_1| - u_{1,\rr}} \left(\sqrt{v_{2,a}}+\sqrt{v_{2,b}}\right),
	\\
	z_{c,\rr}
	=&\,
	-\frac{\sqrt{2} \left(|u_1| +\sqrt{v_{2,a}} \sqrt{v_{2,b}}\right)}{\sqrt{|u_1| -u_{1,\rr} } \left(\sqrt{v_{2,a}}+\sqrt{v_{2,b}}\right)},
	\\
	z_{c,\ii}
	=&\,
	\frac{\sqrt{|u_1| + u_{1,\rr}} \left(\sqrt{v_{2,b}}-\sqrt{v_{2,a}}\right)}{\sqrt{|u_1| - u_{1,\rr}} \left(\sqrt{v_{2,b}}+\sqrt{v_{2,a}}\right)},
	\\
	z_0
	=&\,
	\frac{
	\left\{
	\begin{aligned}
	&\,
	u_{1,\ii}^2+(u_{1,\rr}+v_{2,a}) (u_{1,\rr}+v_{2,b})
	\\&\,
	+
	\sqrt{
	\begin{aligned}
		&\,
		\left[u_{1,\ii}^2+(u_{1,\rr}+v_{2,a})^2\right] 
		\\&\,\times
		\left[u_{1,\ii}^2+(u_{1,\rr}+v_{2,b})^2\right]
	\end{aligned}
	}
	\end{aligned}
	\right\}
	}{
	u_{1,\ii} (v_{2,b}-v_{2,a})
	},
	\\
	N
	=&\,
	\frac{1}{1+z_0^2},
	\\
	\theta
	=&\,
	\left\{
	\begin{aligned}
	&\,
	\left[u_{1,\ii}^2+(u_{1,\rr}+v_{2,a})^2\right] 
	\\&\,\times
	\left[u_{1,\ii}^2+(u_{1,\rr}+v_{2,b})^2\right]
	\end{aligned}
	\right\}^{1/4}.
\end{aligned}
\end{align}
The physical period $T$ of $\Rhot(t)$ is
\begin{align}\label{period-sliv}
	T 
	=
	\frac{4 K(N)}{\theta},
\end{align}
where $K(N)$ is the complete elliptic integral of the first kind (and $N = k^2$).

\section{Green's functions \label{Sec: APP--GFs}}

In this Appendix, we compute single particle Green's functions in the long time limit
for quenches in phase {\bf II} of the diagram in Fig.~\ref{Fig--PhaseDiagBasic}.
Throughout this Appendix, we work in the rotating frame employed in Sec.~\ref{Sec: WindingObs}
such that the 
order parameter
itself (and not only its modulus) asymptotes to a constant.

\subsection{ Post-quench coherence factors }

In phase {\bf II}, the asymptotic spin configuration is given by Eq.~(\ref{GappedPrec}), where
$\gamma_i = \gamma(\e_i)$ in Eq.~(\ref{gammaFinalv2}).
Following the discussion surrounding Eq.~(\ref{PolarRescale}), the spins in the 2D $\vex{k}$-plane
evolving from an initial $p+ip$ state are reconstructed as follows:  
\begin{align}\label{SpinDist}
	2 s^-_{\vex{k}}(t) 
	=&\, 
	\left\{
	\begin{aligned}
	&
	\sqrt{1 - \gamma_k^2} 
	\left[
	\left(\frac{\xi_k}{E_k}\right)
	\cos\left(2 E_k t\right)
	-
	i
	\sin\left(2 E_k t\right)
	\right]
	\\&
	\; +
	\gamma_k
	\left(\frac{k \Dasy}{E_k}\right) 
\end{aligned}
	\right\}
	\nonumber\\
	&\,
	\times
	\exp\left(- i \phi_k \right),
	\nonumber\\
	2 s^z_{\vex{k}}(t) 
	=&\, 
	- 
	\sqrt{1 - \gamma_k^2} 
	\left(\frac{k \Dasy}{E_k}\right)
	\cos\left(2 E_k t\right) 
	+ 
	\gamma_k
	\left(\frac{\xi_k}{E_k}\right),
	\nonumber\\
	\xi_k 
	\equiv&\, 
	{\textstyle{\left(\frac{k^2}{2} - \masy \right)}}, \;\;\;
	E_k 
	\equiv
	\sqrt{\xi_k^2 + k^2 \Dasy^2},
\end{align}
where
$\phi_k$ is the polar angle in momentum space, and $\Dasy$ and $\masy$ refer to the post-quench steady-state (\emph{not} ground state)
values. 
Comparing to Eq.~(\ref{GappedPrec}), we have set the phase shifts $\Theta_i = 0$,
see Sec.~\ref{Sec: SSApproach}.
The pre-quench parameters $\Di$ and $\mui$ enter through the distribution function
\begin{align}
	\gamma_k 
	\equiv&\, 
	\gamma(\e = k^2),
\end{align}
the latter evaluated in Eq.~(\ref{gammaFinalv2}).

In the thermodynamic limit, the many-body wavefunction assumes a BCS product form with
time-dependent coherence factors,
\begin{align}\label{Psi}
	\ket{\Psi(t)}
	=
	\prod_{\vex{k}}'
	\left[
	u_{\vex{k}}(t)
	+
	v_{\vex{k}}(t) 
	\,
	s^{+}_{\vex{k}}
	\right]
	\ket{0},
\end{align}
where $\ket{0}$ is the vacuum (all pseudospins down).
In this state, the expectations of Anderson pseudospin
Schr\"odinger picture operators are given by
\begin{align}\label{SpinExps}
\begin{aligned}
	\bra{\Psi(t)} s^+_{\vex{k}} \ket{\Psi(t)}
	=&\,
	v^*_{\vex{k}} u_{\vex{k}}(t),
	\\
	\bra{\Psi(t)} s^-_{\vex{k}} \ket{\Psi(t)}
	=&\,
	u^*_{\vex{k}} v_{\vex{k}}(t),
	\\
	\bra{\Psi(t)} s^z_{\vex{k}} \ket{\Psi(t)}
	=&\,
	{\textstyle{\frac{1}{2}}}
	\left(|v_{\vex{k}}|^2 - |u_{\vex{k}}|^2\right)(t).
\end{aligned}
\end{align}

The coherence factors solve the same Bogoliubov-de Gennes equation
as the retarded Green's function $\G_{\vex{k}}(t,t')$ 
[Eq.~(\ref{Gasy})]. In the large time limit,
\begin{align}\label{Cohasy}
	i \frac{d }{d t}
	\begin{bmatrix}
	u_{\vex{k}}(t) \\
	v_{\vex{k}}(t)
	\end{bmatrix}
	=&\,
	\begin{bmatrix}
	- \xi_k & \msf{k}^*\Dasy \\ 
	\msf{k} \Dasy & \xi_k 
	\end{bmatrix}
	\begin{bmatrix}
	u_{\vex{k}}(t) \\
	v_{\vex{k}}(t)
	\end{bmatrix},
\end{align}
where $\msf{k} \equiv k^x - i k^y$.
Different from $\G_{\vex{k}}(t,t')$, the coherence factors ``remember'' details
of the pre-quench state through the initial condition at $t = 0$. 
The general solution to Eq.~(\ref{Cohasy}) is
\begin{align}\label{BogAmp}
\begin{aligned}
	u_{\vex{k}}(t)
	=&\,
	(E_k - \xi_k)
	\,
	A_k
	e^{-i E_k t}
	+
	(E_k + \xi_k)
	\,
	B_k
	e^{i E_k t},
	\\
	v_{\vex{k}}(t)
	=&\,
	\Big[
	A_k
	e^{-i E_k t}
	-
	B_k
	e^{i E_k t}
	\Big]
	\Dasy k
	\,
	e^{-i \phi_k},
\end{aligned}
\end{align}
where  
the undetermined complex constants $A_k$ and $B_k$ satisfy
\begin{align}
	1
	=&\,
	\left[(E_k - \xi_k)^2 + k^2 \Dasy^2 \right] |A_k|^2 
	\nonumber\\
	&\,
	+ 
	\left[(E_k + \xi_k)^2 + k^2 \Dasy^2 \right] |B_k|^2.
\end{align}
For the p-wave problem, one coherence factor must have odd parity;
we have chosen $v_{\vex{k}}$ in Eq.~(\ref{BogAmp}).

Using Eq.~(\ref{BogAmp}), one can compute the expectation values 
in Eq.~(\ref{SpinExps}). Comparing the results to that of the Lax calculation in 
Eq.~(\ref{SpinDist}), we determine that 
\begin{align}\label{BogAmpFinal}
\begin{aligned}
	u_{\vex{k}}(t)
	=&\,
	\frac{1}{2}
	\sqrt{\frac{(1+\gamma_k)\left(E_k - \xi_k \right)}{E_k}}
	\,
	e^{-i E_k t + i \Gamma_k}
	\\&\,
	+
	\frac{1}{2}
	\sqrt{\frac{(1 - \gamma_k)\left(E_k + \xi_k\right)}{E_k}}
	\,
	e^{i E_k t + i \Gamma_k},
	\\
	v_{\vex{k}}(t)
	=&\,
	\frac{1}{2}
	\sqrt{\frac{(1+\gamma_k) \left( E_k + \xi_k \right)}{E_k}}
	\,
	e^{-i E_k t -i \phi_k + i \Gamma_k }
	\\&\,
	-
	\frac{1}{2}
	\sqrt{\frac{(1 - \gamma_k)\left(E_k - \xi_k \right)}{E_k}}
	\,
	e^{i E_k t -i \phi_k + i \Gamma_k},
\end{aligned}
\end{align}
where $\Gamma_k$ is an undetermined time-independent phase.

The ground state (zero quench) has $\gamma_k = -1$ for all $k$, leading to 
$
	u_{\vex{k}}(t)
	=
	\sqrt{
	\frac{1}{2}
	\left(
	1
	+
	\xi_k/E_k
	\right)
	}
	\,
	\exp(i E_k t + i \Gamma_k)
$
and
$
	v_{\vex{k}}(t)
	=
	-
	\sqrt{
	\frac{1}{2}
	\left(
	1
	-
	\xi_k/E_k
	\right)
	}
	\,
	\exp(i E_k t -i \phi_k + i \Gamma_k),
$
as expected.

\subsection{One particle Green's functions and structure factors}

The dynamic single particle Green's functions can be computed from the coherence factors.
For example,
\begin{align}\label{GFDyn--1}
	\G_{\vex{k},>}&(t,t')
	\equiv	
	-i 
	\bra{\Psi(t)}
	c_{\vex{k}}^{\phantom{\dagger}}
	e^{-i H (t - t')}
	c_{\vex{k}}^\dagger
	\ket{\Psi(t')}
	\nonumber\\
	=&\,	
	-i 
	u_{\vex{k}}^*(t)
	u_{\vex{k}}(t')
	\nonumber\\
	&\,
	\phantom{-}
	\times\!
	\bra{0}
	c_{\vex{k}}
	\prod_{\vex{q}\neq\vex{k}}'
	\left[
	u_{\vex{q}}^*(t)
	+
	v_{\vex{q}}^*(t) 
	\,
	c_{-\vex{q}}
	c_{\vex{q}}
	\right]
	e^{-i H_{\vex{k}} t}
	\nonumber\\
	&\,
	\phantom{-}
	\times\!
	e^{i H_{\vex{k}} t'}	
	\prod_{\vex{q'}\neq\vex{k}}'
	\left[
	u_{\vex{q'}}(t')
	+
	v_{\vex{q'}}(t') 
	\,
	c^\dagger_{\vex{q'}}
	c^\dagger_{-\vex{q'}}
	\right]
	c_{\vex{k}}^\dagger
	\ket{0},
\end{align}
where $H_{\vex{k}}$ is the interacting (and time-independent) BCS Hamiltonian in Eq.~(\ref{Hchiral}) 
\emph{excluding the mode} $\{\vex{k},-\vex{k}\}$:
\begin{align}\label{HChiralP-wave}
	H_{\vex{k}} = \sum_{\vex{q} \neq \vex{k}}' q^2 s_{\vex{q}}^z 
	- 
	G 
	\sum_{\vex{q_1},\vex{q}_2 \neq \vex{k}}' 
	\msf{q}_1 \msf{q}^{*}_2 \,
	s_{\vex{q}_1}^{+} s_{\vex{q}_2}^{-}.
\end{align}
Above we have used the fact that $e^{i k^2 s^z_{\vex{k}}} c_{\vex{k}}^\dagger \ket{0} = c_{\vex{k}}^\dagger \ket{0}$.
Eq.~(\ref{GFDyn--1}) becomes
\begin{align}
	\G_{\vex{k},>}(t,t')
	=&\,	
	-i 
	u_{\vex{k}}^*(t)
	u_{\vex{k}}(t')
	\nonumber\\
	&\,
	\phantom{-}
	\times
	\bra{0}
	c_{\vex{k}}
	\prod_{\vex{q}\neq\vex{k}}'
	\left[
	u_{\vex{q}}^*(0)
	+
	v_{\vex{q}}^*(0) 
	\,
	c_{-\vex{q}}
	c_{\vex{q}}
	\right]
	\nonumber\\
	&\,
	\phantom{-}
	\times
	\prod_{\vex{q'}\neq\vex{k}}'
	\left[
	u_{\vex{q'}}(0)
	+
	v_{\vex{q'}}(0) 
	\,
	c^\dagger_{\vex{q'}}
	c^\dagger_{-\vex{q'}}
	\right]
	c_{\vex{k}}^\dagger
	\ket{0}
	\nonumber\\
	=&\,	
	-i 
	u_{\vex{k}}^*(t)
	u_{\vex{k}}(t').
\end{align}
In these manipulations, 
we have used the fact that in mean field theory, the many-body BCS state 
can always be expressed as product over modes of either 
a coherent admixture of empty and doubly-occupied levels, or alternatively a singly-occupied (``blocked'') level. 

We thereby obtain the following Green's functions
\bsub\label{AllGFS}
\begin{align}
	i \G_{\vex{k},>}&(t,t')
	=	
	\bra{\Psi_i}
	c^{\phantom{\dagger}}_{\vex{k}}(t)
	\,
	c^\dagger_{\vex{k}}(t')
	\ket{\Psi_i}
	\nonumber\\
	=&\,
	\left(\frac{E_k - \xi_k \gamma_k}{2 E_k}\right)
	\,
	\cos\left[E_k (t - t')\right]
	\nonumber\\&
	+
	i 
	\left(\frac{E_k \gamma_k - \xi_k}{2 E_k}\right)
	\,
	\sin\left[E_k (t - t')\right]
	\nonumber\\&
	+
	\frac{k \Dasy}{2 E_k}
	\sqrt{1 - \gamma_k^2}
	\,
	\cos\left[E_k (t + t')\right],
\end{align}
\begin{align}
	-i \G_{\vex{k},<}&(t,t')
	=	
	\bra{\Psi_i}
	c^\dagger_{\vex{k}}(t') 
	\,
	c^{\phantom{\dagger}}_{\vex{k}}(t)
	\ket{\Psi_i}
	\nonumber\\
	=&\,
	v_{\vex{k}}^*(t') 
	\,
	v_{\vex{k}}(t)
	\nonumber\\
	=&\,
	\left(
	\frac{E_k + \xi_k \gamma_k}{2 E_k}
	\right)
	\,
	\cos\left[E_k (t - t')\right]
	\nonumber\\&\,
	-
	i
	\left(
	\frac{E_k  \gamma_k + \xi_k}{2 E_k}
	\right)
	\,
	\sin\left[E_k (t - t')\right]
	\nonumber\\&\,
	-
	\frac{k \Dasy}{2 E_k}
	\sqrt{1 - \gamma_k^2}
	\,
	\cos\left[E_k (t + t')\right],
\end{align}
\begin{align}
	\G_{\vex{k}}^+&(t,t')
	=	
	\bra{\Psi_i}
	c^\dagger_{\vex{k}}(t) 
	\,
	c^\dagger_{-\vex{k}}(t') 
	\ket{\Psi_i}
	\nonumber\\
	& =	
	v_{\vex{k}}^*(t) 
	\,
	u_{\vex{k}}(t')
	\nonumber\\
	& =	
	\left\{
	\begin{aligned}
	&
	\frac{\xi_k}{2 E_k}
	\sqrt{1 - \gamma_k^2}
	\,
	\cos\left[E_k (t+t')\right]
	\\&
	+
	\gamma_k
	\frac{k \Dasy}{2 E_k}
	\,
	\cos\left[E_k (t - t')\right]
	\\&
	+
	\frac{i}{2}
	\sqrt{1 - \gamma_k^2}
	\,
	\sin\left[E_k (t+t')\right]
	\\&
	+
	i
	\frac{k \Dasy}{2 E_k}
	\,
	\sin\left[E_k (t - t')\right]
	\end{aligned}
	\right\}
	e^{i \phi_k},
\end{align}
\begin{align}
	\G_{\vex{k}}^-(t,t')
	=&\,	
	\bra{\Psi_i}
	c_{-\vex{k}}(t) 
	\,
	c_{\vex{k}}(t') 
	\ket{\Psi_i}
	\nonumber\\
	=&\,	
	u_{\vex{k}}^{*}(t)
	\,
	v_{\vex{k}}(t')
	=
	\left[\G_{\vex{k}}^+(t',t)\right]^*.
\end{align}
\esub
In these equations, $\ket{\Psi_i}$ denotes the initial pre-quench BCS, BEC, or quantum
critical state.

Using these results, the retarded Green's function in Eq.~(\ref{GDef})
evaluates to
\bsub\label{GEval}
\begin{align}
	\G_{\vex{k}}(t,t')
	=
	-i
	\begin{bmatrix}
	\G_{\vex{k}}^{\pup{1}}(t - t')
	&
	\G_{\vex{k}}^{\pup{2}}(t - t')
	\\
	-\left[\G_{\vex{k}}^{\pup{2}}(t - t')\right]^{*}
	&
	\left[\G_{\vex{k}}^{\pup{1}}(t - t')\right]^{*}
	\end{bmatrix}
	\,
	\theta(t - t'),
\end{align}
where
\begin{align}
\begin{aligned}
	\G_{\vex{k}}^{\pup{1}}(t)
	\equiv&\,
	\cos\left(E_k t \right)
	+
	i 
	\left(\frac{\xi_k}{E_k}\right)
	\,
	\sin\left(E_k t \right),
	\\
	\G_{\vex{k}}^{\pup{2}}(t)
	\equiv&\,
	-
	i
	\left[\frac{(k^x + i k^y)\Dasy}{E_k}\right]
	\sin\left(E_k t\right).
\end{aligned}
\end{align}
\esub
The retarded function in Eq.~(\ref{GEval}) is a function only of the time difference $(t - t')$, and is 
independent of $\gamma_k$. It satisfies Eq.~(\ref{Gasy}) with the initial condition in Eq.~(\ref{GasyIC}).
By contrast, the other Green's functions in Eq.~(\ref{AllGFS}) depend upon both the relative
and average $(t + t')$ times, and upon the non-thermal Cooper pair distribution function $\gamma_k$.


\begin{thebibliography}{99}
\bibitem{TISC}
	M. Z. Hasan and C. L. Kane, 
	Rev. Mod. Phys. {\bf 82}, 3045 (2010);
	X.-L. Qi and S.-C. Zhang, 
	\textit{ibid.} {\bf 83}, 1057 (2011).
\bibitem{TopClassesDirty}
	A. P. Schnyder, S. Ryu, A. Furusaki, and A. W. W. Ludwig,
	Phys. Rev. B {\bf 78}, 195125 (2008);
	A. Kitaev,
	AIP Conf. Proc. No. 1134 
	(AIP, New York, 2009), p. 22.
\bibitem{Bloch02}
	M. Greiner, O. Mandel, T. W. Hansch, and I. Bloch,
	Nature {\bf 419}, 51 (2002).
\bibitem{Weiss06}
	T. Kinoshita, T. Wenger, and D. S. Weiss,
	Nature {\bf 440}, 900 (2006).
\bibitem{Stamper-Kurn06}
	L. E. Sadler, J. M. Higbie, S. R. Leslie, M. Vengalattore, and D. M. Stamper-Kurn,
	Nature {\bf 443}, 312 (2006).
\bibitem{Barankov04}
	R. A. Barankov, L. S. Levitov, and B. Z. Spivak,
	Phys. Rev. Lett. {\bf 93}, 160401 (2004);
	R. A. Barankov and L. S. Levitov,
	Phys. Rev. A {\bf 73}, 033614 (2006).
\bibitem{amin}  
	M. H. S. Amin, E. V. Bezuglyi, A. S. Kijko, A. N. Omelyanchouk,  
	Low Temp. Phys. {\bf 30}, 661 (2004).
\bibitem{simons} 
	M. H. Szymanska, B. D. Simons, and K. Burnett,  
	Phys. Rev. Lett. {\bf 94} , 170402 (2005).
\bibitem{WarnerLeggett05}
	G. L. Warner and A. J. Leggett,
	Phys. Rev. B {\bf 71}, 134514 (2005).
\bibitem{YuzbashyanAltshuler05}
	E. A. Yuzbashyan, B. L. Altshuler, V. B. Kuznetsov, and V. Z. Enolskii,
	Phys. Rev. B {\bf 72}, 220503(R) (2005);
	J. Phys. A {\bf 38}, 7831 (2005).
\bibitem{YKA05}
	E. A. Yuzbashyan, V. B. Kuznetsov, and B. L. Altshuler,
	Phys. Rev. B {\bf 72}, 144524 (2005).
\bibitem{YuzbashyanAltshuler06}
	E. A. Yuzbashyan, O. Tsyplyatyev, and B. L. Altshuler,
	Phys. Rev. Lett. {\bf 96}, 097005 (2006).
\bibitem{BarankovLevitov06}
	R. A. Barankov and L. S. Levitov,
	Phys. Rev. Lett. {\bf 96}, 230403 (2006).
\bibitem{DzeroYuzbashyan06}
	E. A. Yuzbashyan and M. Dzero,
	Phys. Rev. Lett. {\bf 96}, 230404 (2006).
\bibitem{Chien2010}
	C.-C. Chien and B. Damski,
	Phys. Rev. A {\bf 82}, 063616 (2010).
\bibitem{Rigol}
	M. Rigol, V. Dunjko, V. Yurovsky, and M. Olshanii,
	Phys. Rev. Lett. {\bf 98}, 050405 (2007);
	M. Rigol, V. Dunjko, and M. Olshanii, 
	Nature (London) {\bf 452}, 854 (2008).
\bibitem{CardyCalabrese06}
	P. Calabrese and J. Cardy,
	Phys. Rev. Lett. {\bf 96}, 136801 (2006);
	J. Stat. Mech. P06008 (2007).
\bibitem{Kollath07}
	C. Kollath, A. M. L\"auchli, and E. Altman,
	Phys. Rev. Lett. {\bf 98}, 180601 (2007).
\bibitem{Cincio07}
	L. Cincio, J. Dziarmaga, M. M. Rams, and W. H. Zurek,
	Phys. Rev. A {\bf 75}, 052321 (2007).
\bibitem{Polk11}
	A. Polkovnikov, K. Sengupta, A. Silva, and M. Vengalattore, 
	Rev. Mod. Phys. {\bf 83}, 863 (2011). 
\bibitem{BlochRMP08}
	I. Bloch, J. Dalibard, and W. Zwerger, 
	Rev. Mod. Phys. {\bf 80}, 885 (2008).
\bibitem{galperin} 
	Yu. M. Gal'perin, V. I. Kozub, and B. Z. Spivak, 
	Zh. Eksp. Teor. Fiz. {\bf 81}, 2118 (1981)
	[Sov. Phys. JETP {\bf 54}, 1126 (1981)].
\bibitem{shumeiko} 
	V. S. Shumeiko, 
	{\it Dynamics of electronic system with off-diagonal order parameter 
	and non-linear resonant phenomena in superconductors}, Doctoral Thesis, Kharkov, 1990.
\bibitem{Volovik}
	G. E. Volovik, Zh. Eksp. Teor. Fiz {\bf 94}, 123 (1988), 
	[Sov. Phys. JETP {\bf 67}, 1804 (1988)];
	\textit{The Universe in a Helium Droplet} 
	(Oxford University Press, Oxford, 2003).
\bibitem{ReadGreen2000}
	N. Read and D. Green,
	Phys. Rev. B {\bf 61}, 10267 (2000).
\bibitem{GurarieFB-1}
	V. Gurarie, L. Radzihovsky, and A. V. Andreev, 
	Phys. Rev. Lett. {\bf 94}, 230403 (2005).
\bibitem{GurarieFB-2}
	V. Gurarie and L. Radzihovsky,
	Ann. Phys. {\bf 322}, 2 (2009).
\bibitem{Gaebler07}
	J. P. Gaebler, J. T. Stewart, J. L. Bohn, and D. S. Jin,
	Phys. Rev. Lett. {\bf 98}, 200403 (2007).
\bibitem{Fuchs08}
	J. Fuchs, C. Ticknor, P. Dyke, G. Veeravalli, E. Kuhnle,
	W. Rowlands, P. Hannaford, and C. J. Vale, 
	Phys. Rev. A {\bf 77}, 053616 (2008).
\bibitem{Inada08}
	Y. Inada, M. Horikoshi, S. Nakajima, M. Kuwata-Gonokami, 
	M. Ueda, and T. Mukaiyama, 
	Phys. Rev. Lett. {\bf 101}, 100401 (2008).
\bibitem{Jone08}
	M. Jona-Lasinio, L. Pricoupenko, and Y. Castin, 
	Phys. Rev. A {\bf 77}, 043611 (2008).
\bibitem{LCG08}
	J. Levinsen, N. R. Cooper, and V. Gurarie, 
	Phys. Rev. A {\bf 78}, 063616 (2008).
\bibitem{ZhangTewari08}
	C. Zhang, S. Tewari, R. M. Lutchyn, and S. Das Sarma,
	Phys. Rev. Lett. {\bf 101}, 160401 (2008).
\bibitem{Sato09}
	M. Sato, Y. Takahashi, and S. Fujimoto,
	Phys. Rev. Lett. {\bf 103}, 020401 (2009).
\bibitem{SauSensarma11}
	J. D. Sau, R. Sensarma, S. Powell, I. B. Spielman, and S. Das Sarma,
	Phys. Rev. B {\bf 83}, 140510(R) (2011).
\bibitem{Zhu11}
	S.-L. Zhu, L.-B. Shao, Z. D. Wang, and L.-M. Duan,
	Phys. Rev. Lett. {\bf 106}, 100404 (2011).
\bibitem{LiuJiang12}
	X.-J. Liu, L. Jiang, H. Pu, and H. Hu,
	Phys. Rev. A {\bf 85}, 021603(R) (2012).
\bibitem{CooperShlyapnikov11}
	N. R. Cooper and G. V. Shlyapnikov,
	Phys. Rev. Lett. {\bf 103}, 155302 (2009).
\bibitem{Richardson02}
	R. W. Richardson,
	arXiv:cond-mat/0203512 (unpublished).
\bibitem{Skrypnyk09}
	T. Skrypnyk, 
	J. Math. Phys. {\bf 50}, 033504 (2009).
\bibitem{Sierra09}
	M. Iba\~nez, J. Links, G. Sierra, and S.-Y. Zhao,
	Phys. Rev. B {\bf 79}, 180501(R) (2009).
\bibitem{Sierra10}
	C. Dunning, M. Iba˜nez, J. Links, G. Sierra, and S.-Y. Zhao,
	J. Stat. Mech. P08025 (2010).
\bibitem{Ortiz10}
	S. M. A. Rombouts, J. Dukelsky, and G. Ortiz,
	Phys. Rev. B {\bf 82}, 224510 (2010).
\bibitem{GurarieRGF}
	V. Gurarie,
	Phys. Rev. B {\bf 83}, 085426 (2011).
\bibitem{EssinGurarie11}
	A. M. Essin and V. Gurarie, 
	Phys. Rev. B {\bf 84}, 125132 (2011).
\bibitem{PwaveLett}
	M. S. Foster, V. Gurarie, M. Dzero, and E. A. Yuzbashyan,
	arXiv:1307.2256.
\bibitem{Floquet-1}
	N. H. Lindner, G. Refael, and V. Galitski, 
	Nat. Phys. {\bf 7}, 490 (2011).
\bibitem{Floquet-2}
	T. Kitagawa, T. Oka, A. Brataas, L. Fu, and E. Demler, 
	Phys. Rev. B {\bf 84}, 235108 (2011).
\bibitem{Floquet-3}
	Z. Gu, H. A. Fertig, D. P. Arovas, and A. Auerbach, 
	Phys. Rev. Lett. {\bf 107}, 216601 (2011).
\bibitem{Floquet-4}
	M. S. Rudner, N. H. Lindner, E. Berg, and M. Levin (2012), 
	arXiv:1212:3324.
\bibitem{AndBCS58}
	P. W. Anderson,
	Phys. Rev. {\bf 112}, 1900 (1958).
\bibitem{Schrieffer}
	J. R. Schrieffer,
	\textit{Theory of Superconductivity}
	(Perseus Books, Reading, Massachusettes, 1983).
\bibitem{TKNN}
	D. J. Thouless, M. Kohmoto, M. P. Nightingale, and M. den Nijs,
	Phys. Rev. Lett. {\bf 49}, 405 (1982);
	Q. Niu, D. J. Thouless, and Y.-S. Wu,
	Phys. Rev. B {\bf 31}, 3372 (1985).
\bibitem{KaneFisher}
	C. L. Kane and M. P. A. Fisher,
	Phys. Rev. B {\bf 55}, 15832 (1997).
\bibitem{Capelli02}
	A. Capelli, M. Huerta, and G. Zemba, 
	Nucl. Phys. B {\bf 636}, 568 (2002).
\bibitem{Richardson} 
	R.W. Richardson and N. Sherman, 
	{\it Nucl. Phys.} {\bf 52}, 221 (1964); 
	{\bf 52}, 253 (1964).
\bibitem{Gaudin} 
	M. Gaudin, Note CEA 1559, 1 (1972); 
	J. Phys. (Paris) 37, 1087 (1976);
	\emph{La fonction d'onde de Bethe}, (Masson, Paris, 1983).
\bibitem{Dukelsky} 
	J. Dukelsky, S. Pittel, G. Sierra, 
	Rev. Mod. Phys. {\bf 76} 643 (2004).	
\bibitem{footnote--RotFrameOrbit}
	The orbits A and B shown in Fig.~\ref{Fig--Dasy}(b) were obtained from the
	isolated roots $u_{1,\pm}$ and $u_{2,\pm}$ of the corresponding quenches,
	see Eqs.~(\ref{u1pmDef}), (\ref{u2pmDef}), and the surrounding discussion.
	For a given quench, the orbit plotted is 
	[Eq.~(\ref{DeltaPolarRT})]
	\[
		\Delta(t) 
		=
		\sqrt{\Rhot(t)}
		\exp[-i \phi(t) + 2 i \bar{\mu}_\infty t],
		\;\;
		0 \leq t \leq T.
	\]
	In this equation, $\Rhot(t)$ has the explicit form in Eqs.~(\ref{yDefs})--(\ref{ySolRootParams}),
	the period $T$ is given by Eq.~(\ref{period}),
	and the phase $\phi(t)$ is obtained by numerically integrating Eq.~(\ref{PhaseEOM}).
	Finally, the parameter $2 \bar{\mu}_\infty \equiv [\phi(T) - \phi(0)]/T$;
	the orbits in Fig.~\ref{Fig--Dasy}(b) are thus plotted in the frame rotating with
	frequency $2 \bar{\mu}_\infty$. 
\bibitem{footnote--NotAlignGND}
	In a BCS ground state (which includes the Fermi liquid as a
	special case), the Anderson pseudospins are aligned to the
	magnetic field that incorporates the chemical potential.
	The ground state configuration has
	\[
		2 s_i^a = \frac{b_i^a}{|\vec{b}_i|},
		\;\;
		\vec{b}_i = \vec{B}_i + 2 \mu \hat{z},
	\]
	with $\mu$ determined by Eq.~(\ref{muGND}).
	The equation of motion 
	in Eq.~(\ref{spinEOM})
	instead involves $\vec{B}_i$.
	Due to the mismatch, all spins precess in the ground state 
	around $\hat{z}$ with frequency $2 \mu$; similarly, $\Delta(t) = \Do \exp(-2 i \mu t)$.
	One can eliminate this spurious evolution 
	(and align the spins to the field)
	by boosting to the rotating frame
	$s_i^-(t) \rightarrow \exp(2 i \mu t) \, s_i^-(t)$.
\bibitem{Perfetto2013}
	E. Perfetto,
	Phys. Rev. Lett. {\bf 110}, 087001 (2013).
\bibitem{DzeroYuzbashyan07}
	M. Dzero, E. A. Yuzbashyan, B. L. Altshuler, and P. Coleman,
	Phys. Rev. Lett. {\bf 99}, 160402 (2007).
\bibitem{sklyanin} 
	E. K. Sklyanin,  
	J. Sov. Math. {\bf 47}, 2473 (1989); 
	Progr. Theoret. Phys. Suppl. {\bf 118}, 35 (1995).
\bibitem{vadim} 
	V. B. Kuznetsov,  
	J. Math. Phys. {\bf 33}, 3240, (1992).
\bibitem{footnote--LaxIBP}
	In fact, there is a point of stationary phase for $0 < \Dasy^2 < \masy$ which obstructs this 
	procedure. Nevertheless, the corresponding saddle point gives a term that decays as $t^{-1/2}$,
	and this can be neglected at $t = \infty$. 
	See also Sec.~\ref{Sec: SSApproach}.
\bibitem{VolkovKogan74}
	A. F. Volkov and Sh. M. Kogan,
	Zh. Eksp. Teor. Fiz. {\bf 65}, 2038 (1973)
	[JETP {\bf 38}, 1018 (1974)].
\end{thebibliography}
\end{document}